%% file: Main.tex
\documentclass[3p,12pt]{elsarticle}
\usepackage[fleqn]{amsmath}
\usepackage{cases}
\usepackage{booktabs}
\usepackage{multirow}
\usepackage{xcolor}
\usepackage[normalem]{ulem}
\usepackage{tabularx}
\usepackage{siunitx}
\usepackage[caption=false]{subfig}
\usepackage{comment}
\usepackage{threeparttable}
\usepackage[ruled, linesnumbered]{algorithm2e}
\usepackage{algpseudocode}
\usepackage{grffile}
\usepackage{rotating}
\usepackage{lineno}
\usepackage{hyperref}
\biboptions{sort&compress}

\usepackage{breqn}

\usepackage{graphicx,enumerate}
\usepackage{amssymb}
\allowdisplaybreaks[4]
\usepackage{bm,amsmath}
\usepackage[caption=false]{subfig}
\usepackage{caption,color}
\usepackage{subeqnarray}
\usepackage{multirow}
\usepackage{lscape}
\usepackage{rotating}
\usepackage{graphics}
\usepackage{epstopdf}
\journal{X} %

\newcommand{\bc}{\bm c}

\newcommand{\bn}{\bm n}

\newcommand{\bq}{\bm q}

\newcommand{\bu}{\bm u}

\newcommand{\bx}{\bm x}

\newcommand{\bxi}{\bm \xi}

\newcommand{\Kn}{\text{Kn}}

\newcommand{\myd}{\;\mathrm{d} }

\allowdisplaybreaks[4]
\begin{document}

\input{Abstract}
\input{Introduction}

\input{Governing_equations}

\input{Iteration_Framework}
\input{Convergence_rate}
\input{Asymptotic}

\input{Numerical_schemes}
\input{Testcase}
\input{Conclusions}

\appendix
\input{Appendix_sourceterm_treatment}

\fontsize{11pt}{11pt}\selectfont 
\bibliographystyle{elsarticle-num}
\bibliography{ref}
\end{document}

%% file: Abstract.tex
\begin{frontmatter}
\title{Accelerated simulation of multiscale gas-radiation coupling flows via a general synthetic iterative scheme}

\author[SUSTech]{Jianan Zeng}
\author[SYSU]{Qi Li \corref{mycorrespondingauthor}}
\ead{liqi99@mail.sysu.edu.cn}
\author[SUSTech]{Yanbing Zhang}
\author[HKUST1,HKUST2]{Wei Su}
\author[SUSTech]{Lei Wu \corref{mycorrespondingauthor}}
\ead{wul@sustech.edu.cn}

\cortext[mycorrespondingauthor]{Corresponding authors}

\address[SUSTech]{Department of Mechanics and Aerospace Engineering, Southern University of Science and Technology, Shenzhen 518055, China}
\address[SYSU]{School of Aeronautics and Astronautics, Sun Yat-sen University, Shenzhen, 518107, China}
\address[HKUST1]{Division of Emerging Interdisciplinary Areas, Academy of Interdisciplinary Studies, The Hong Kong University of Science and Technology, Hong Kong SAR, China}
\address[HKUST2]{Department of Mathematics, The Hong Kong University of Science and Technology, Hong Kong SAR, China}

\begin{abstract}
Gas-radiation coupling critically influences hypersonic reentry flows, where extreme temperatures induce pronounced non-equilibrium gas and radiative heat transport. Accurate and efficient simulation of radiative gas dynamics is therefore indispensable for reliable design of thermal protection systems for atmospheric entry vehicles. 
In this study, a Boltzmann-type kinetic model for radiative gas flows is solved across a broad spectrum of flow and radiation transport regimes using the general synthetic iterative scheme (GSIS). The approach integrates an unstructured finite-volume discrete velocity method with a set of macroscopic synthetic equations. Within this framework, the kinetic model provides high-order closures for the constitutive relations in the synthetic equations. Simultaneously, the macroscopic synthetic equations drive the evolution of the mesoscopic kinetic system, significantly accelerating steady-state convergence in near-continuum regimes, as substantiated by linear Fourier stability analysis. Crucially, the algorithm is proven to be asymptotic-preserving, correctly recovering the continuum and optically thick limits—represented by the radiative Navier--Stokes--Fourier equations governing distinct translational, rotational, vibrational, and radiative temperatures—on coarse meshes independent of the mean free path.
Numerical simulations of challenging benchmarks, including three-dimensional hypersonic flow over an Apollo reentry capsule, demonstrate that GSIS achieves orders-of-magnitude speedup over conventional iterative schemes in multiscale simulations of radiative gas flows while accurately capturing non-equilibrium effects and radiative heat transfer in hypersonic environments.

\end{abstract}


\end{frontmatter}

%% file: Introduction.tex

\section{Introduction}\label{sec:introduction}

Thermal non-equilibrium effects are encountered in multiple engineering disciplines, notably aerospace engineering~\cite{ivanov1998computational, kustova2011comparison} and inertial confinement fusion~\cite{ICF_radiation}. During atmospheric re-entry~\cite{votta2013hypersonic}, forebody compression generates extreme thermodynamic conditions that precipitate intense thermal radiation via molecular electronic and vibrational transitions~\cite{le2011investigation,cruden2014measurement}. In the afterbody, rapid expansion induces pronounced thermal non-equilibrium between vibrational/electronic and translational modes due to insufficient collisional relaxation~\cite{kelly2021emission,brandis2023new}, introducing substantial uncertainty in radiative flux predictions~\cite{johnston2015features,lino2011contribution}. For example, for hypersonic nitrogen expansion flows, predicted spectral intensities exceed measurements by factors of three to four~\cite{zhang2025nitrogen,xu2025chemiluminescence}. Meanwhile, Mars 2020 mission data reveal that afterbody radiative heat fluxes are comparable to convective contributions~\cite{brandis2020radiative,tibere2023examination}. Accurate modeling of non-equilibrium radiative processes is therefore essential for thermal protection system design. In inertial confinement fusion, radiative gas dynamics governs the interplay among radiation transport, high-temperature plasma evolution, and rarefied flow effects during capsule implosion~\cite{lindl2004physics,scott2010advances}. Radiation not only mediates energy transfer and preheating of the fuel but also strongly influences ablation dynamics and hydrodynamic stability at the capsule surface. These processes are inherently multiscale and often far from equilibrium.

Radiative gas dynamics considers both gas dynamics and radiation effects~\cite{vincenti1971coupling,shestakov2000radiation}. Conventionally, the coupling between gas internal energy and radiation is treated by incorporating radiative heat transfer as a source term in the total energy conservation equation~\cite{gnoffo1989conservation,armenise2016advanced,kustova2011comparison,cruden2014updates,johnston2008spectrum}. However, the Navier--Stokes--Fourier (NSF) equations for gas dynamics are derived under the assumption of small Knudsen number (Kn, the mean free path of gas molecules over the characteristic flow length), thus, their linear constitutive relations lose accuracy in strongly non-equilibrium regimes. In contrast, the Boltzmann equation for monatomic gas and the Wang--Chang and Uhlenbeck (WCU) equation for polyatomic gas provide a mesoscopic description of gas evolution and are, in principle, valid across all flow regimes~\cite{wangcs1951transport}. However, the WCU equation is very difficult to solve because the velocity distribution function is defined in an eight-dimensional space—spanning time, space, molecular velocity, and internal energy—while the collision operator is a five-dimensional nonlinear integral that must be summed over the internal energy levels; with $N$ internal energy levels, the computational cost scales as $N^4$ times that of the Boltzmann collision operator for a monatomic gas. This prohibitive cost has motivated the development of simplified kinetic models. For example, Groppi and Spiga~\cite{groppi1999kinetic} proposed a kinetic model in which radiation-induced transitions between energy levels are described through photon absorption and emission. However, the complexity of this formulation and its computational cost, exceeding that of the original Boltzmann collision operator, limit its practical applicability. To address this limitation, Li \textit{et al.}~\cite{li2023kinetic} developed a kinetic model in which gas flow and radiation are coupled self-consistently. As an extension of the relaxation-type kinetic model for molecules with rotational and vibrational modes, this formulation further accounts for radiative transitions between vibrational energy states. It preserves the essential physics while being considerably simpler than the Boltzmann collision operator, making it a promising candidate for practical computations.

Several numerical schemes have been developed to solve kinetic equations, where the discrete velocity method (DVM)~\cite{broadwell1964study,BOSCHERI2021114180} and direct simulation Monte Carlo (DSMC) method~\cite{bird1994molecular} are the most widely used. Because of the splitting of streaming and collision processes, these conventional schemes require the spatial cell size and time step to be smaller than the molecular mean free path and mean collision time, respectively, which makes them prohibitively expensive in the continuum flow regime. When gas--photon coupling is considered, two distinct Knudsen numbers must be introduced: the gas Knudsen number and the photon Knudsen number (defined as the mean free path of photons over the characteristic flow length). In realistic applications, both Knudsen numbers may span several orders of magnitude, leading to regimes where rarefied and continuum effects coexist. Therefore, developing unified simulations based on appropriate multiscale methods is essential.

In recent years, substantial progress has been made in the development of multiscale methods. For instance, the (discrete) unified gas kinetic scheme~\cite{xu2010unified,guo2013discrete} treats particle streaming and collision simultaneously. This approach possesses the asymptotic-preserving property and recovers NSF solutions accurately without requiring spatial cell sizes and time steps smaller than the molecular mean free path and mean collision time, respectively. It has been successfully applied to rarefied gas flow problems~\cite{zhu2016implicit, zhu2019unified, yuan2021multi, yuan2021novel} and photon transport~\cite{sun2015grayradiative, sun2015grayunstructured, sun2017frequency}. The general synthetic iteration scheme (GSIS) employs a different multiscale strategy, wherein the kinetic equation and corresponding synthetic equations are solved alternately~\cite{su2020can,zhu2021general}. The constitutive relations in the synthetic equations include both the linear terms employed in the NSF equations and higher-order terms evaluated directly from the velocity distribution function. Rigorous mathematical analyses and extensive numerical tests demonstrate that GSIS significantly reduces the number of iterations required for convergence~\cite{su2020fast}. Furthermore, when the Knudsen number is small, the higher-order constitutive relations vanish, ensuring asymptotic-preserving behavior even when the spatial cell size is much larger than the mean free path. Conversely, at large Knudsen numbers, these higher-order constitutive relations naturally capture rarefaction effects. This mesoscopic--macroscopic coupling concept has been successfully extended to problems in rarefied gas dynamics~\cite{su2021multiscale,zeng2023general,zeng2024general,zeng2025gsis,LUO2026118508} and phonon transport~\cite{Zhang2021,liu2022fast}.

However, radiative gas flow problems require self-consistent coupling between gas dynamics and radiative heat transfer. Existing mesoscopic coupling approaches rely primarily on DSMC-photon Monte Carlo methods~\cite{Sohn2012JTHT}, which suffer from computational inefficiency in multiscale regimes and instability arising from statistical fluctuations in the flow field that are amplified through their coupling with radiation rates~\cite{Prem2019Icarus}. DVM-based multiscale methods offer superior performance for these problems, but most existing studies have been limited to pure gas dynamics or pure photon transport, without demonstrating their potential for handling gas-radiation interaction problems within a multiscale framework. Hence, the present work extends GSIS to radiative rarefied flows, covering wide ranges of Knudsen numbers for both gas molecules and photons.

The remainder of this paper is organized as follows. Section \ref{sec:equations} provides a comprehensive overview of the kinetic model equations for radiative gas flows and derives the corresponding macroscopic synthetic equations. Section \ref{sec:framework} presents the GSIS coupling strategy that combines mesoscopic and macroscopic solvers. Section \ref{sec:convergence} illustrates rapid convergence of the proposed method through rigorous Fourier stability analysis, and Section \ref{sec:asymptotic} analyses the asymptotic preserving property. Section \ref{sec:implementation} details the numerical implementation employed in GSIS, including the schemes adopted to solve both the mesoscopic and macroscopic equations. Section \ref{sec:cases} validates the method through a comprehensive suite of challenging numerical examples, including three-dimensional hypersonic flow over a reentry capsule, demonstrating orders-of-magnitude acceleration over conventional methods. Finally, the main conclusions are summarized in Section \ref{sec:conclusion}.

%% file: Governing_equations.tex
\section{Governing equations for GSIS}\label{sec:equations}

For non-equilibrium radiative gas flows, we adopt the kinetic model of Li \textit{et al}.~\cite{li2023kinetic}, which extends the relaxation time approximation model for molecular gas with internal energy~\cite{rykov1975model, LeiJFM2015, li2021uncertainty} and bidirectionally couples gas dynamics with photon transport through vibrational radiative transitions. Crucially, model parameters can be adjusted to accurately reproduce key relaxation rates and transport coefficients, including shear viscosity, bulk viscosity, and thermal conductivities of individual energy modes. By taking moments of this kinetic model without approximating stress or heat flux, we obtain macroscopic synthetic equations for radiative gases that remain valid across all flow regimes.

\subsection{Kinetic model for radiative gas flows}

We consider a molecular gas with translational, rotational, and vibrational energy modes that undergo vibrational-radiative transitions, thereby coupling with photon transport. In kinetic theory, the state of a molecular gas is described by the distribution function $f(t, \bm{x}, \bm{\xi}, \varepsilon_r, \varepsilon_v)$, where $t$ denotes the time, $\bm{x}\in\mathbb{R}_3$ the spatial coordinate, $\bm{\xi}\in\mathbb{R}_3$ the molecular velocity, and $\varepsilon_r$ and $\varepsilon_v$ are the rotational and vibrational energies, respectively. To reduce computational cost, three reduced velocity distribution functions are introduced via integration over internal energies,
\begin{equation}\label{eq:reduced_distribution}
(f_0, f_1, f_2)=\iint_0^{\infty}(1,\varepsilon_r,\varepsilon_v)f,\mathrm{d}\varepsilon_r\mathrm{d}\varepsilon_v.
\end{equation}
For the radiation field, the total radiative intensity is represented by $I_R(t, \bm{x}, \bm{\Omega})$, which measures the total energy flux per unit solid angle carried by photons propagating along the direction $\bm{\Omega}$, summed over all frequencies.

The model of Li \textit{et al}.~\cite{li2023kinetic} employs a gray model approximation for photons, which neglects spectral resolution but enables study of overall radiative effects through an effective frequency-independent absorption coefficient $k_{\mathrm{eff}}$.  The evolution of the reduced distribution functions and radiative intensity is governed by the coupled kinetic equations,
\begin{equation}\label{eq:kinetic_equation}
\begin{aligned}
\frac{\partial f_0}{\partial t}+ \bxi\cdot \frac{\partial f_0}{\partial \bx} =& \frac{ g_{0t}- f_0}{ \tau}+\frac{ g_{0r}- g_{0t}}{Z_r \tau}+\frac{ g_{0v}- g_{0t}}{Z_v \tau}, \\
\frac{\partial f_1}{\partial  t}+ \bxi\cdot \frac{\partial f_1}{\partial \bx} =& \frac{ g_{1t}- f_1}{ \tau}+\frac{ g_{1r}- g_{1t}}{Z_r\tau}+\frac{ g_{1v}- g_{1t}}{Z_v \tau}, \\
\frac{\partial f_2}{\partial t}+\bxi\cdot \frac{\partial f_2}{\partial \bx} =& \frac{ g_{2t}- f_2}{\tau}+\frac{ g_{2r}- g_{2t}}{Z_r \tau}+\frac{ g_{2v}- g_{2t}}{Z_v\tau} 
-\frac{ f_0}{ \rho}k_{\mathrm{eff}}
\int_{4\pi}\left(B_R(T_v) - I_R\right)\myd \Omega,\\
\frac{1}{ c_l}\frac{\partial  I_R}{\partial t}+  \bm{\Omega} \cdot \frac{\partial  I_R}{\partial {\bm{x}}}=&k_{\mathrm{eff}}\left(B_R(T_v)- I_R\right),
\end{aligned}
\end{equation}
where $\tau$ denotes the translational relaxation time, $Z_r$ and $Z_v$ are the rotational and vibrational collision numbers; $B_R(T)=\sigma_R T^4/\pi$ is the frequency-integrated Planck function; The reference distribution functions $g$ are expanded about the equilibrium distributions in a series of orthogonal polynomials,
\begin{equation}\label{eq:reference_distribution}
\begin{aligned}
g_{0t}&= {\rho} f_0^{\mathrm{eq}}(T_t   )\left[1+\frac{2\bq_t \cdot \bc}{15 T_t p_t      }\left(\frac{ {c}^2}{2 T_t   }-\frac{5}{2}\right)\right],\\
g_{0r}&= {\rho} f_0^{\mathrm{eq}}(T_{tr})\left[1+\frac{2\bq_0 \cdot \bc}{15 T_{tr} p_{tr}}\left(\frac{ {c}^2}{2 T_{tr}}-\frac{5}{2}\right)\right],\\
g_{0v}&= {\rho} f_0^{\mathrm{eq}}(T_{tv})\left[1+\frac{2\bq_0 \cdot \bc}{15 T_{tv} p_{tv}}\left(\frac{ {c}^2}{2 T_{tv}}-\frac{5}{2}\right)\right],\\
g_{1t}&=\frac{d_r}{2}T_r g_{0t}+\frac{\bq_r  \cdot \bc}{T_t   } f_0^{\mathrm{eq}}(T_t   ), \quad
g_{2t}=\frac{d_v(T_v)}{2}T_v g_{0t}+\frac{\bq_v \cdot \bc}{T_t   } f_0^{\mathrm{eq}}(T_t   ), \\
g_{1r}&=\frac{d_r}{2}T_{tr} g_{0r}+\frac{\bq_1  \cdot \bc}{T_{tr}} f_0^{\mathrm{eq}}(T_{tr}), \quad
g_{2r}=\frac{d_v(T_v)}{2}T_v g_{0r}+\frac{\bq_2 \cdot \bc}{T_{tr}} f_0^{\mathrm{eq}}(T_{tr}), \\
g_{1v}&=\frac{d_r}{2}T_r g_{0v}+\frac{\bq_1  \cdot \bc}{T_{tv}} f_0^{\mathrm{eq}}(T_{tv}), \quad
g_{2v}=\frac{d_v(T_{tv})}{2}T_{tv} g_{0v}+\frac{\bq_2 \cdot \bc}{T_{tv}} f_0^{\mathrm{eq}}(T_{tv}), \\
\end{aligned}
\end{equation}
with the local equilibrium distribution
\begin{equation}\label{eq:equilibrium_distribution}
\begin{aligned}
f_0^{\mathrm{eq}}(T)&=\left(\frac{1}{2\pi T}\right)^{3/2}\exp\left(-\frac{c^2}{2T}\right).
\end{aligned}
\end{equation}
Here, $\rho$ is the mass density, $\bm{c}=\bm{\xi}-\bm{u}$ is the peculiar velocity with $\bm{u}$ being the flow velocity, $T$ denotes the temperatures and $\bm{q}$ represents the heat flux. Subscripts $t, r$ and $v$ indicate the translational, rotational and vibrational components, respectively. The equilibrium temperatures between the energy modes $T_{tr}$ (translational-rotational), $T_{tv}$ (translational-vibrational), and the total temperature $T$ are defined as,
\begin{equation}
\begin{aligned}
&T_{tr} = \frac{3T_t + d_rT_r}{3+d_r},\quad
T_{tv} = \frac{3T_t + d_v(T_v)T_v}{3+d_v(T_{tv})},\quad
T = \frac{3T_t + d_rT_r + d_v(T_v)T_v}{3+d_r+d_v(T)},
\end{aligned}
\end{equation}
with corresponding pressure components,
\begin{equation}
\begin{aligned}
&(p_t, p_r, p_v, p, p_{tr}, p_{tv}) = \rho (T_t, T_r, T_v, T, T_{tr}, T_{tv}).
\end{aligned}
\end{equation}

Note that all variables in the above equations are made dimensionless using reference scales: the reference length $L_0$, temperature $T_0$, and mass density $\rho_0$. The reference velocity is chosen as $v_0=\sqrt{k_B T_0/m}$ with $k_B$ being the Boltzmann constant and $m$ the molecular mass. Then the reference energy is $k_B T_0$ and the reference pressure is $\rho_0 k_B T_0/m$. Then, the macroscopic quantities of interest, including mass density $\rho$, flow velocity $\bm{u}$, shear stress tensor $\bm{\sigma}$, temperatures $T$ and heat fluxes $\bm{q}$, are obtained by taking velocity moments of the distribution functions,
\begin{equation}\label{eq:getmoment}
\begin{aligned}
(\rho,\rho\bu,\bm{\sigma})&=\int\left(1,\bxi, \bm{c}\bm{c}-\frac{c^2}{3}\mathbf{I}\right) f_0\myd{\bxi},\\
\left(\frac{3}{2}\rho T_t, \frac{d_r}{2}\rho T_r, \frac{d_v(T_v)}{2}\rho T_v\right)&=\int\left(\frac{c^2}{2}f_0,f_1,f_2\right)\myd \bxi,\quad
4\sigma_R T_{R}^4=\int I_R \myd \bm{\Omega},\\
(\bq_t,\bq_r,\bq_v)&=\int\bc\left(\frac{1}{2}c^2f_0, f_1, f_2\right)\myd \bxi,\quad
\bq_{R}=\int \bm{\Omega}I_R \myd \bm{\Omega},
\end{aligned}
\end{equation}
where $\mathbf{I}$ is a $3\times 3$ identity matrix, and ${\sigma}_R$ is the dimensionless Stefan-Boltzmann constant normalized by $n_0 k_B v_0/T_0^3$; the rotational degrees of freedom $d_r$ is fixed, and the vibrational degrees of freedom $d_v(T_v)$ depends on the vibrational temperature through,
\begin{equation}
    d_v(T_v) = \frac{2T_v^{\text{ref}}/T_v}{\exp(T_v^{\text{ref}}/T_v)-1},
\end{equation}
where $T_v^{\text{ref}}$ is the characteristic vibrational temperature. 

For radiative gas flows, the degree of non-equilibrium is characterized by two Knudsen numbers: the gas Knudsen number $\mathrm{Kn}_{\mathrm{gas}}$ and the photon Knudsen number $\mathrm{Kn}_{\mathrm{photon}}$. These dimensionless parameters enter the kinetic model equation \eqref{eq:kinetic_equation} through the dimensionless translational relaxation time $\tau$ and photon absorption coefficient $k_{\text{eff}}$ as,
\begin{equation}
\begin{aligned}
    \tau=\sqrt{\frac{2}{\pi}} \frac{{T_t}^{\omega-1}}{\rho} \mathrm{Kn}_{\mathrm{gas}}, \quad
    k_{\text{eff}} = \frac{1}{\mathrm{Kn}_{\mathrm{photon}}},
\end{aligned}
\end{equation}
where $\omega$ is the index in a power-law relation of viscosity $\mu(T) = \mu_0T^{\omega}$, with $\mu_0$ being the viscosity at reference temperature. This parameterization enables the kinetic model to reproduce the correct gas shear viscosity and effective radiative absorption characteristics.

Beyond the relaxation time and photon absorption coefficient, the kinetic model requires specification of the auxiliary heat fluxes $\bm{q}_0$, $\bm{q}_1$ and $\bm{q}_2$ that ensure accurate relaxation of the heat fluxes governed by the dimensionless matrix $A$,
\begin{equation}
   \frac{\mathrm{d}}{\mathrm{d} t} 
   \left[ 
\begin{array}{ccc} 
\bq_t \\
\bq_r \\
\bq_v
\end{array}
\right]=-\frac{p_t}{\mu}
\left[ 
\begin{array}{ccc} 
A_{tt} & A_{tr} & A_{tv} \\
A_{rt} & A_{rr} & A_{rv} \\
A_{vt} & A_{vr} & A_{vv} 
\end{array}
\right]
\left[ 
\begin{array}{ccc} 
\bq_t \\
\bq_r \\
\bq_v
\end{array}
\right].
\end{equation}
Thus, these auxiliary fluxes are linear combinations of $\bm{q}_t$, $\bm{q}_r$ and $\bm{q}_v$ as \cite{li2021uncertainty},
\begin{equation}
\begin{aligned}
\left[ 
\begin{array}{ccc} 
\bq_0 \\
\bq_1 \\
\bq_2 
\end{array}
\right]
&=  
\left[ 
\begin{array}{ccc} 
(2-3A_{tt})Z_{int}+1 & -3A_{tr}Z_{int} & -3A_{tv}Z_{int} \\
-A_{rt}Z_{int} & 1-A_{rr}Z_{int} & -A_{rv}Z_{int} \\
-A_{vt}Z_{int} & -A_{vr}Z_{int} & 1-A_{vv}Z_{int} 
\end{array}
\right]
\left[ 
\begin{array}{ccc} 
\bq_t \\
\bq_r \\
\bq_v
\end{array}
\right],\\
\end{aligned}
\end{equation}
where $Z_{int} = (1/Z_r + 1/Z_v)^{-1}$. The thermal relaxation rate matrix $A$ cannot be fully determined by the thermal conductivities of the gas, while it can be extracted from molecular dynamics when the intermolecular interaction is given, or DSMC simulations when the collision model is adopted \cite{li2021uncertainty, li2023kinetic}. Also, the values of $A$ can be approximately derived from the asymptotic expansion of the WCU equation as~\cite{Mason1962},
\begin{equation}
\begin{aligned}
    A
=&
\left[ {
\begin{array}{ccc} 
\frac{2}{3}+\frac{5}{6}\left(\frac{d_r}{(3+d_r)Z_r}+\frac{d_v}{(3+d_v)Z_v}\right) & -\frac{5}{2(3+d_r)Z_r} & -\frac{5}{2(3+d_v)Z_v} \\
-\frac{d_r}{(3+d_r)Z_r} & \text{Sc}+\frac{3}{2(3+d_r)Z_r} & 0 \\
-\frac{d_v}{(3+d_v)Z_v} & 0 & \text{Sc}+\frac{3}{2(3+d_v)Z_v} 
\end{array}}
\right],
\end{aligned}
\end{equation}
where $\text{Sc}$ is the Schmidt number. This parameterization ensures that the kinetic model correctly reproduces not only the gas thermal conductivities in the continuum flow regime but also thermal relaxation behavior across all flow regimes.


\subsection{Macroscopic synthetic equations}

To construct macroscopic equations applicable across all flow regimes, we derive exact moment equations directly from the kinetic model \eqref{eq:kinetic_equation} without invoking near-equilibrium approximations. Taking velocity moments yields a set of equations that enforces the conservation laws of mass, momentum, and total energy, and incorporates the relaxation mechanisms for internal energy exchange and gas--photon energy exchange,
\begin{equation}\label{eq:macroscopic_equation}
	\begin{aligned}
		\frac{\partial{\rho}}{\partial{t}} + \nabla\cdot\left(\rho\bm{u}\right)  &= 0, \\
		\frac{\partial}{\partial{t}}\left(\rho\bm{u}\right) + \nabla\cdot\left(\rho\bm{u}\bm{u}\right) + \nabla\cdot\left(p_t\mathbf{I}+\bm{\sigma}\right) &= 0, \\
		\frac{\partial e}{\partial{t}} + \nabla\cdot\left(\bm{u}e_{\text{gas}}\right) + \nabla\cdot\left(p_t\bm{u}+\bm{\sigma}\cdot\bm{u}+\bm{q}\right) &= 0, \\
        \frac{\partial e_r}{\partial{t}} + \nabla\cdot\left(\bm{u} e_r\right) + \nabla\cdot\bm{q}_r &= \frac{e_{tr}-e_r}{Z_r\tau},\\
        \frac{\partial e_v}{\partial{t}} + \nabla\cdot\left(\bm{u} e_v\right) + \nabla\cdot\bm{q}_v &= \frac{e_{tv}-e_v}{Z_v\tau}-k_{\text{eff}}\left(e_{vR}-e_R\right),\\
        \frac{\partial e_R}{\partial{t}} +  \nabla\cdot\bm{q}_R &= k_{\text{eff}}\left(e_{vR}-e_R\right),
	\end{aligned}
\end{equation}
where the energy components are,
\begin{equation}
    \begin{aligned}
    & e_t = \frac{3}{2}\rho T_t + \frac{1}{2}\rho u^2, \quad e_r = \frac{d_r}{2}\rho T_r, \quad e_v = \frac{d_v(T_v)}{2}\rho T_v, \quad e_R = 4{\sigma}_R T_R^4, \\
    & e_{tr} = \frac{d_r}{2}\rho T_{tr}, \quad e_{tv} = \frac{d_v(T_{tv})}{2}\rho T_{tv}, \quad e_{Rv} = 4{\sigma}_R T_v^4,
    \end{aligned}
\end{equation}
and $e_{\text{gas}} = e_t + e_r + e_v$ denotes the gas energy, $e=e_{\text{gas}} + e_R$ is the total energy includes radiative component; $\bm{q}= \bq_t + \bq_r + \bq_v + \bq_R$ is the total heat flux.

The central challenge is that Eq.~\eqref{eq:macroscopic_equation} gives an unclosed system, as the exact shear stress $\bm{\sigma}$ and the heat fluxes $\bm{q}_{t}, \bm{q}_{r}, \bm{q}_{v}, \bm{q}_{R}$ are not specified. The Chapman-Enskog expansion~\cite{chapman1990mathematical} to the first order of the Knudsen number provides the NSF constitutive relations for the gas and the diffusion approximation for the photon transport,
\begin{equation}\label{eq:NSF_constitutive}
    \begin{aligned}
        \bm{\sigma}^{\text{NSF}} &= -\mu \left(\nabla\bm{u}+\nabla\bm{u}^{\top}-\frac{2}{3}\nabla\cdot\bm{u}\mathbf{I}\right),\\
        \left[\bm{q}_{t}^{\text{NSF}}, \bm{q}_{r}^{\text{NSF}}, \bm{q}_{v}^{\text{NSF}} \right]^{\top} &= -\bm{\kappa}_{\text{gas}} \nabla \left[T_{t}, T_{r}, T_{v} \right]^{\top},  \\
        \bm{q}_{R}^{\text{NSF}} & = -\kappa_R\nabla e_{R},
    \end{aligned}
\end{equation}
where the coefficients are the transport properties: the shear viscosity $\mu=p_t\tau$, the thermal conductivity matrix of gas
\begin{equation}\label{eq:NSF_thermal_conductivities}
 \begin{aligned}
	&\bm{\kappa}_{\text{gas}} = 
    \begin{bmatrix}
		\kappa_{tt} & \kappa_{tr} & \kappa_{tv} \\
		\kappa_{rt} & \kappa_{rr} & \kappa_{rv} \\
		\kappa_{vt} & \kappa_{vr} & \kappa_{vv}
	\end{bmatrix}
    = \frac{p_t\tau}{2}
	\left[ 
      \begin{array}{ccc} 
        A_{tt} & A_{tr} & A_{tv}\\ A_{rt} & A_{rr} & A_{rv} \\ A_{vt} & A_{vr} & A_{vv}
      \end{array}
    \right]^{-1}
    \left[ 
      \begin{array}{ccc} 
        5 & 0 & 0 \\ 0 & d_r & 0 \\ 0 & 0 & d_v(T_v)
      \end{array}
    \right],
\end{aligned}
\end{equation}
and the thermal transport coefficient for radiation $\kappa_R = {1}/{3 k_{\text{eff}}}$.
However, these linear constitutive relations are only valid in the near-continuum regime of gas flow and the diffusion limit of photon transport. Thus, Eq. \eqref{eq:NSF_thermal_conductivities} is inadequate for describing highly non-equilibrium systems. 

The key point of the GSIS method is to solve the macroscopic equations with the exact shear stress tensor and heat fluxes without any approximation. Thus, the solutions of macroscopic synthetic equations will be consistent with those of kinetic equations. Therefore, we split the exact shear stress tensor and heat fluxes into the linear constitutive relations and higher-order terms (HoTs):
\begin{equation}\label{eq:split_constitutive}
    \begin{aligned}
        \bm{\sigma} &= -\mu \left(\nabla\bm{u}+\nabla\bm{u}^{\top}-\frac{2}{3}\nabla\cdot\bm{u}\mathbf{I}\right) + \text{HoT}_{\bm{\sigma}},\\
        \left[\bm{q}_{t}, \bm{q}_{r}, \bm{q}_{v} \right]^{\top} &= -\bm{\kappa}_{\text{gas}} \nabla \left[T_{t}, T_{r}, T_{v} \right]^{\top} + \left[\text{HoT}_{\bm{q}_{t}}, \text{HoT}_{\bm{q}_{r}}, \text{HoT}_{\bm{q}_{v}} \right]^{\top},  \\
        \bm{q}_{R} &= -\kappa_R\nabla e_R + \text{HoT}_{\bm{q}_{R}}.
    \end{aligned}
\end{equation}

Critically, the HoTs, which capture the non-equilibrium effects beyond the NSF approximation, are evaluated explicitly by subtracting the NSF approximation from the exact moments obtained from the kinetic distribution functions:
\begin{equation}\label{eq:getHoTs}
    \begin{aligned}
        \text{HoT}_{\bm{\sigma}} &= \int \left(\bm{c}\bm{c}-\frac{c^2}{3}\mathbf{I}\right)f_0 \mathrm{d}\bm{\xi} -\bm{\sigma}^{\text{NSF}},\\
        \text{HoT}_{\bm{q}_{t}} &= \int \frac{1}{2}\bm{c}c^2f_0 \myd\bxi -\bm{q}_{t}^{\text{NSF}},\quad
        \text{HoT}_{\bm{q}_{r}} = \int \bm{c}f_1 \myd\bxi  -\bm{q}_{r}^{\text{NSF}},\\
        \text{HoT}_{\bm{q}_{v}} &= \int \bm{c}f_2 \myd\bxi  -\bm{q}_{v}^{\text{NSF}},\quad
        \text{HoT}_{\bm{q}_{R}} = \int \bm{\Omega}I_R \myd\bm{\zeta}  -\bm{q}_{R}^{\text{NSF}},
    \end{aligned}
\end{equation}
where $\bm{\sigma}^{\text{NSF}}$ and $\bm{q}_{t/r/v/R}^{\text{NSF}}$ are evaluated using the macroscopic properties obtained from the distribution function moments based on Eq.~\eqref{eq:NSF_constitutive}. Therefore, the macroscopic equations \eqref{eq:macroscopic_equation} are closed with the exact constitutive relations containing non-equilibrium effects, and form the basis for the GSIS method together with the kinetic model \eqref{eq:kinetic_equation}.

%% file: Iteration_Framework.tex
\section{GSIS iteration framework}\label{sec:framework}

Conventional iterative schemes (CIS) for solving kinetic equations rely on successive updates of the distribution functions alone. Since the reference distributions, relaxation times, and radiative properties all depend on the velocity moments of the distribution functions, CIS must iterate until these quantities converge self-consistently. It will be rigorously proven in Section \ref{sec:convergence} that the CIS is efficient for highly rarefied gas flow, however, becomes prohibitively slow in near-continuum regimes.

GSIS overcomes this limitation by introducing a tight kinetic--macroscopic coupling that utilizes the macroscopic synthetic equations to accelerate convergence. Rather than waiting for the kinetic solution to converge in isolation, GSIS alternates between a single kinetic step (to capture non-equilibrium details) and an implicit macroscopic step (to propagate bulk information efficiently). This coupled iteration ensures that the high-order terms applied in the macroscopic evolution are continuously updated by kinetic solution, while the macroscopic state in turn provides improved reference distributions for the next kinetic step, creating a rapidly convergent feedback loop.


The complete GSIS procedure for advancing the solution from iteration $k$ to $k+1$ is given as follows:
\begin{enumerate}
\item Mesoscopic (kinetic) step: 
Starting from the distribution function $f^k=\left(f_0^k,f_1^k,f_2^k,I_R^k\right)$, calculate the macroscopic state $M^k=\left(\rho^k,\bm{u}^k,T_t^k,T_r^k,T_v^k,T_R^k\right)$ via moment relations \eqref{eq:getmoment}. Solve the kinetic equations \eqref{eq:kinetic_equation} once (equivalent to a single CIS iteration) to obtain the intermediate distribution $f^{k+1/2}$. The superscript $k+1/2$ denotes an uncorrected kinetic solution that retains non-equilibrium features but has not yet been modified by the macroscopic solver.
\item Macroscopic (synthetic) step: 
Using distribution functions $f^{k+1/2}$, evaluate the intermediate macroscopic state $M^{k+1/2}$ and the corresponding HoTs via \eqref{eq:getmoment} and \eqref{eq:getHoTs}, respectively. With the HoTs fixed as explicit source terms, iterate the macroscopic synthetic equations \eqref{eq:macroscopic_equation} to convergence, yielding the updated macroscopic fields $M^{k+1}$.
\item Distribution function synchronization: 
Correct the intermediate distribution function $f^{k+1/2}$ to be consistent with $M^{k+1}$ using:
\begin{equation}
    f^{k+1} = f^{k+1/2}+[f^{\text{eq}}(M^{k+1})-f^{\text{eq}}(M^{k+1/2})],
\end{equation}
where $f^{\mathrm{eq}}(M)$ denotes the local equilibrium distribution reconstructed from the macroscopic state. This step ensures that the distribution function for the next iteration respects the updated conservation laws while preserving the non-equilibrium content captured by the kinetic step.
\item Convergence assessment: 
Repeat the cycle until the residual of the macroscopic variables falls below a prescribed tolerance. At convergence, the final solution satisfies both the kinetic and macroscopic descriptions simultaneously.
\end{enumerate}

This iterative design fundamentally equips GSIS with fast convergence and asymptotic-preserving properties. In the continuum limit, the implicit macroscopic solver dominates, ensuring accurate and fast radiative hydrodynamic solutions without kinetic-scale resolution, while in high Knudsen number regimes, the kinetic step captures essential non-equilibrium effects. Consequently, the method delivers robust performance across the entire spectrum of gas and photon Knudsen numbers. A rigorous Fourier stability analysis quantifying the convergence acceleration is presented in Section \ref{sec:convergence}, and the analysis of asymptotic preserving property is provided in Section \ref{sec:asymptotic}.

%% file: Convergence_rate.tex
\section{Convergence acceleration of GSIS}  \label{sec:convergence}

The convergence degradation of CIS in near-continuum regimes is well-demonstrated for pure gas dynamics \cite{su2020fast,su2021multiscale}, while its performance in multi-physics systems with self-consistent radiation coupling remains unexplored. In this section, we present a systematic linear Fourier stability analysis to rigorously quantify the convergence acceleration of GSIS over conventional methods for coupled gas--radiation systems. To maintain analytical tractability while preserving the essential physics, we linearize the kinetic equations about a uniform equilibrium and treat spatial coordinates as continuous. The convergence factors are obtained that contrast the stability and efficiency of CIS and GSIS,  particularly in the near-continuum regimes.

\subsection{Linearized kinetic model and perturbed variables} 

For a system of radiative gas that deviates slightly from a global equilibrium state, the velocity distribution functions, $f_l~(l=0,1,2)$, and the radiative intensity, $I_R$, are expressed as the sum of their equilibrium components and perturbations: $f_l =  f_{l}^{\mathrm{eq}} + h_l$ and $I_R =  I_{R}^{\mathrm{eq}} + h_R$, respectively. Here, $f_{l}^{\mathrm{eq}}$ and $I_R^{\mathrm{eq}}$ are the equilibrium distribution functions,
\begin{equation}
	\begin{aligned}[b]
		f_{0}^{\mathrm{eq}} &= \left(\frac{1}{2\pi}\right)^{3/2}\exp{\left(-\frac{\xi^2}{2}\right)}, \quad
		f_{1}^{\mathrm{eq}} &= \frac{d_r}{2}f_{0}^{\mathrm{eq}}, \quad
		f_{2}^{\mathrm{eq}} &= \frac{d_v}{2}f_{0}^{\mathrm{eq}}, \quad
		I_R^{\mathrm{eq}} &= \frac{{\sigma}_R}{\pi},
	\end{aligned}
\end{equation}
and $h_l$ and $h_R$ are the perturbations.

The steady-state solutions can be obtained by iteratively solving the following linearized equations,
\begin{equation}\label{eq:steady_state_linear_kinetic_equation}
	\begin{aligned}[b]
		{\bm{\xi}} \cdot \nabla {{h}_{0}^{k+1}} =& \sqrt{\frac{\pi}{2}}\frac{1}{\mathrm{Kn}_{\mathrm{gas}}}\left[\left({g'}_{0t}^{k}-{h}_0^{k+1}\right)+\frac{1}{Z_r}\left({g'}_{0r}^{k}-{g'}_{0t}^{k}\right)+\frac{1}{Z_v}\left({g'}_{0v}^{k}-{g'}_{0t}^{k}\right)\right], \\
		{\bm{\xi}} \cdot \nabla {{h}_{1}^{k+1}} =& \sqrt{\frac{\pi}{2}}\frac{1}{\mathrm{Kn}_{\mathrm{gas}}}\left[\left({g'}_{1t}^{k}-{h}_1^{k+1}\right)+\frac{1}{Z_r}\left({g'}_{1r}^{k}-{g'}_{1t}^{k}\right)+\frac{1}{Z_v}\left({g'}_{1v}^{k}-{g'}_{1t}^{k}\right)\right], \\
		{\bm{\xi}} \cdot \nabla {{h}_{2}^{k+1}} =& \sqrt{\frac{\pi}{2}}\frac{1}{\mathrm{Kn}_{\mathrm{gas}}}\left[\left({g'}_{2t}^{k}-{h}_2^{k+1}\right)+\frac{1}{Z_r}\left({g'}_{2r}^{k}-{g'}_{2t}^{k}\right)+\frac{1}{Z_v}\left({g'}_{2v}^{k}-{g'}_{2t}^{k}\right)\right] \\
		&-\frac{1}{\mathrm{Kn}_{\mathrm{photon}}}f_{0}^{\mathrm{eq}}\left(16{\sigma}_R\Delta T_v^{k}-\Delta e_R^{k}\right), \\
		\bm{\Omega} \cdot \nabla {{h}_{R}^{k+1}} =& \frac{1}{\mathrm{Kn}_{\mathrm{photon}}}\left(\frac{4{\sigma}_R}{\pi}\Delta T_v^{k}-{h}_{R}^{k+1}\right),
	\end{aligned}
\end{equation}
where the superscript $k$ represents the iteration step, and the perturbed reference distributions $g'_{0t}$ in the linearized collision operators are expressed as,
\begin{equation}\label{eq:linear_perturbed_g}
	\begin{aligned}
		g'_{0t} &= f_{0}^{\mathrm{eq}}\left[\Delta{\rho} + {\bm{u}}\cdot\bm{\xi} + \Delta{T_{t}}\left(\frac{1}{2}\xi^2-\frac{3}{2}\right) + \frac{2\bm{q}_{t}\cdot\bm{\xi}}{15}\left(\frac{1}{2}\xi^2-\frac{5}{2}\right)\right], \\
		g'_{0r} &= f_{0}^{\mathrm{eq}}\left[\Delta{\rho} + {\bm{u}}\cdot\bm{\xi} + \Delta{T_{tr}}\left(\frac{1}{2}\xi^2-\frac{3}{2}\right) + \frac{2\bm{q}_{0}\cdot\bm{\xi}}{15}\left(\frac{1}{2}\xi^2-\frac{5}{2}\right)\right], \\
		g'_{0v} &= f_{0}^{\mathrm{eq}}\left[\Delta{\rho} + {\bm{u}}\cdot\bm{\xi} + \Delta{T_{tv}}\left(\frac{1}{2}\xi^2-\frac{3}{2}\right) + \frac{2\bm{q}_{0}\cdot\bm{\xi}}{15}\left(\frac{1}{2}\xi^2-\frac{5}{2}\right)\right], \\
		g'_{1t} &= f_{1}^{\mathrm{eq}}\left[\Delta{\rho} + {\bm{u}}\cdot\bm{\xi} + \Delta{T_{t}}\left(\frac{1}{2}\xi^2-\frac{3}{2}\right) + \frac{2\bm{q}_{t}\cdot\bm{\xi}}{15}\left(\frac{1}{2}\xi^2-\frac{5}{2}\right)+\Delta T_r\right] + f_{0}^{\mathrm{eq}}\bm{q}_{r}\cdot\bm{\xi}, \\
		g'_{1r} &= f_{1}^{\mathrm{eq}}\left[\Delta{\rho} + {\bm{u}}\cdot\bm{\xi} + \Delta{T_{tr}}\left(\frac{1}{2}\xi^2-\frac{3}{2}\right) + \frac{2\bm{q}_{0}\cdot\bm{\xi}}{15}\left(\frac{1}{2}\xi^2-\frac{5}{2}\right)+\Delta T_{tr}\right] + f_{0}^{\mathrm{eq}}\bm{q}_{1}\cdot\bm{\xi}, \\
		g'_{1v} &= f_{1}^{\mathrm{eq}}\left[\Delta{\rho} + {\bm{u}}\cdot\bm{\xi} + \Delta{T_{tv}}\left(\frac{1}{2}\xi^2-\frac{3}{2}\right) + \frac{2\bm{q}_{0}\cdot\bm{\xi}}{15}\left(\frac{1}{2}\xi^2-\frac{5}{2}\right)+\Delta T_{r}\right] + f_{0}^{\mathrm{eq}}\bm{q}_{1}\cdot\bm{\xi}, \\
		g'_{2t} &= f_{2}^{\mathrm{eq}}\left[\Delta{\rho} + {\bm{u}}\cdot\bm{\xi} + \Delta{T_{t}}\left(\frac{1}{2}\xi^2-\frac{3}{2}\right) + \frac{2\bm{q}_{t}\cdot\bm{\xi}}{15}\left(\frac{1}{2}\xi^2-\frac{5}{2}\right)+\Delta T_v\right] + f_{0}^{\mathrm{eq}}\bm{q}_{v}\cdot\bm{\xi}, \\
		g'_{2r} &= f_{2}^{\mathrm{eq}}\left[\Delta{\rho} + {\bm{u}}\cdot\bm{\xi} + \Delta{T_{tr}}\left(\frac{1}{2}\xi^2-\frac{3}{2}\right) + \frac{2\bm{q}_{0}\cdot\bm{\xi}}{15}\left(\frac{1}{2}\xi^2-\frac{5}{2}\right)+\Delta T_{v}\right] + f_{0}^{\mathrm{eq}}\bm{q}_{2}\cdot\bm{\xi}, \\
		g'_{2v} &= f_{2}^{\mathrm{eq}}\left[\Delta{\rho} + {\bm{u}}\cdot\bm{\xi} + \Delta{T_{tv}}\left(\frac{1}{2}\xi^2-\frac{3}{2}\right) + \frac{2\bm{q}_{0}\cdot\bm{\xi}}{15}\left(\frac{1}{2}\xi^2-\frac{5}{2}\right)+\Delta T_{tv}\right] + f_{0}^{\mathrm{eq}}\bm{q}_{2}\cdot\bm{\xi}. \\
	\end{aligned}
\end{equation}
These reference distributions are linear combinations of the macroscopic perturbed variables, $\Delta{\rho}$, $\bm{u}$, $\Delta{T}$ and $\bm{q}$, from the previous iteration step $k$, used to calculate the gain part of the collision operator.

These macroscopic quantities, which represent the deviation from their corresponding equilibrium values, are calculated by taking the velocity moments of the perturbed distribution functions as,
\begin{equation}\label{eq:linear_perturbed_macroscopic_variable}
	\begin{aligned}
		&\Delta{\rho} = \int{h_0}\mathrm{d}\bm{\xi}, \quad
        \bm{u}       = \int{\bm{\xi}h_0}\mathrm{d}\bm{\xi}, \quad
		\Delta T_{t} = \int{\left(\frac{1}{3}\xi^2-1\right)h_0}\mathrm{d}\bm{\xi}, \\
		&\Delta T_{r} = \int{\left(\frac{2}{d_r}h_1-h_0\right)}\mathrm{d}\bm{\xi}, \quad
		\Delta T_{v} = \int{\left(\frac{2}{d_v}h_2-h_0\right)}\mathrm{d}\bm{\xi}, \quad
		\Delta e_{R} = \int{h_R}\mathrm{d}\bm{\Omega}, \\
		&\bm{\sigma} = \int{\left(\bm{\xi}\bm{\xi}-\frac{1}{3}\xi^2\bm{\mathrm{I}}\right)h_0}\mathrm{d}\bm{\xi}, \quad	
		\bm{q}_t 	 = \int{\left(\frac{1}{2}\xi^2-\frac{5}{2}\right)\bm{\xi}h_0}\mathrm{d}\bm{\xi}, \\
		&\bm{q}_r 	 = \int{\bm{\xi}\left(h_1-\frac{d_r}{2}h_0\right)}\mathrm{d}\bm{\xi}, \quad
		\bm{q}_v 	 = \int{\bm{\xi}\left(h_2-\frac{d_v}{2}h_0\right)}\mathrm{d}\bm{\xi}, \quad
		\bm{q}_R 	 = \int{\bm{\Omega}h_R}\mathrm{d}\bm{\Omega},
	\end{aligned}
\end{equation}

\subsection{Convergence analysis of CIS}

To quantitatively assess the convergence rate of the CIS, we employ a linear stability analysis based on the evolution of the iteration error. We define the error functions for the distribution functions ($Y_l, Y_R$) as the difference between two consecutive iterative steps,
\begin{equation}\label{eq:error_function_h}
	\begin{aligned}[b]
		Y_l^{k+1}\left(\bm{x},\bm{\xi}\right) &\equiv h_l^{k+1}\left(\bm{x},\bm{\xi}\right)-h_l^{k}\left(\bm{x},\bm{\xi}\right), \quad l=0,1,2., \\
		Y_R^{k+1}\left(\bm{x},\bm{\Omega}\right) &\equiv h_R^{k+1}\left(\bm{x},\bm{\Omega}\right)-h_R^{k}\left(\bm{x},\bm{\Omega}\right),
	\end{aligned}
\end{equation}

Similarly, the error functions for the macroscopic quantities $\Phi\left(\bm{x}\right)$ are defined for the full set of 17 perturbed variables $M$,
\begin{equation}
	\begin{aligned}[b]
		M=\left[\Delta{\rho},\bm{u},\Delta T_t,\Delta T_r,\Delta T_v,\Delta e_R,\bm{q}_t,\bm{q}_r,\bm{q}_v \right]
	\end{aligned}
\end{equation}
The macroscopic error vector $\Phi^{k+1}\left(\bm{x}\right)$ is obtained by taking the moments of the distribution error functions,
\begin{equation}\label{eq:error_function_M}
	\begin{aligned}[b]
		\Phi^{k+1}\left(\bm{x}\right) &= \left[\Phi_{\Delta{\rho}}^{k+1},\Phi_{\bm{u}}^{k+1},\Phi_{\Delta{T}_t}^{k+1},\Phi_{\Delta{T}_r}^{k+1},\Phi_{\Delta{T}_v}^{k+1},\Phi_{\Delta{e}_R}^{k+1},\Phi_{\bm{q}_t}^{k+1},\Phi_{\bm{q}_r}^{k+1},\Phi_{\bm{q}_v}^{k+1}\right] \\
		&\equiv M^{k+1}\left(\bm{x}\right)-M^{k}\left(\bm{x}\right) \\
		&= \sum_{l=0,1,2}\int{\phi_l\left(\bm{\xi}\right) Y_l^{k+1}\left(\bm{x},\bm{\xi}\right)}\mathrm{d}\bm{\xi} + \int{\phi_R\left(\bm{\Omega}\right) Y_R^{k+1}\left(\bm{x},\bm{\Omega}\right)}\mathrm{d}\bm{\Omega},
	\end{aligned}
\end{equation}
where $\phi_l\left(\bm{\xi}\right),~(l=0,1,2)$ and $\phi_R\left(\bm{\Omega}\right)$ are $1\times17$ moment vectors, with $\phi_l\left(\bm{\xi}\right)$ extracting density, velocity, gas temperature/heat flux components, and $\phi_R\left(\bm{\Omega}\right)$ extracting the radiative energy density,
\begin{equation}\label{eq:error_function_M_phi}
	\begin{aligned}[b]
		\phi_0\left(\bm{\xi}\right) &= \left[1,\bm{\xi},\left(\frac{1}{3}\xi^2-1\right),-1,-1,0,\left(\frac{1}{2}\xi^2-\frac{5}{2}\right)\bm{\xi},-\frac{d_r}{2}\bm{\xi},-\frac{d_v}{2}\bm{\xi}\right], \\
		\phi_1\left(\bm{\xi}\right) &= \left[0,\bm{0},0,\frac{2}{d_r},0,0,\bm{0},\bm{\xi},\bm{0}\right], \\
		\phi_2\left(\bm{\xi}\right) &= \left[0,\bm{0},0,0,\frac{2}{d_v},0,\bm{0},\bm{0},\bm{\xi}\right], \\
		\phi_R\left(\bm{\Omega}\right) &= \left[0,\bm{0},0,0,0,1,\bm{0},\bm{0},\bm{0}\right], \\
	\end{aligned}
\end{equation}

By subtracting the kinetic equations \eqref{eq:steady_state_linear_kinetic_equation} at iteration $k$ from those at $k+1$, and substituting the error definitions \eqref{eq:error_function_h} and \eqref{eq:error_function_M}, the evolution of the distribution function errors is governed by the following system of linear equations,
\begin{equation}\label{eq:error_function_propagation}
	\begin{aligned}[b]
		\left[1+\sqrt{\frac{2}{\pi}}\mathrm{Kn}_{\mathrm{gas}}\bm{\xi}\cdot\nabla \right] Y_0^{k+1} = &
		f_{0}^{\mathrm{eq}} 
		\left\{ \Phi_{\Delta{\rho}}^{k}
		+ \Phi_{\bm{u}}^{k} \cdot \bm{\xi} 
		+ \Phi_{\Delta{T}_t}^{k}\left(1-\frac{d_r}{(3+d_r)Z_r}-\frac{d_v}{(3+d_v)Z_v}\right) \left(\frac{1}{2}\xi^2-\frac{3}{2}\right) \right. \\ 
		&\left. + \left[\frac{d_r}{(3+d_r)Z_r}\Phi_{\Delta{T}_r}^{k} +\frac{d_v}{(3+d_v)Z_v}\Phi_{\Delta{T}_v}^{k}\right] \left(\frac{1}{2}\xi^2-\frac{3}{2}\right) \right. \\ 
		&\left. + \left[\left(3-3A_{tt}\right)\Phi_{\bm{q}_t}^{k} -3A_{tr}\Phi_{\bm{q}_r}^{k} -3A_{tv}\Phi_{\bm{q}_v}^{k}\right] \cdot \frac{2}{15}\left(\frac{1}{2}\xi^2-\frac{5}{2}\right)\bm{\xi} \right\}, \\
		\left[1+\sqrt{\frac{2}{\pi}}\mathrm{Kn}_{\mathrm{gas}}\bm{\xi}\cdot\nabla \right] Y_1^{k+1} = &
		f_{1}^{\mathrm{eq}} 
		\left\{ \Phi_{\Delta{\rho}}^{k}
		+ \Phi_{\bm{u}}^{k} \cdot \bm{\xi} 
		+ \Phi_{\Delta{T}_t}^{k}\left(1-\frac{d_r}{(3+d_r)Z_r}-\frac{d_v}{(3+d_v)Z_v}\right) \left(\frac{1}{2}\xi^2-\frac{3}{2}\right) \right. \\ 
		&\left. + \left[\frac{d_r}{(3+d_r)Z_r}\Phi_{\Delta{T}_r}^{k} +\frac{d_v}{(3+d_v)Z_v}\Phi_{\Delta{T}_v}^{k}\right] \left(\frac{1}{2}\xi^2-\frac{3}{2}\right) \right. \\ 
		&\left. + \left[\frac{3}{(3+d_r)Z_r}\Phi_{\Delta{T}_t}^{k} +\left(1-\frac{3}{(3+d_r)Z_r}\right)\Phi_{\Delta{T}_r}^{k} \right] \right. \\ 
		&\left. + \left[\left(3-3A_{tt}\right)\Phi_{\bm{q}_t}^{k} -3A_{tr}\Phi_{\bm{q}_r}^{k} -3A_{tv}\Phi_{\bm{q}_v}^{k}\right] \cdot \frac{2}{15}\left(\frac{1}{2}\xi^2-\frac{5}{2}\right)\bm{\xi} \right. \\
		&\left. + \left[-A_{rt}\Phi_{\bm{q}_t}^{k} +\left(1-A_{rr}\right)\Phi_{\bm{q}_r}^{k} -A_{rv}\Phi_{\bm{q}_v}^{k}\right] \cdot \frac{2}{d_r}\bm{\xi}\right\}, \\
		\left[1+\sqrt{\frac{2}{\pi}}\mathrm{Kn}_{\mathrm{gas}}\bm{\xi}\cdot\nabla \right] Y_2^{k+1} = &
		f_{2}^{\mathrm{eq}} 
		\left\{ \Phi_{\Delta{\rho}}^{k}
		+ \Phi_{\bm{u}}^{k} \cdot \bm{\xi} 
		+ \Phi_{\Delta{T}_t}^{k}\left(1-\frac{d_r}{(3+d_r)Z_r}-\frac{d_v}{(3+d_v)Z_v}\right) \left(\frac{1}{2}\xi^2-\frac{3}{2}\right) \right. \\ 
		&\left. + \left[\frac{d_r}{(3+d_r)Z_r}\Phi_{\Delta{T}_r}^{k} +\frac{d_v}{(3+d_v)Z_v}\Phi_{\Delta{T}_v}^{k}\right] \left(\frac{1}{2}\xi^2-\frac{3}{2}\right) \right. \\ 
		&\left. + \left[\frac{3}{(3+d_v)Z_v}\Phi_{\Delta{T}_t}^{k} +\left(1-\frac{3}{(3+d_v)Z_v}\right)\Phi_{\Delta{T}_v}^{k} \right] \right. \\ 
		&\left. + \left[\left(3-3A_{tt}\right)\Phi_{\bm{q}_t}^{k} -3A_{tr}\Phi_{\bm{q}_r}^{k} -3A_{tv}\Phi_{\bm{q}_v}^{k}\right] \cdot \frac{2}{15}\left(\frac{1}{2}\xi^2-\frac{5}{2}\right)\bm{\xi} \right. \\
		&\left. + \left[-A_{vt}\Phi_{\bm{q}_t}^{k} -A_{vr}\Phi_{\bm{q}_r}^{k} +\left(1-A_{vv}\right)\Phi_{\bm{q}_v}^{k}\right] \cdot \frac{2}{d_v}\bm{\xi}\right\} \\
		&-\sqrt{\frac{2}{\pi}}\frac{\mathrm{Kn}_{\mathrm{gas}}}{\mathrm{Kn}_{\mathrm{photon}}}f_{0}^{\mathrm{eq}}\left(16{\sigma}_R\Phi_{\Delta{T}_v}^{k}-\Phi_{\Delta{e}_R}^{k}\right), \\
		\left[1+{\mathrm{Kn}_{\mathrm{photon}}}\bm{\Omega}\cdot\nabla \right]{{Y}_{R}^{k+1}} = & \frac{4{\sigma}_R}{\pi}\Phi_{\Delta{T}_v}^{k}.
	\end{aligned}
\end{equation}

These equations show the characteristic error propagation mechanism of the CIS, that the distribution function error at the next step $Y^{k+1}$ depends linearly on the macroscopic error from the previous step $\Phi^{k}$. 

To determine the convergence rate, we perform a linear Fourier stability analysis. The error functions are decomposed into Fourier modes,
\begin{equation}\label{eq:error_function_fourier}
	\begin{aligned}
		Y_l^{k+1}\left(\bm{x},\bm{\xi}\right) &= e^k y_l\left(\bm{\xi}\right)\exp{\left(\mathrm{i}\bm{\theta}\cdot\bm{x}\right)}, \quad l=0,1,2, \\ 
		Y_R^{k+1}\left(\bm{x},\bm{\Omega}\right) &= e^k y_R\left(\bm{\Omega}\right)\exp{\left(\mathrm{i}\bm{\theta}\cdot\bm{x}\right)}, \\
		\Phi^{k+1}\left(\bm{x}\right) &= e^{k+1}\alpha_{M}\exp{\left(\mathrm{i}\bm{\theta}\cdot\bm{x}\right)},
	\end{aligned}
\end{equation}
where $\mathrm{i}$ is the imaginary unit, $\bm{\theta}$ is the wave vector representing the spatial frequency of the perturbation, and $e$ is the amplification factor, which quantifies the reduction in error magnitude per iteration. Note that the factor $e^k$, rather than $e^{k+1}$, emerges in the expression of error function $Y^{k+1}$, reflecting that $Y^{k+1}$ is driven by the macroscopic error of the $k$-th iteration.

By substituting the Fourier expansions \eqref{eq:error_function_fourier} into the error definition \eqref{eq:error_function_M} and the error propagation equations \eqref{eq:error_function_propagation}, we obtain the following $17 \times 17$ linear algebraic eigenvalue problem for the wave vector $\bm{\theta}$ and the amplification factor $e$:
\begin{equation}\label{eq:eigenfunction_cis}
	\begin{aligned}[b]
	e \alpha_{M}^{\mathsf{T}} &= C_{17\times17}\alpha_{M}^{\mathsf{T}} 
	= \sum_{l=0,1,2}\int{\phi_l^{\mathsf{T}}\left(\bm{\xi}\right) y_l\left(\bm{\xi}\right)}\mathrm{d}\bm{\xi} + \int{\phi_R^{\mathsf{T}}\left(\bm{\Omega}\right) y_R\left(\bm{\Omega}\right)}\mathrm{d}\bm{\Omega}.
	\end{aligned}
\end{equation}
The convergence rate is determined by the spectral radius, which is the maximum absolute value of the eigenvalues $e$. The iterative scheme is stable if $e_{\max} \leq 1$.

To establish a baseline for the coupled system, we first consider the convergence properties of the decoupled fields. It is known that the CIS is convergent for both the gas kinetic equation and the radiative transfer equation. However, the spectral radius of the CIS approaches unity in the near-continuum regime ($\mathrm{Kn}_{\mathrm{gas}} \rightarrow 0$), indicating severely degraded and impractically slow convergence. Similarly, for the radiative field, slow convergence is observed in the optically thick region ($\mathrm{Kn}_{\mathrm{photon}} \rightarrow 0$).

\begin{figure}[t]
\centering
    {\includegraphics[width=0.5\textwidth, clip = true]{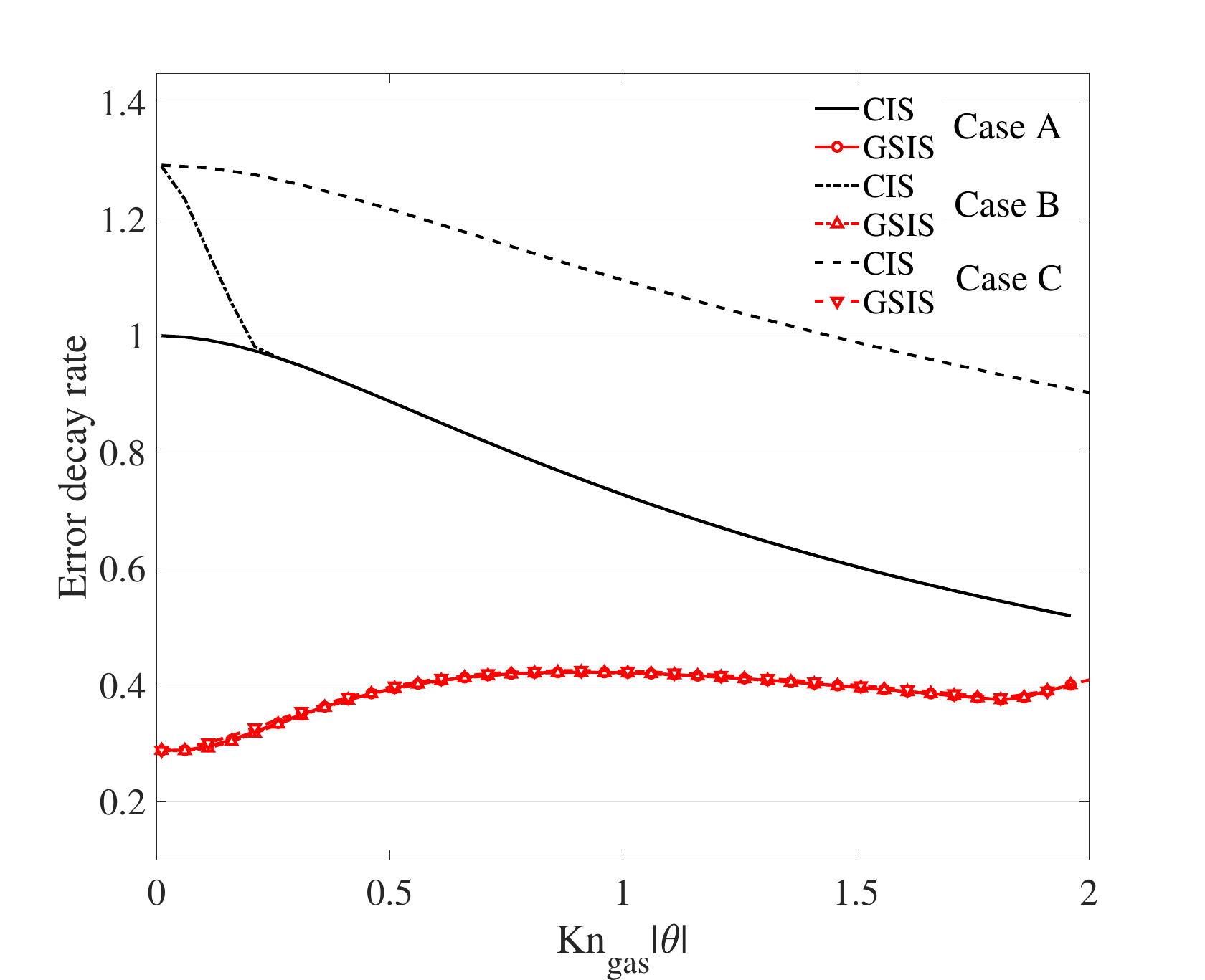}}
    \caption{The error decay rate as a function of $\mathrm{Kn}_{\mathrm{gas}}|\theta|$ in CIS and GSIS. Case A represents a collection of case parameters: $\left(\mathrm{Kn}_{\mathrm{gas}}/\mathrm{Kn}_{\mathrm{photon}}\right) \times {\sigma}_R < 0.03,~\left({\mathrm{Kn}_{\mathrm{gas}}}/{\mathrm{Kn}_{\mathrm{photon}}} \in [10^{-3},1],~{\sigma_R} \in [10^{-3},10^{-1}] \right)$; Case B: $\left(\mathrm{Kn}_{\mathrm{gas}}/\mathrm{Kn}_{\mathrm{photon}}\right) \times {\sigma}_R =0.05,~{\mathrm{Kn}_{\mathrm{gas}}}/{\mathrm{Kn}_{\mathrm{photon}}} =0.1,~{\sigma_R} =0.5$; Case C: $\left(\mathrm{Kn}_{\mathrm{gas}}/\mathrm{Kn}_{\mathrm{photon}}\right) \times {\sigma}_R =0.05,~{\mathrm{Kn}_{\mathrm{gas}}}/{\mathrm{Kn}_{\mathrm{photon}}} =10,~{\sigma_R} =0.005$. The rotational and vibrational degrees of freedom are $d_r=2$ and $d_v=1$, respectively, and the collision numbers are $Z_r=2.67$ and $Z_v=26.7$ in all the cases.}
    \label{fig:Fourier_stability_analysis}
\end{figure}

For the convergence of the coupled radiative gas flow problem, two crucial dimensionless parameters emerge from the analysis: the ratio $\mathrm{Kn}_{\mathrm{gas}}/\mathrm{Kn}_{\mathrm{photon}}$ and the parameter ${\sigma}_R$. $\mathrm{Kn}_{\mathrm{gas}}/\mathrm{Kn}_{\mathrm{photon}}$ is equivalent to the ratio of the radiative absorption cross-section to the molecular collision cross-section, determined by the gas interaction potential and molecular structure at a given temperature. The parameter ${\sigma}_R$, which is the inverse of the Boltzmann number, quantifies the relative strength of radiative energy transport to the thermal energy transport of gases. We calculate the CIS convergence rate for various combinations of the parameters $\mathrm{Kn}_{\mathrm{gas}}/\mathrm{Kn}_{\mathrm{photon}}$ and ${\sigma}_R$, as shown in Fig. \ref{fig:Fourier_stability_analysis}. A critical limitation is found that the CIS for the coupled system exhibits divergence (spectral radius $e_{\max} > 1$) in the near-continuum regime, when $\left(\mathrm{Kn}_{\mathrm{gas}}/\mathrm{Kn}_{\mathrm{photon}}\right) \times {\sigma}_R$ is approximately larger than $0.03$.

The primary reason for this instability is that the coupling terms between the gas and radiation fields are treated explicitly (i.e., based on the macroscopic quantities $\Phi_{\Delta T_v}^{k}$ and $\Phi_{\Delta e_R}^{k}$ from the previous step $k$), preventing the full-field implicit update. This coupling strength factor, $\left(\mathrm{Kn}_{\mathrm{gas}}/\mathrm{Kn}_{\mathrm{photon}}\right) \times {\sigma}_R$, appearing in the off-diagonal blocks of the characteristic matrix $C_{17\times17}$, acts like a positive feedback loop in the eigenvalue system. When this product becomes sufficiently large, the coupling effect dominates the iterative process, causing the spectral radius of the matrix to exceed unity and leading to the divergence of the CIS. This critical finding motivates the need for an implicit, accelerated scheme.

\subsection{Convergence analysis of GSIS}

GSIS is designed to accelerate the convergence of the coupled gas-photon system by implicitly solving macroscopic synthetic equations alongside the kinetic equations. To be specific, when the velocity distribution function $h^k$ has been obtained from a previous iteration step $k$, the next iteration involves a kinetic step to update the distribution functions to $h^{k+1/2}$, and a subsequent macroscopic solver, where the macroscopic quantities $M^{k+1}$ are obtained by solving the derived synthetic equations. This macroscopic system, closed by the constitutive relations with high-order terms, is essential for rapid information propagation and high efficiency across all flow regimes.

The macroscopic quantities $M^{k+1}$ at the $(k+1)$-th iteration step are obtained by solving the following set of linearized synthetic equations, which implicitly enforce the conservation of mass, momentum, and energy across the coupled system,
\begin{equation}\label{eq:steady_state_linear_macroscopic_equation}
	\begin{aligned}
		\nabla  \cdot \bm{u}^{k+1} = 0, \\
		\nabla \left(\Delta{\rho}^{k+1}\right) +\nabla \left(\Delta{T}_t^{k+1}\right) +\nabla \cdot\bm{\sigma}^{k+1} = 0, \\
		\nabla  \cdot \left(\bm{q}_t^{k+1}+\bm{q}_r^{k+1}+\bm{q}_v^{k+1}+\bm{q}_R^{k+1}\right) = 0, \\
		\nabla  \cdot \bm{q}_r^{k+1} = \sqrt{\frac{\pi}{2}}\frac{1}{\mathrm{Kn}_{\mathrm{gas}}}\frac{3d_r}{2(3+d_r)Z_r}\left(\Delta{T}_t^{k+1}-\Delta{T}_r^{k+1}\right), \\
		\nabla  \cdot \bm{q}_v^{k+1} = \sqrt{\frac{\pi}{2}}\frac{1}{\mathrm{Kn}_{\mathrm{gas}}}\frac{3d_v}{2(3+d_v)Z_v}\left(\Delta{T}_t^{k+1}-\Delta{T}_v^{k+1}\right) + \frac{1}{\mathrm{Kn}_{\mathrm{photon}}}\left(\Delta{e}_R^{k+1}-16{\sigma}_R\Delta{T}_v^{k+1}\right), \\
		\nabla  \cdot \bm{q}_R^{k+1} = \frac{1}{\mathrm{Kn}_{\mathrm{photon}}}\left(16{\sigma}_R\Delta{T}_v^{k+1}-\Delta{e}_R^{k+1}\right).
	\end{aligned}
\end{equation} 
These are closed using the constitutive relations,
\begin{equation}\label{eq:linear_constitutive}
	\begin{aligned}
		\bm{\sigma}^{k+1} &= \int{\left(\bm{\xi}\bm{\xi}-\frac{1}{3}\xi^2\mathrm{I}\right)h_0^{k+1/2}}\mathrm{d}\bm{\xi} \\
		&+ \left(\frac{2}{\pi}\right)^{1/2}\text{Kn}_{\mathrm{gas}}\left(\nabla\bm{u}^{k+1/2}+\left(\nabla\bm{u}^{k+1/2}\right)^{\mathrm{T}}-\frac{2}{3}\nabla\cdot\bm{u}^{k+1/2}\bm{\mathrm{I}}\right) \\
		&- \left(\frac{2}{\pi}\right)^{1/2}\text{Kn}_{\mathrm{gas}}\left(\nabla\bm{u}^{k+1}+\left(\nabla\bm{u}^{k+1}\right)^{\mathrm{T}}-\frac{2}{3}\nabla\cdot\bm{u}^{k+1}\bm{\mathrm{I}}\right), \\
		\bm{q}_t^{k+1} &= \int{\left(\frac{1}{2}\xi^2-\frac{5}{2}\right)\bm{\xi}h_0^{k+1/2}}\mathrm{d}\bm{\xi} 
		+ \sum_{l=t,r,v}\kappa_{tl} \nabla\left(\Delta{T}_l^{k+1/2}-\Delta{T}_l^{k+1}\right), \\
		\bm{q}_r^{k+1} &= \int{\bm{\xi}\left(h_1^{k+1/2}-\frac{d_r}{2}h_0^{k+1/2}\right)}\mathrm{d}\bm{\xi}
		+ \sum_{l=t,r,v}\kappa_{rl} \nabla\left(\Delta{T}_l^{k+1/2}-\Delta{T}_l^{k+1}\right), \\
		\bm{q}_v^{k+1} &= \int{\bm{\xi}\left(h_2^{k+1/2}-\frac{d_v}{2}h_0^{k+1/2}\right)}\mathrm{d}\bm{\xi}
		+ \sum_{l=t,r,v}\kappa_{vl} \nabla\left(\Delta{T}_l^{k+1/2}-\Delta{T}_l^{k+1}\right), \\
		\bm{q}_R^{k+1} &= \int{\bm{\Omega}h_R^{k+1/2}}\mathrm{d}\bm{\Omega}
		+ \kappa_R \nabla\left(\Delta{e}_R^{k+1/2}-\Delta{e}_R^{k+1}\right), \\
	\end{aligned}
\end{equation}
where the thermal transport parameters $\bm{\kappa}_{\text{gas}}$ and $\kappa_R$ are measured at equilibrium state due to linearization.

To calculate the convergence rate of GSIS, the error functions $Y^{k+1/2}$ are defined over consecutive half-steps and decomposed into Fourier modes:
\begin{equation}\label{eq:error_function_gsis}
	\begin{aligned}[b]
		Y_l^{k+1/2}\left(\bm{x},\bm{\xi}\right) &\equiv h_l^{k+1/2}\left(\bm{x},\bm{\xi}\right)-h_l^{k-1/2}\left(\bm{x},\bm{\xi}\right) = e^k y_l\left(\bm{\xi}\right)\exp{\left(\mathrm{i}\bm{\theta}\cdot\bm{x}\right)}, \quad l=0,1,2, \\
		Y_R^{k+1/2}\left(\bm{x},\bm{\Omega}\right) &\equiv h_R^{k+1/2}\left(\bm{x},\bm{\Omega}\right)-h_R^{k-1/2}\left(\bm{x},\bm{\Omega}\right) = e^k y_R\left(\bm{\Omega}\right)\exp{\left(\mathrm{i}\bm{\theta}\cdot\bm{x}\right)}.
	\end{aligned}
\end{equation}

By substituting these error functions $Y^{k+1/2}$ and the macroscopic error $\Phi^{k+1} = e^{k+1}\alpha_{M}\exp{\left(\mathrm{i}\bm{\theta}\cdot\bm{x}\right)}$ into the macroscopic synthetic and constitutive equations, we obtain the following set of linear algebraic equations for the eigenvalue $e$,
\begin{equation}\label{eq:eigenfunction_gsis}
	\begin{aligned}
		e&\left[\mathrm{i}\bm{\theta}\cdot{\alpha_{\bm{u}}}\right] = S_{1}, \\		
		e&\left[\mathrm{i}\bm{\theta}\left(\alpha_{\Delta{\rho}}+\alpha_{\Delta{T}_t}\right) + \left(\frac{2}{\pi}\right)^{1/2}\text{Kn}_{\mathrm{gas}}\theta^2{\alpha_{\bm{u}}} \right] = S_{2-4}, \\
		e&\left[\mathrm{i}\bm{\theta}\cdot\left({\alpha_{\bm{q}_t}+\alpha_{\bm{q}_r}+\alpha_{\bm{q}_v}}\right) + \theta^2\kappa_R{\alpha_{\Delta{e}_R}}\right] = S_{5}, \\
		e&\left[\mathrm{i}\bm{\theta}\cdot\alpha_{\bm{q}_r} + \sqrt{\frac{\pi}{2}}\frac{1}{\mathrm{Kn}_{\mathrm{gas}}}\frac{3d_r}{2(3+d_r)Z_r}\left(\alpha_{\Delta{T}_r}-\alpha_{\Delta{T}_t}\right)\right]	= S_{6}, \\	
		e&\left[\mathrm{i}\bm{\theta}\cdot\alpha_{\bm{q}_v} + \sqrt{\frac{\pi}{2}}\frac{1}{\mathrm{Kn}_{\mathrm{gas}}}\frac{3d_v}{2(3+d_v)Z_v}\left(\alpha_{\Delta{T}_v}-\alpha_{\Delta{T}_t}\right) + \frac{1}{\mathrm{Kn}_{\mathrm{photon}}}\left(16{\sigma}_R\alpha_{\Delta{T}_v}-\alpha_{\Delta{e}_R}\right)\right]	= S_{7}, \\
		e&\left[{\theta}^2 \kappa_R \alpha_{\Delta{e}_R} + \frac{1}{\mathrm{Kn}_{\mathrm{photon}}}\left(\alpha_{\Delta{e}_R}-16{\sigma}_R\alpha_{\Delta{T}_v}\right)\right] = S_{8}, \\
		e&\left[\alpha_{\bm{q}_t} + \mathrm{i}\bm{\theta} \kappa_t \alpha_{\Delta{T}_t}\right] = S_{9-11}, \\
		e&\left[\alpha_{\bm{q}_r} + \mathrm{i}\bm{\theta} \kappa_r \alpha_{\Delta{T}_r}\right] = S_{12-14}, \\
		e&\left[\alpha_{\bm{q}_v} + \mathrm{i}\bm{\theta} \kappa_v \alpha_{\Delta{T}_v}\right] = S_{15-17}, \\
	\end{aligned}
\end{equation}
where the source terms $S_{1-17}$ are,
\begin{equation}\label{eq:eigenfunction_gsis_S}
	\begin{aligned}
		S_{1} &= 0, \\
		S_{2-4} &= \int {\left[-\mathrm{i}\bm{\theta}\cdot\left(\bm{\xi}\bm{\xi}-\frac{1}{3}\xi^2\mathrm{I}\right) + \left(\frac{2}{\pi}\right)^{1/2}\text{Kn}_{\mathrm{gas}}\left(\theta^2\bm{\xi}+\frac{1}{3}\bm{\theta}\left(\bm{\theta}\cdot\bm{\xi}\right)\right)\right]y_0} \mathrm{d}\bm{\xi}, \\
		S_{5} &= \int{\left[-\mathrm{i}\bm{\theta}\cdot\bm{\Omega} + \kappa_R\theta^2\right] y_R}\mathrm{d}\bm{\Omega}, \\
		S_{6} &= 0, \\
		S_{7} &= 0, \\
		S_{8} &= \int{\left[-\mathrm{i}\bm{\theta}\cdot\bm{\Omega} + \kappa_R\theta^2\right] y_R}\mathrm{d}\bm{\Omega}, \\
		S_{9-11} &= \int \left[\left(\frac{1}{2}\xi^2-\frac{5}{2}\right)\bm{\xi} +\mathrm{i}\bm{\theta} \kappa_t\left(\frac{1}{3}\xi^2-1\right)\right]y_0 \mathrm{d}\bm{\xi}, \\
		S_{12-14} &= \int \left[\left(\bm{\xi}+\mathrm{i}\bm{\theta}\frac{2 \kappa_r}{d_r}\right)y_1 - \left(\frac{d_r}{2}\bm{\xi}+\mathrm{i}\bm{\theta} \kappa_r\right)y_0\right]\mathrm{d}\bm{\xi}, \\
		S_{15-17} &= \int \left[\left(\bm{\xi}+\mathrm{i}\bm{\theta}\frac{2 \kappa_v}{d_v}\right)y_2 - \left(\frac{d_v}{2}\bm{\xi}+\mathrm{i}\bm{\theta} \kappa_v\right)y_0\right]\mathrm{d}\bm{\xi}. \\
	\end{aligned}
\end{equation}
Equation \eqref{eq:eigenfunction_gsis} with the source terms \eqref{eq:eigenfunction_gsis_S} can be rewritten in the matrix form as,
\begin{equation}\label{eq:eigenfunction_gsis_matrix}
	\begin{aligned}[b]
		eL_{17\times17}\alpha_{M}^{\mathsf{T}} = R_{17\times17}\alpha_{M}^{\mathsf{T}}.
	\end{aligned}
\end{equation}
The convergence rate is obtained by numerically computing the eigenvalues $e$ of the matrix $(L^{-1}_{17\times17}R_{17\times17})$ and determining its spectral radius.

Based on the Fourier stability analysis, we calculated the convergence rate (spectral radius $e_{\max}$) of the GSIS for various coupling parameters $\mathrm{Kn}_{\mathrm{gas}}/\mathrm{Kn}_{\mathrm{photon}}$ and ${\sigma}_R$, and compared the results directly with the previously analyzed CIS (Fig. \ref{fig:Fourier_stability_analysis}). The results demonstrate that the GSIS successfully overcomes the limitations of the CIS. In the near-continuum regime ($\mathrm{Kn}_{\mathrm{gas}}\rightarrow 0, \mathrm{Kn}_{\mathrm{photon}} \rightarrow 0$), where the CIS exhibits slow convergence ($e_{\max} \rightarrow 1$) or even divergence ($e_{\max} > 1$) for strongly coupled cases, the GSIS spectral radius remains significantly less than unity. Even when the coupling strength factor $\left(\mathrm{Kn}_{\mathrm{gas}}/\mathrm{Kn}_{\mathrm{photon}}\right) \times {\sigma}_R$ is large, which corresponds to high radiation intensity, strong coupling, near-continuum gas flow, and optically thin radiative transport conditions, the spectral radius of the GSIS remains below 0.5. In other words, the number of iterations required to reduce the solution error by three orders of magnitude is consistently reduced to ten iterations. This performance contrasts with the instability observed in the CIS under similar conditions. This high convergence efficiency means that the GSIS provides a robust and rapid solution method for multiscale, coupled gas-radiation problems.

%% file: Asymptotic.tex
\section{Asymptotic-preserving property of GSIS}\label{sec:asymptotic}

Conventional iteration solvers for coupled gas--radiation systems impose severe grid-resolution constraints: spatial cells must resolve both gas and photon mean free paths to prevent numerical dissipation from contaminating physical transport. This requirement becomes prohibitive in near-continuum regimes where both $\mathrm{Kn}_{\mathrm{gas}}$ and $\mathrm{Kn}_{\mathrm{photon}}$ approach zero. We demonstrate that GSIS eliminates this restriction by establishing its asymptotic-preserving property for the spatially discretized kinetic equations. To be specific, a numerical scheme is considered asymptotic-preserving if it recovers the correct macroscopic limits, that is, the radiative NSF equations in this case, as the Knudsen numbers vanish, while maintaining spatial resolution independent of the mean free path.

The discretized kinetic equation on a mesh with spacing $\Delta x$ take the form,
\begin{equation}\label{eq:ap_equation}
    \begin{aligned}
        \bm{\xi}\cdot\nabla_{\delta}f_l &= \mathcal{J}_l, \quad l=0,1,2 \\
        \bm{\Omega}\cdot\nabla_{\delta}I_R &= \mathcal{J}_R,
    \end{aligned}
\end{equation}
where $\nabla_{\delta} = \nabla + \mathcal{O}\left(\Delta x^n\right)$ denotes the discrete gradient operator of order $n$ accuracy, and $\mathcal{J}$ represents the collision terms given in Eq. \eqref{eq:kinetic_equation}.

Applying the Chapman--Enskog expansion directly to this discretized system, we express the distribution functions and radiative intensity as power series in the small parameters,
\begin{equation}\label{eq:ap_expansion}
\begin{aligned}
f_l &= f_l^{(0)} + \mathrm{Kn}_{\mathrm{gas}} f_l^{(1)} + \mathrm{Kn}_{\mathrm{gas}}^2 f_l^{(2)} +\cdots , \quad l=0,1,2 \\
I_R &= I_R^{(0)} + \mathrm{Kn}_{\mathrm{photon}} I_R^{(1)} + \mathrm{Kn}_{\mathrm{photon}}^2 I_R^{(2)} +\cdots .
\end{aligned}
\end{equation}
The stress and heat flux admit analogous expansions in terms of Knudsen numbers, while the conservative variables $\left[\rho, \bm{u}, T_{t/r/v/R}\right]$ are determined solely by the zeroth-order components $f_l^{(0)}$ and $I_R^{(0)}$.
Note that the expansion now explicitly includes the numerical error term: the higher order corrections contain both physical contributions from kinetic theory and discretization artifacts of order $\Delta x^n$.

Since $\mathrm{Kn}_{\mathrm{gas}}/\mathrm{Kn}_{\mathrm{photon}}$ is equivalent to the ratio of radiative absorption to molecular collision cross-section, the two Knudsen numbers approach to zero at comparable rates, and we assume $\mathrm{Kn}_{\mathrm{gas}} \sim \mathrm{Kn}_{\mathrm{photon}}$. Furthermore, internal energy relaxation is assumed substantially slower than elastic collisions, yielding $Z_r \sim Z_v \sim \mathcal{O}\left(\mathrm{Kn}_{\mathrm{gas}}^{-1}\right)$.

To demonstrate asymptotic behavior, we select $\Delta x \sim \mathcal{O}\left(1\right)$, which implies that the spatial cell size is independent of the mean free paths. Substituting expansions \eqref{eq:ap_expansion} into Eq. \eqref{eq:ap_equation}, and collecting terms of order $\mathcal{O}\left(\mathrm{Kn}_{\mathrm{gas}}^{-1}\right)$ and $\mathcal{O}\left(\mathrm{Kn}_{\mathrm{photon}}^{-1}\right)$, we obtain the leading-order solution recovers the equilibrium distribution functions with distinct temperatures for each energy mode,
\begin{equation}\label{eq:ap_1}
\begin{aligned}
f_0^{(0)} = \rho\left(\frac{1}{2\pi T_t}\right)^{3/2}\exp\left(-\frac{c^2}{2T_t}\right), \quad 
f_1^{(0)} = \frac{d_r}{2}f_0^{(0)}, \quad
f_2^{(0)} = \frac{d_v(T_v)}{2}f_0^{(0)}, \quad
I_R^{(0)} = \frac{1}{\pi}\sigma_R T_v^4.
\end{aligned}
\end{equation}

At the next order $\mathcal{O}\left(\mathrm{Kn}_{\mathrm{gas}}^{0}\right)$ and $\mathcal{O}\left(\mathrm{Kn}_{\mathrm{photon}}^{0}\right)$, the corrections are determined by the discretized transport equations,
\begin{equation}\label{eq:ap_2}
    \begin{aligned}
        f_0^{(1)} &= \left(1-\frac{1}{Z_r}-\frac{1}{Z_v}\right)g_{0t}^{(1)}-\frac{1}{Z_r}g_{0r}^{(1)}-\frac{1}{Z_v}g_{0v}^{(1)} -\bm{\xi}\cdot\nabla_{\delta}f_0^{(0)}, \\
        f_1^{(1)} &= \left(1-\frac{1}{Z_r}-\frac{1}{Z_v}\right)g_{1t}^{(1)}-\frac{1}{Z_r}g_{1r}^{(1)}-\frac{1}{Z_v}g_{1v}^{(1)} -\bm{\xi}\cdot\nabla_{\delta}f_1^{(0)}, \\
        f_2^{(0)} &= \left(1-\frac{1}{Z_r}-\frac{1}{Z_v}\right)g_{2t}^{(1)}-\frac{1}{Z_r}g_{2r}^{(1)}-\frac{1}{Z_v}g_{2v}^{(1)} +\frac{f_0^{(0)}}{\rho} \int{I_R^{(1)}\mathrm{d}\bm{\Omega}} -\bm{\xi}\cdot\nabla_{\delta}f_2^{(0)}, \\
        I_R^{(1)} &= -\bm{\Omega}\cdot\nabla_{\delta}I_R^{(0)}.
    \end{aligned}
\end{equation}
Calculating moments of $f_l^{(0)}+\mathrm{Kn}_{\mathrm{gas}} f_l^{(1)},~(l=0,1,2)$ and $I_R^{(0)}+\mathrm{Kn}_{\mathrm{photon}}I_R^{(1)}$ produces the constitutive relations containing both physical and numerical errors,
\begin{equation}\label{eq:ap_constitutive}
    \begin{aligned}
        \bm{\sigma} =& -\left(\frac{2}{\pi}\right)^{1/2}\text{Kn}_{\mathrm{gas}} \left(\nabla\bm{u}+\nabla\bm{u}^{\top}-\frac{2}{3}\nabla\cdot\bm{u}\mathbf{I} + \mathcal{O}\left(\Delta x^n\right)\right) + \mathcal{O}\left(\mathrm{Kn}_{\mathrm{gas}}^{2}\right), \\
        \begin{bmatrix}
		    \bm{q}_t \\
		    \bm{q}_r \\
		    \bm{q}_v
        \end{bmatrix}
        =& -\left(\frac{1}{2\pi}\right)^{1/2}\text{Kn}_{\mathrm{gas}} 
        \begin{bmatrix}
		    A_{tt} & A_{tr} & A_{tv} \\
		    A_{rt} & A_{rr} & A_{rv} \\
		    A_{vt} & A_{vr} & A_{vv}
        \end{bmatrix}^{-1}
        \begin{bmatrix}
		    5 & 0 & 0 \\
		    0 & d_r & 0 \\
		    0 & 0 & d_v(T_v)
        \end{bmatrix}
        \begin{bmatrix}
		    \nabla T_t \\
		    \nabla T_r \\
		    \nabla T_v
        \end{bmatrix} \\
    &- \left(\frac{1}{2\pi}\right)^{1/2}\text{Kn}_{\mathrm{gas}} \times \mathcal{O}\left(\Delta x^n\right) + \mathcal{O}\left(\mathrm{Kn}_{\mathrm{gas}}^{2}\right) , \\
    \bm{q}_{R} & = -\frac{4{\sigma}_R}{3} \mathrm{Kn}_{\mathrm{photon}} \left(\nabla T_v^4 
+ \mathcal{O}\left(\Delta x^n\right) \right)+\mathcal{O}\left(\mathrm{Kn}_{\mathrm{photon}}^{2}\right),
    \end{aligned}
\end{equation}

For conventional schemes, the $\mathcal{O}(\Delta x^n)$ error terms prevent recovery of the correct macroscopic limits unless $\Delta x\sim\mathcal{O}(\mathrm{Kn}_{\mathrm{gas}}^{1/n})$ and $\Delta x\sim\mathcal{O}(\mathrm{Kn}_{\mathrm{photon}}^{1/n})$.
In GSIS, the decomposition of constitutive relations (Eq. \eqref{eq:split_constitutive}) constructs the higher-order terms as,
\begin{equation}\label{eq:HoT_discrete}
    \begin{aligned}
        \text{HoT}_{\bm{\sigma}} =& \bm{\sigma} + \left(\frac{2}{\pi}\right)^{1/2}\text{Kn}_{\mathrm{gas}}  \left(\nabla_{\delta}\bm{u}+\nabla_{\delta}\bm{u}^{\top}-\frac{2}{3}\nabla_{\delta}\cdot\bm{u}\mathbf{I}\right), \\
        \begin{bmatrix}
		    \text{HoT}_{\bm{q}_t} \\
		    \text{HoT}_{\bm{q}_r} \\
		    \text{HoT}_{\bm{q}_v}
        \end{bmatrix}
        =& 
        \begin{bmatrix}
		    \bm{q}_t \\
		    \bm{q}_r \\
		    \bm{q}_v
        \end{bmatrix}
        + \left(\frac{1}{2\pi}\right)^{1/2}\text{Kn}_{\mathrm{gas}} 
        \begin{bmatrix}
		    A_{tt} & A_{tr} & A_{tv} \\
		    A_{rt} & A_{rr} & A_{rv} \\
		    A_{vt} & A_{vr} & A_{vv}
        \end{bmatrix}^{-1}
        \begin{bmatrix}
		    5 & 0 & 0 \\
		    0 & d_r & 0 \\
		    0 & 0 & d_v(T_v)
        \end{bmatrix}
        \begin{bmatrix}
		    \nabla_{\delta} T_t \\
		    \nabla_{\delta} T_r \\
		    \nabla_{\delta} T_v
        \end{bmatrix}, \\
        \text{HoT}_{\bm{q}_{R}} =& \bm{q}_{R} + \frac{4{\sigma}_R}{3} \mathrm{Kn}_{\mathrm{photon}} \nabla_{\delta} T_v^4,
    \end{aligned}
\end{equation}
Since both kinetic moments and linear constitutive relations employ the identical discrete operator $\nabla_{\delta}$, the leading-order discretization errors ($\mathcal{O}(\Delta x^n\mathrm{Kn}_{\mathrm{gas}})$, $\mathcal{O}(\Delta x^n\mathrm{Kn}_{\mathrm{photon}})$) cancel identically, yielding,
\begin{equation}
    \text{HoT}_{\bm{\sigma}} \sim \text{HoT}_{\bm{q}_{t/r/v}} \sim \mathcal{O}(\mathrm{Kn}_{\mathrm{gas}}^2), \quad
    \text{HoT}_{\bm{q}_R} \sim \mathcal{O}(\mathrm{Kn}_{\mathrm{photon}}^2).
\end{equation}

In the double limit $\mathrm{Kn}_{\mathrm{gas}}\to0$ and $\mathrm{Kn}_{\mathrm{photon}}\to0$, the higher-order contributions to constitutive relations vanish, and the constitutive relations reduce to their radiative NSF forms with truncation errors independent of $\Delta x$. Consequently, GSIS correctly recovers the macroscopic equations even when $\Delta x\sim\mathcal{O}(1)$, which is the essence of the asymptotic-preserving property for coupled radiative gas flows.

The analysis assumes smooth macroscopic fields, while steep gradients such as shock fronts or Knudsen layers require local mesh refinement to resolve the underlying physics, irrespective of asymptotic-preserving considerations.

%% file: Numerical_schemes.tex
\section{Numerical implementation of GSIS} \label{sec:implementation}

This section details the numerical implementation of the GSIS method through the solution of two coupled equation sets: the mesoscopic kinetic equations~\eqref{eq:kinetic_equation} governing distribution function evolution, and the macroscopic synthetic equations~\eqref{eq:macroscopic_equation} describing conserved macroscopic quantities. The latter constitute a generalized radiative NSF system augmented by non-equilibrium source terms, that is, the higher-order terms extracted from kinetic moments. Both systems are discretized using established computational fluid dynamics techniques. Notably, the modular structure allows independent optimization of each solver. In the present work, a second-order unstructured finite volume method is applied and provides flexible spatial discretization. The specific numerical formulation and iterative procedures for both the mesoscopic and macroscopic equations are detailed below.

Since there are four kinetic equations in Eq.~\eqref{eq:kinetic_equation}, which differ only in the form of their collision terms. To simplify the presentation, we use $f$ to denote both the gas velocity distribution functions and the photon intensity, while $g$ represents their corresponding reference distribution functions.

The integral forms of the mesoscopic and macroscopic governing equations, applied over a control volume $V$ with its boundary $\partial V$, can be written as,
\begin{equation}
\begin{aligned}
    &\frac{\partial }{\partial t}\int_{V}f\myd V + \oint_{\partial V} \bxi f\cdot \myd \bm{S}= \int_{V}\frac{g-f}{\tau}\myd V,\\
    &\frac{\partial }{\partial t}\int_{V}\bm{W}\myd V + \oint_{\partial V} (\bm{F}_c+\bm{F}_v)\cdot \myd \bm{S} =\int_{V}\bm{Q}\myd V, 
\end{aligned}
\end{equation}
where the velocity vector $\bm{\xi}$ for the distribution function of gas will change to $\bm{\Omega}$ when it applies to the equation of $I_R$; similarly, the characteristic time $\tau$ will be $c_l \mathrm{Kn}_{\mathrm{photon}}$. The surface element is $\myd\bm{S}=\bm{n} \myd S$, with $\bm{n}$ as the outward unit normal vector and $\myd S$ is the area of the surface element. 

The macroscopic conservative variables $\bm{W}$, convective fluxes $\bm{F}_c$, viscous fluxes $\bm{F}_v$, and source terms $\bm{Q}$ are defined as vectors corresponding to mass, momentum, total energy, rotational energy, vibrational energy, and radiation energy,
\begin{equation}
    \begin{aligned}
\bm{W}&=\left[
    \rho,~
	\rho \bm{u},~
	e,~
    e_{r},~
    e_{v},~
    e_R
\right]^{\mathrm{T}} \\
\bm{F}_c&=\left[
    \rho \bm{u},~
	\rho \bm{u}\bm{u}+p_{t} \mathbf{I},~
	\bm{u} (e_{\text{gas}}+p_{t}),~
    \bm{u} e_{r},~
    \bm{u} e_{v},~
    0
\right]^{\mathrm{T}} \\
\bm{F}_v&=\left[
    0,~
	\bm{\sigma},~
	\bm{\sigma}\cdot\bm{u}+\bm{q},~
    \bm{q}_r,~
    \bm{q}_v,~
    \bm{q}_R
\right]^{\mathrm{T}} \\
\bm{Q}&=\left[
    0,~
	0,~
	0,~
    {\frac{e_{tr}-e_{r}}{Z_r\tau}},~
    {\frac{e_{tv}-e_{v}}{Z_v\tau}}-{\frac{e_{vR}-e_{R}}{\mathrm{Kn}_{\mathrm{photon}}}},~
    {\frac{e_{vR}-e_{R}}{\mathrm{Kn}_{\mathrm{photon}}}}
\right]^{\mathrm{T}}.
    \end{aligned}
\end{equation}

For steady-state problems, the governing equations are discretized in time using the first-order backward Euler scheme and reformulated in an incremental form to construct a robust implicit solver. This approach allows for large time steps $\Delta t^k = t^{k+1} - t^{k}$, accelerating convergence by overcoming the restrictive explicit Courant--Friedrichs--Lewy (CFL) condition.

Let $\Delta f^{k} = f^{k+1} - f^{k}$ and $\Delta \bm{W}^{k} = \bm{W}^{k+1} - \bm{W}^{k}$ be the increments of the distribution function and the conservative variables, respectively. The implicit incremental forms of the mesoscopic and macroscopic equations are,
\begin{subequations}\label{eq:dis_micro}
\begin{align}
    \left( \frac{1}{\Delta t_i} + \frac{1}{\tau_i^k}\right)\Delta f_i^{k} + \frac{1}{V_i}\sum_{j\in N(i)} \xi_n\Delta f_{ij}^{k}S_{ij}&=\underbrace{ \frac{g^{k}_{i}-f^{k}_{i}}{\tau^{k}_i}-\frac{1}{V_i}\sum_{j\in N(i)} \xi_nf_{ij}^{k}S_{ij}}_{r_i^k},\label{eq:delta_micro}\\
    \left(\frac{1}{\Delta t_i}- \frac{\partial \bm{Q}_i}{\partial \bm{W}}\right)\Delta \bm{W}_i^{k} + \frac{1}{V_i} \sum_{j\in N(i)} \Delta\bm{F}_{ij}^{k}S_{ij}&=\underbrace{-\frac{1}{V_i}\sum_{j\in N(i)} \bm{F}_{ij}^{k}S_{ij}+\bm{Q}_i^{k}}_{R_i^k},\label{eq:delta_macro}
\end{align}
\end{subequations}
where the subscripts $i$ and $j$ denote the global index of a control volume and the local index of its neighboring cell, respectively. $N(i)$ represents the set of neighboring cells of cell $i$, and the subscript $ij$ denotes the interface connecting cells $i$ and $j$. For example, $f_i$ and $\bm{W}_i$ denote the cell-averaged distribution function and conservative variables in control volume $i$, respectively. $V_i$ is the volume of cell $i$, $S_{ij}$ is the area of the interface $ij$, and $\xi_n = \bxi\cdot\bn_{ij}$, where $\bm{n}_{ij}$ is the unit normal vector pointing from cell $i$ to cell $j$. The right-hand side of the equation is referred as the residual term, where $r^k$ and $R^k$ represent the residuals of the mesoscopic and macroscopic equations, respectively, and $f_{ij}$ and $\bm{F}_{ij}$ denote the mesoscopic and macroscopic fluxes at interface $ij$. When the governing equations converge, the increments vanish.

The schemes for the flux increments $\Delta f_{ij}$ and $\Delta \bm{F}_{ij}$ are formulated as first-order for simplicity and robustness in the implicit operator,
\begin{enumerate}
	\item Mesoscopic flux increment $\Delta f_{ij}$: a first-order upwind scheme is used,
	\begin{equation}
		\Delta f_{ij} = \xi_n^+\Delta f_i + \xi_n^-\Delta f_j,
	\end{equation}
	where $\xi_n^{\pm}=[\xi_n \pm |\xi_n|]/2$ denotes the upwind direction with respect to the interface normal direction.
	\item Macroscopic flux increment $\Delta \bm{F}_{ij}$: an Euler-like scheme is used, incorporating a robust numerical viscosity $\Gamma_{ij}$,
	\begin{equation}
		\Delta \bm{F}_{ij} = \frac{1}{2}\left[\Delta \bm{F}_i + \Delta \bm{F}_j+\Gamma_{ij}(\Delta\bm{W}_i - \Delta\bm{W}_j)\right],
	\end{equation}
	where $\Gamma_{ij}=|u_n| + c_s + {2\mu}/{\rho|\bm{n}_{ij}\cdot(\bm{x}_j)-\bm{x}_i|}$ serves as the approximate spectral radius, containing a combined convective $|u_n|$ ( product of the velocity vector and the interface normal vector), acoustic $c_s$ (sound speed), and viscous component for stability.
\end{enumerate}

For the residual terms $r_i^k$ and $R_i^k$ on the right-hand side, a second-order accuracy is maintained using the Venkatakrishnan limiter $\phi$ ~\cite{venkatakrishnan1995convergence} for interface interpolation,
\begin{equation}\label{eq:deltaflux_micro}
\begin{aligned}
    f_{ij} &= \xi_n^+\Delta f_L + \xi_n^-\Delta f_R, \quad f_{L/R} = f_{i/j} + \phi_f \nabla f_{i/j}\cdot \bm{x},\\
    \bm{F}_{ij} &= \mathbb{F}(W_L, W_R, \myd S_{ij}), \quad \bm{W}_{L/R} = \bm{W}_{i/j} + \phi_{\bm{W}} \nabla \bm{W}_{i/j}\cdot \bm{x},
\end{aligned}
\end{equation}
where $\Delta \bm{x}$ is the distance vector from the cell center to the interface center, and $\mathbb{F}$ is the numerical flux function. Since $\sum_{j\in N(i)}\bm{F}_{i}A_{ij}=0$ is satisfied over the control volume, the macroscopic flux increment can be written as $\Delta \bm{F}_{ij}=[\Delta \bm{F}_j+\Gamma_{ij}(\Delta \bm{W}_i - \Delta \bm{W}_j)]/2$. 

Substituting the flux increment definitions into the incremental equations \eqref{eq:dis_micro} results in a set of coupled linear algebraic equations
\begin{equation} 
	\begin{aligned} 
		d_i\Delta f_i^{k} + \frac{1}{2V_i}\sum_{j\in N(i)} \bm{\xi}_n^-S_{ij}\Delta f_j^{k}&=r_i^k,\\
		D_i\Delta \bm{W}_i^{k} + \frac{1}{2V_i}\sum_{j\in N(i)} \left(\Delta\bm{F}_{j}^{k} - \Gamma_{ij}\Delta \bm{W}_j^k\right)S_{ij}&=\bm{R}_i^k,
	\end{aligned} 
\end{equation}
where the macroscopic flux increments are $\Delta \bm{F}_j=\bm{F}(\bm{W}_j + \Delta \bm{W}_j) - \bm{F}(\bm{W}_j)$, and the matrix elements are given by,
\begin{equation}
    \begin{aligned}
d_i&=\frac{1}{\Delta t_i} + \frac{1}{\tau_i^k} + \frac{1}{V_i}\sum_{j\in N(i)}\bm{\xi}_n^+S_{ij},\\
D_i &= \frac{1}{\Delta t_i}+\frac{1}{V_i}\sum_{j\in N(i)}\Gamma_{ij}S_{ij}- \left(\frac{\partial \bm{Q}}{\partial \bm{W}}\right)^k_i.
    \end{aligned}
\end{equation}

The above equations are solved using a point relaxation method, analogous to the Lower-Upper Symmetric-Gauss-Seidel (LU-SGS) method. This method typically involves a forward and backward sweep per iteration. The local pseudo time step is evaluated as $\Delta t = \text{CFL}\times \Delta x /(|u_n| + c_s)$. To maximize efficiency, different iteration counts and CFL numbers are used for the two sets of equations: one point relaxation sweep is performed for the mesoscopic equations, while six iterations are carried out for the macroscopic equations. The CFL numbers are set to $10^3$ for the macroscopic equations and $10^5$ for the mesoscopic equations.

%% file: Testcase.tex
\section{Numerical tests}\label{sec:cases}

This section presents several demanding cases to systematically validate the accuracy, efficiency, and asymptotic-preserving capabilities of the GSIS method in multiscale radiative gas flows. We progress from canonical one-dimensional benchmark to progressively complex environments that typify aerospace applications. Specifically, four representative cases are examined: (i) a normal shock wave to probe accuracy and efficiency in strong non-equilibrium, (ii) a lid-driven cavity flow to assess asymptotic-preserving property, (iii) supersonic flow past a cylinder to evaluate performance of multiscale coupling in canonical configurations, and (iv) hypersonic flow over a three-dimensional Apollo reentry capsule to demonstrate robustness on industrial-grade geometries. These benchmarks collectively demonstrate the method's ability to capture non-equilibrium effects of radiative gas flows, verify convergence acceleration exceeding two orders of magnitude over conventional CIS in near-continuum regimes, and provide insights into radiative heat transfer dominance in atmospheric entry conditions.

The gas parameters used across these tests are set as follows: the rotational and vibrational degrees of freedom are $d_r = 2$ and $d_v = 1.16$, respectively. The characteristic collision numbers are $Z_r = 2.67$ for rotational relaxation and $Z_v = 26.7$ for vibrational relaxation, unless otherwise specified in the individual test cases. The viscosity index is $\omega = 0.74$. The gas Knudsen number $\mathrm{Kn}_{\mathrm{gas}}$, the photon Knudsen number $\mathrm{Kn}_{\mathrm{photon}}$, and the relative strength of radiation ${\sigma}_R$ are varied to demonstrate the capability of the GSIS across different rarefaction and coupling regimes.

The convergence criterion for the overall iterative loop is based on the volume-weighted relative change of the key macroscopic moments (density, flow velocity, and total temperature) between two consecutive iterations, which must be less than a specified small value,
\begin{equation}
    \epsilon^k = \frac{\sqrt{\sum_i(\phi_i^k - \phi_i^{k-1})^2\myd \Omega}}{\sqrt{\sum_i(\phi_i^{k-1})^2\myd \Omega}}\Big|_{max}, \quad \phi\in(\rho, \bm{u}, T).
\end{equation}
On the other hand, the convergence criterion for each inner macroscopic solver step in GSIS is set to a tolerance of $10^{-7}$, with a maximum number of inner iterations limited to 100.

The numerical simulation environment involves a workstation equipped with Intel(R) Core(TM) i7-9700K CPU@3.60GHz processors, performing calculations in double precision. For large-scale problems, partitions in the physical domain velocity space are utilized. The macroscopic solver uses physical domain partitioning for parallel acceleration, while the mesoscopic solver utilizes both physical domain and velocity space partitioning for parallel acceleration~\cite{zhang2024efficient}.

\subsection{Normal shock wave}

\begin{figure}[p]
\centering
\begin{minipage}{0.3\linewidth}
\vspace{1pt}
\centerline{\includegraphics[width=1.2\textwidth,trim = 10 10 10 40, clip = true]{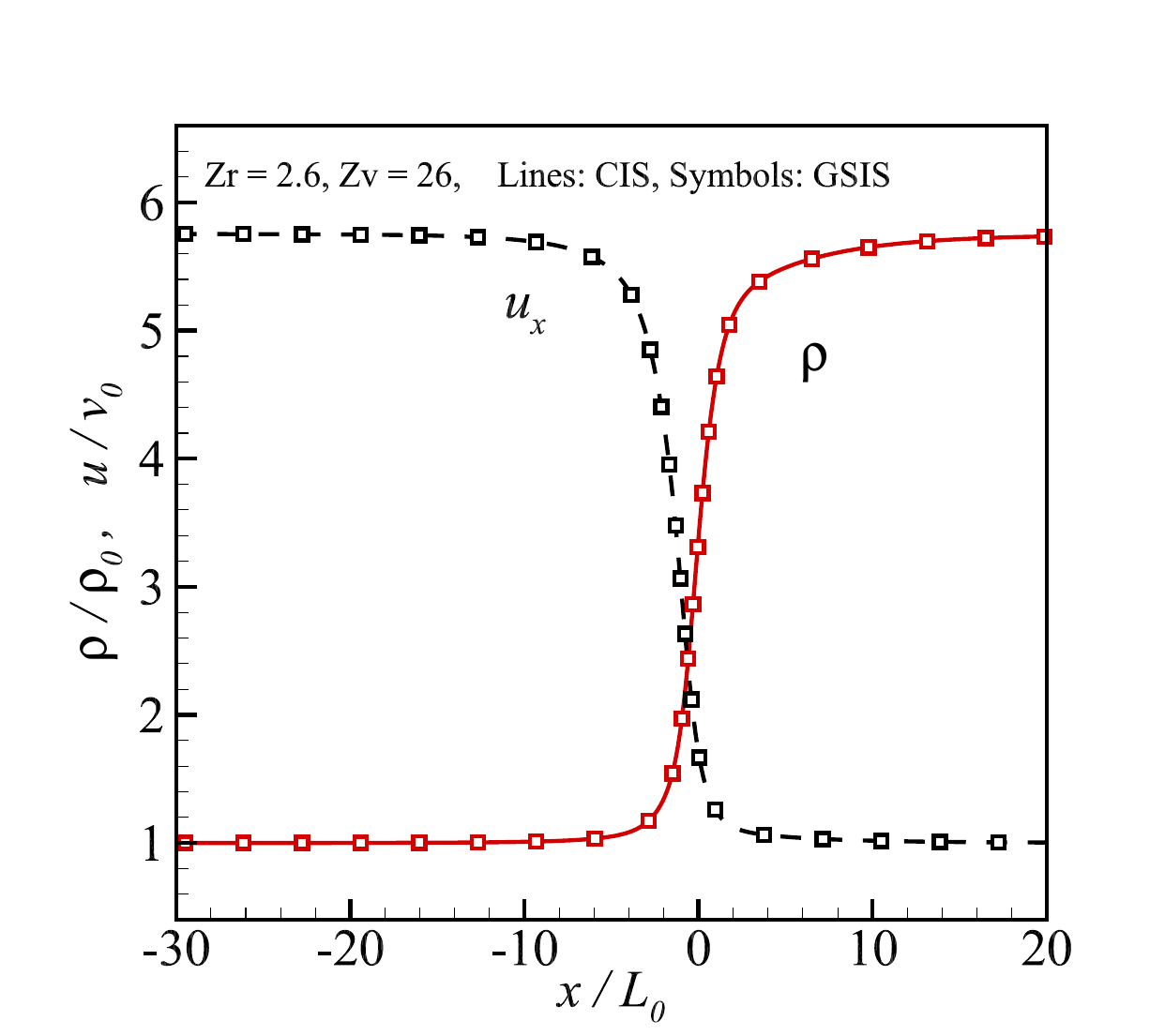}}
\vspace{1pt}
\centerline{\includegraphics[width=1.2\textwidth,trim = 10 10 10 40, clip = true]{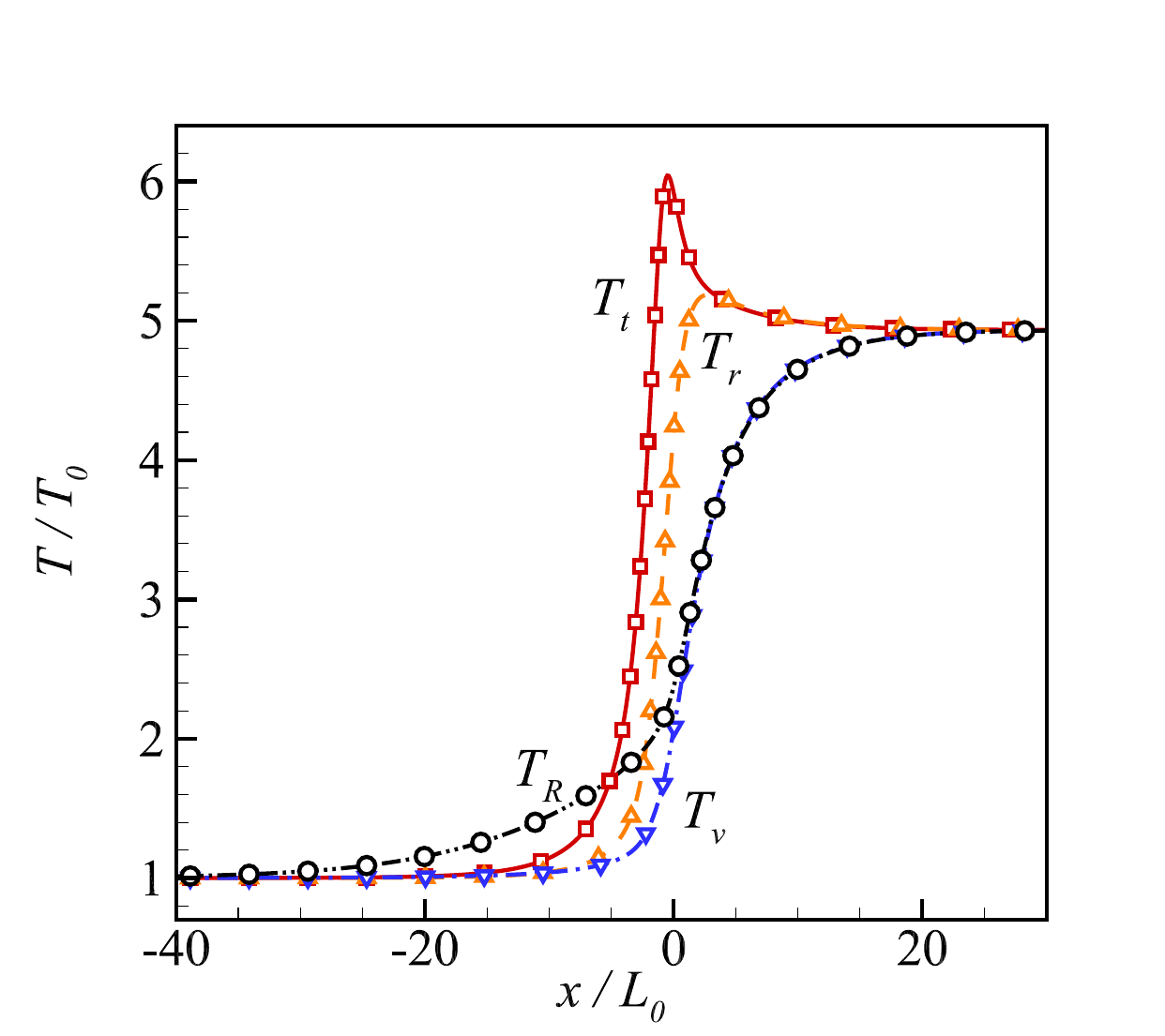}}
\vspace{1pt}
\centerline{\includegraphics[width=1.2\textwidth,trim = 10 10 10 40, clip = true]{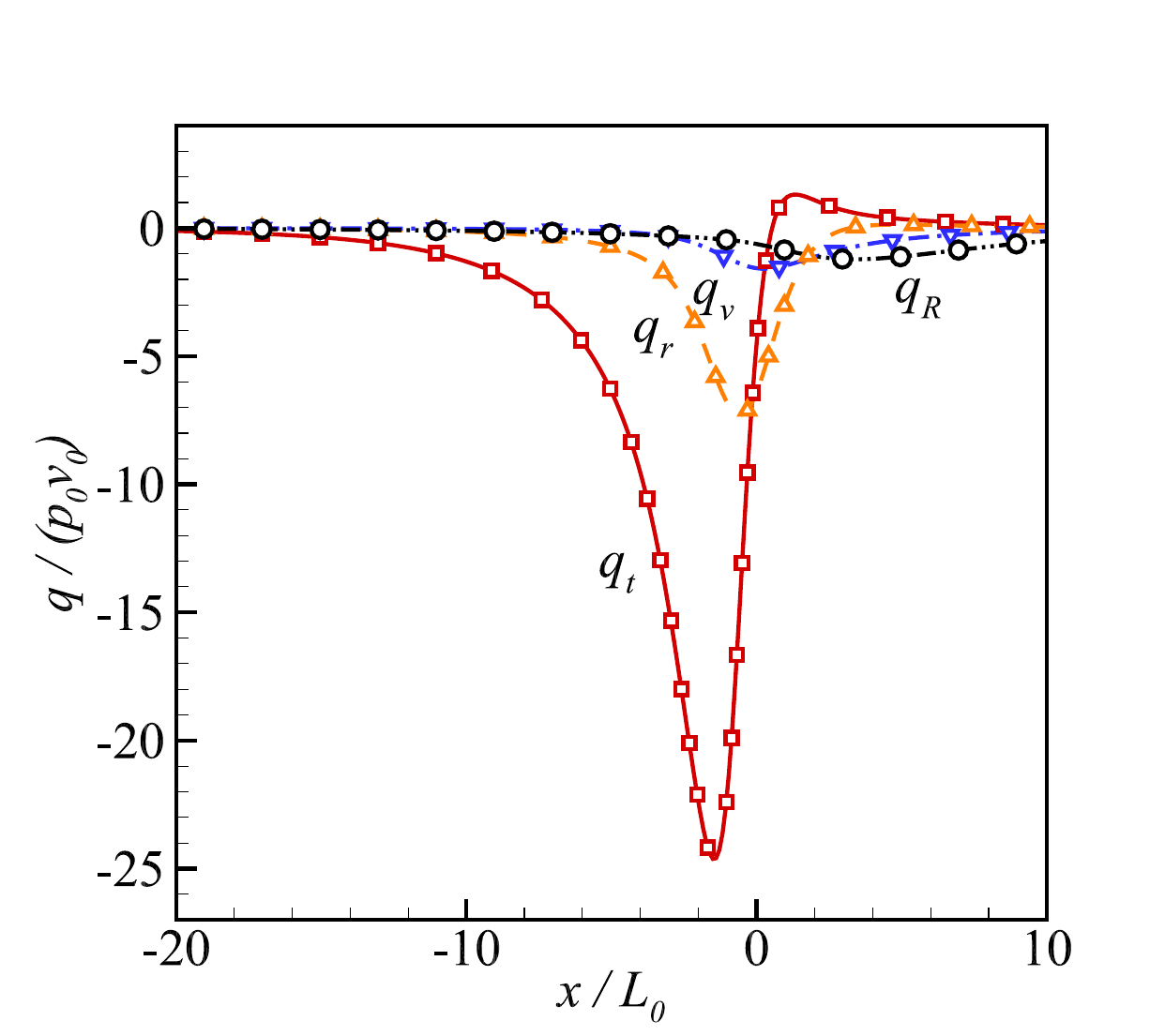}}
\vspace{1pt}
\centerline{\includegraphics[width=1.2\textwidth,trim = 10 10 10 40, clip = true]{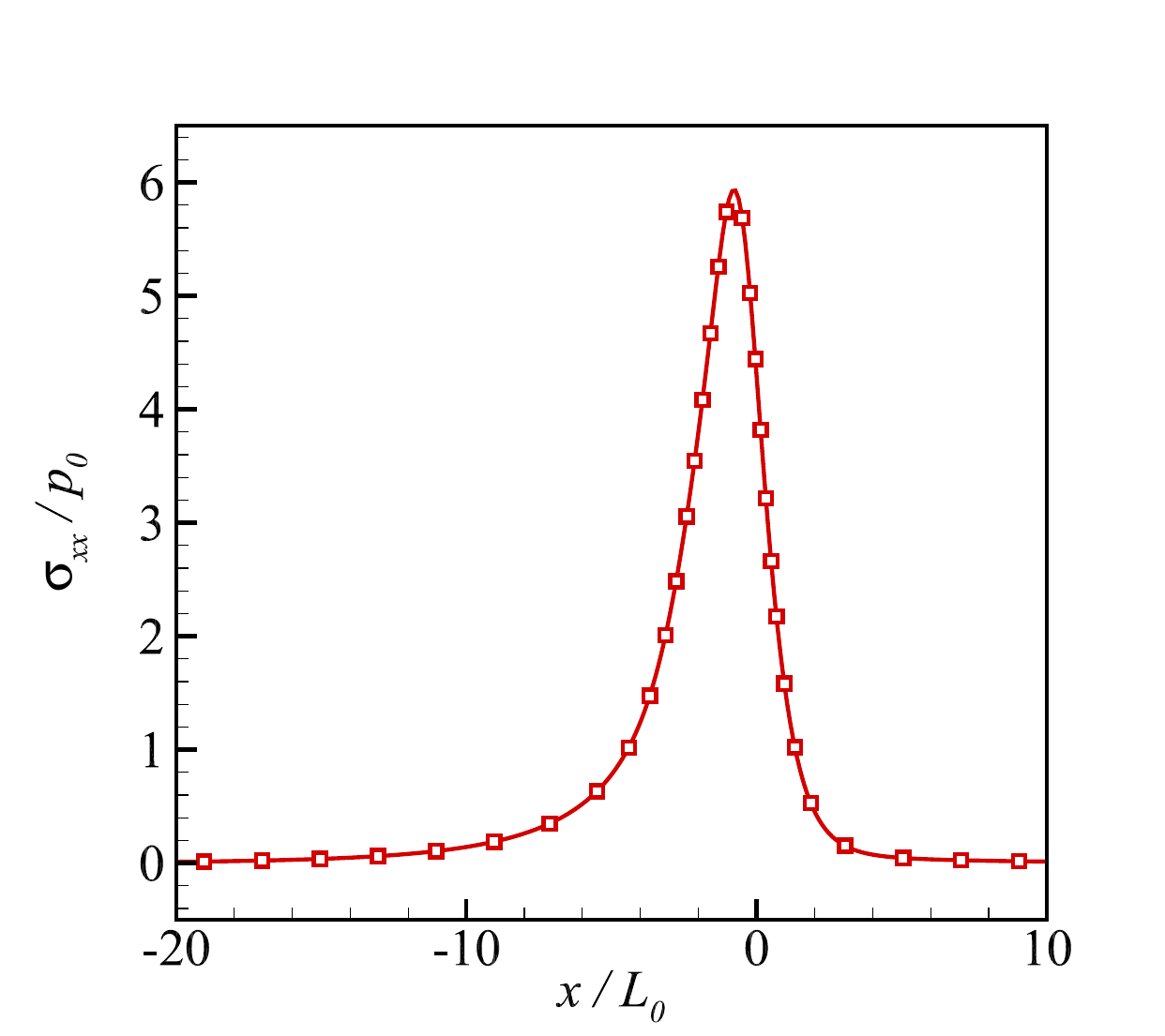}}
\end{minipage}\quad\quad
\begin{minipage}{0.3\linewidth}
\vspace{1pt}
\centerline{\includegraphics[width=1.2\textwidth,trim = 10 10 10 40, clip = true]{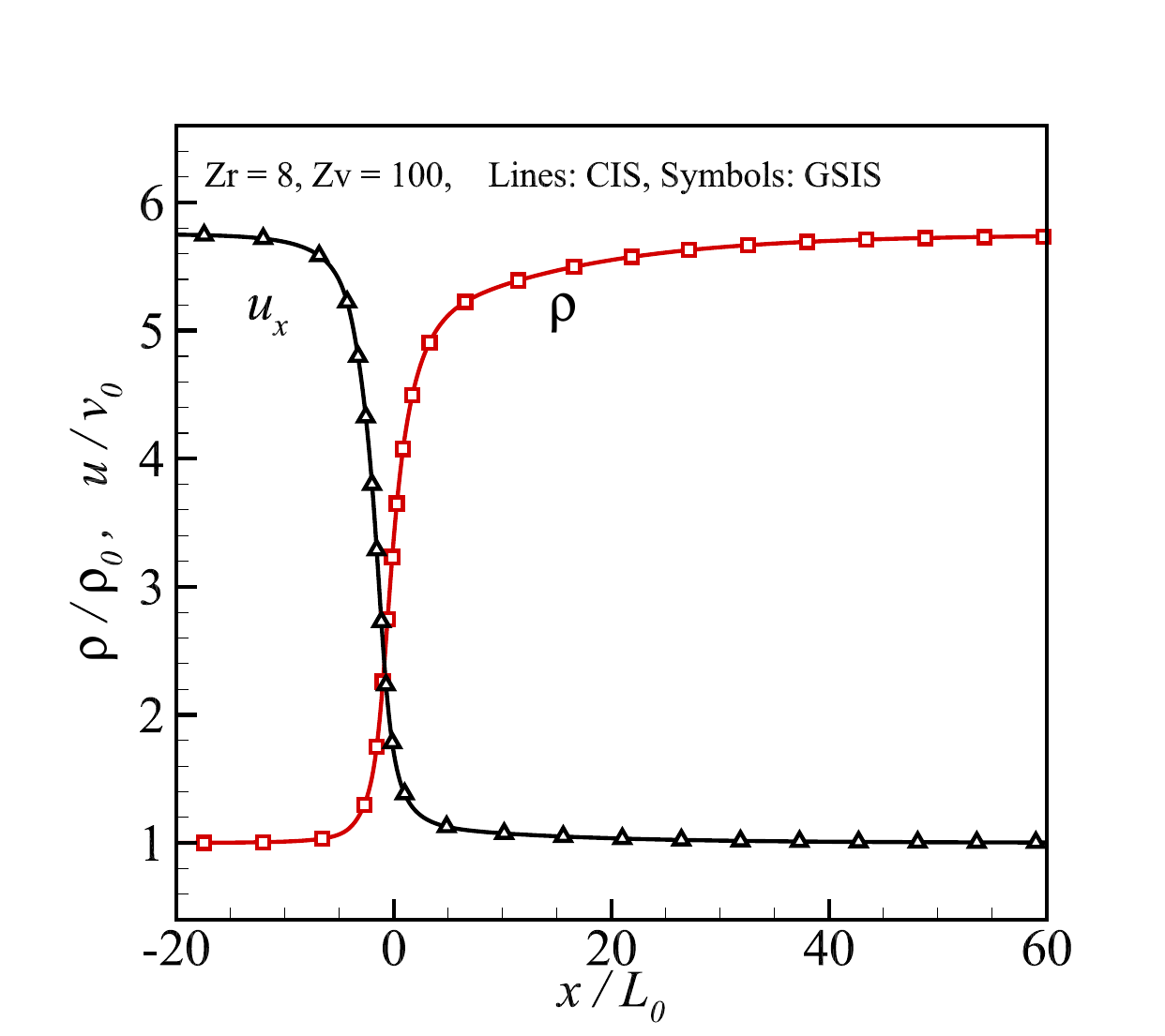}}
\vspace{1pt}
\centerline{\includegraphics[width=1.2\textwidth,trim = 10 10 10 40, clip = true]{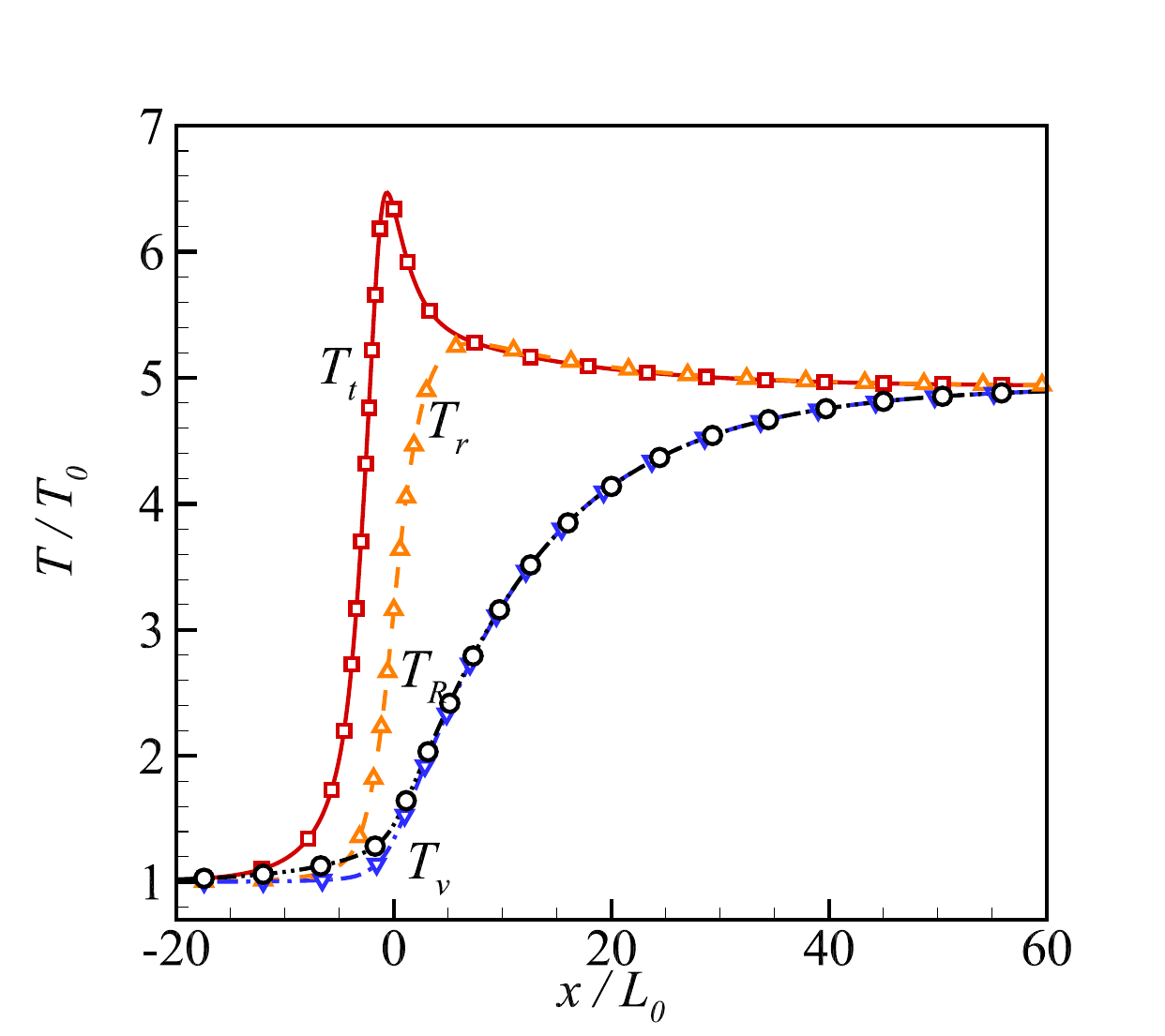}}
\vspace{1pt}
\centerline{\includegraphics[width=1.2\textwidth,trim = 10 10 10 40, clip = true]{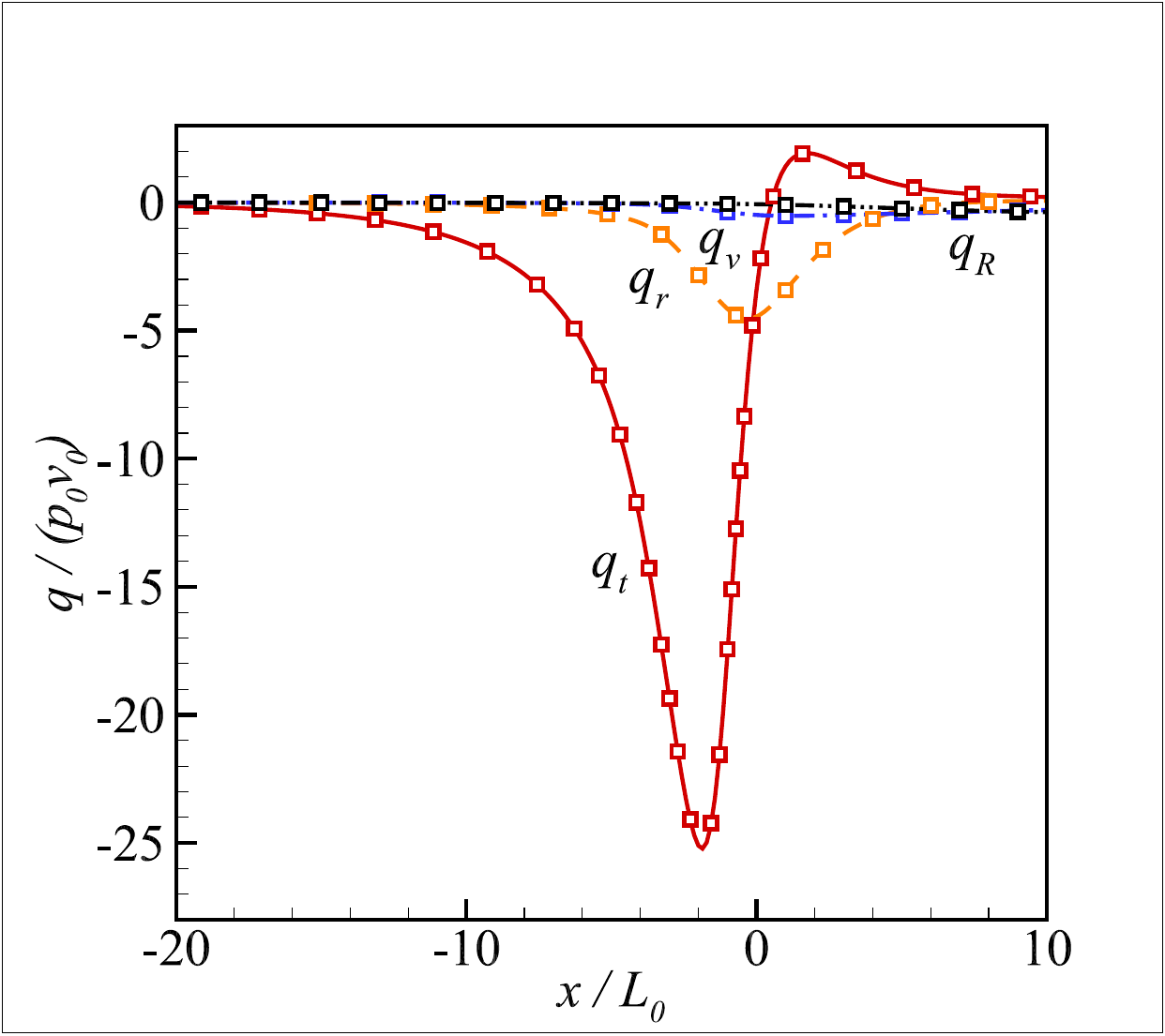}}
\vspace{1pt}
\centerline{\includegraphics[width=1.2\textwidth,trim = 10 10 10 40, clip = true]{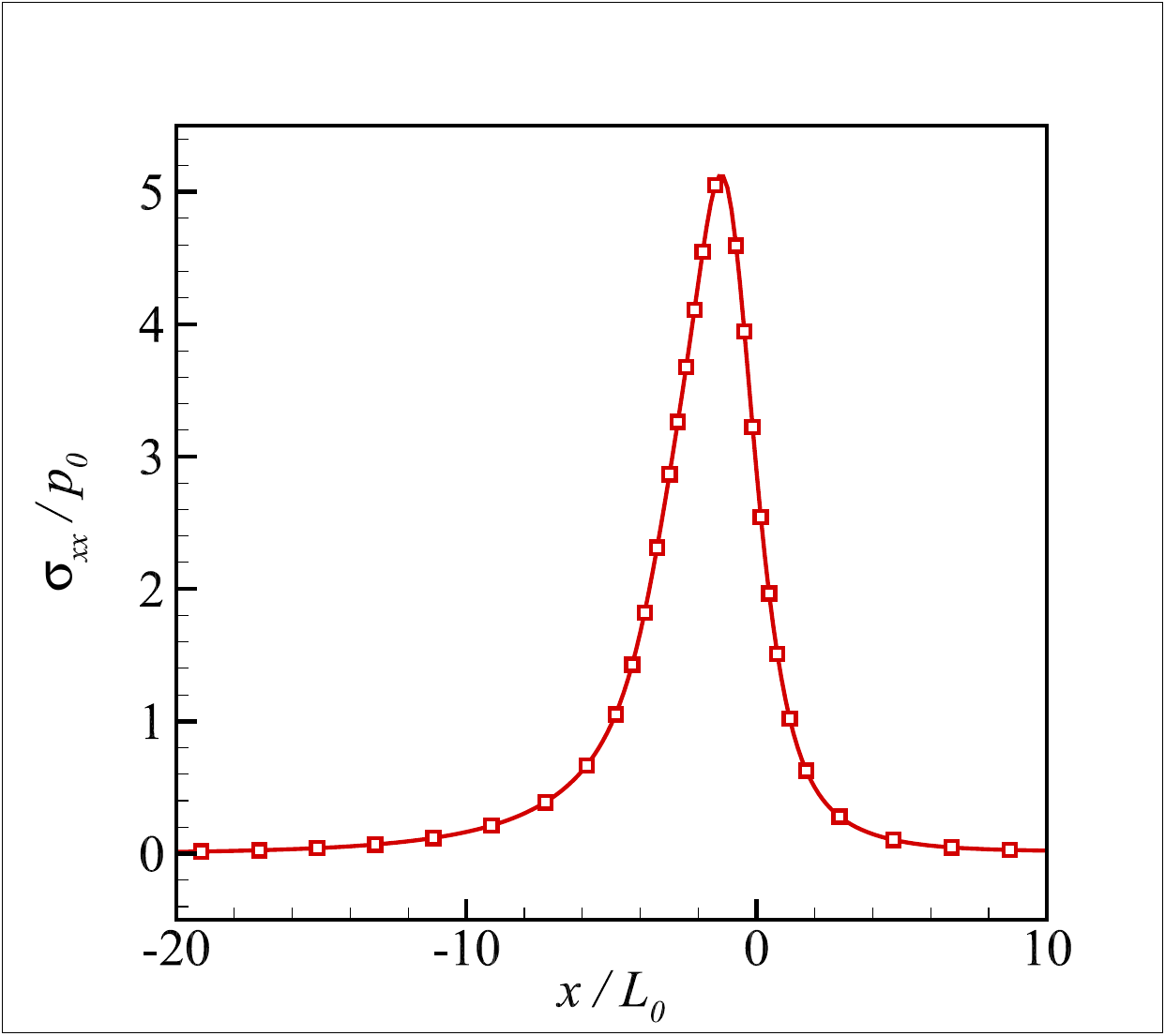}}
\end{minipage}
\caption{Comparison of the dimensionless density, velocity, temperature, heat flux and shear stress between the results obtained by CIS (lines) and GSIS (symbols) for the normal shock wave at $\text{Ma}=5$. The left column: $Z_r = 2.6,,~Z_v = 26$, the right column: $Z_r = 8,,~Z_v =100$.}
\label{fig:nswMa5_macro}
\end{figure}

We consider a gas flow propagating in the $x$ direction with a Mach number $\text{Ma} = 5$. This case serves as a classical and rigorous test to validate the accuracy and convergence efficiency of the GSIS algorithm, particularly for flows exhibiting strong non-equilibrium. The upstream conditions, density $\rho_u$ and temperature $T_u$, are chosen as the reference values $\rho_0$ and $T_0$, respectively. The characteristic length $L_0$ is defined based on the mean free path such that $L_0=16\mu(T_0)/(5\rho_0\sqrt{2\pi R T_0})$. Accordingly, the gas Knudsen number is fixed at $\mathrm{Kn}_{\mathrm{gas}}=5\pi/16$. The radiation characteristics are specified by a photon Knudsen number $\mathrm{Kn}_{\mathrm{photon}}=10$ and a relative radiation strength of ${\sigma}_R = 0.01$.

Two distinct sets of collision numbers are examined: $(Z_r,Z_v)=(2.6,26)$ and $(8,100)$. The case with larger collision numbers corresponds to a slower translational-internal relaxation commonly observed in molecular gases. This is used to demonstrate the efficiency of GSIS, as the slower relaxation process spans a relatively larger physical space and thus poses a greater challenge for traditional solvers. Due to the thermal non-equilibrium between the translational and internal modes, a long-tail structure emerges downstream of the shock. Consequently, the computational domain is set to a total length ranging from $200 L_0$ to $350 L_0$, depending on the chosen collision numbers.

The simulation is initialized with a discontinuity located at $x=0$. The initial gas molecular distribution functions ($f_0, f_1, f_2$) and the radiative intensity $I_R$ are set to the corresponding equilibrium states for the upstream ($x \leq 0$) and downstream ($x \geq 0$) regions,
\begin{equation}
\left\lbrace
\begin{aligned}
    x \leq 0&: f_0 = \rho_u f_0^{\mathrm{eq}}(T_u),\quad f_1 = \frac{d_r}{2}T_uf_0,\quad f_2 = \frac{d_v}{2}T_uf_0,\quad I_R = \frac{1}{\pi}{\sigma}_R T_u^4,\\
    x > 0&:f_0 = \rho_d f_0^{\mathrm{eq}}(T_d),\quad f_1 = \frac{d_r}{2}T_df_0,\quad f_2 = \frac{d_v}{2}T_df_0,\quad I_R = \frac{1}{\pi}{\sigma}_R T_d^4.
\end{aligned}
\right.
\end{equation}
The upstream macroscopic quantities are $\rho_u = 1$, $u_u =\sqrt{\gamma}Ma$, and $T_u = 1$. The downstream macroscopic quantities are determined by the Rankine--Hugoniot relations for a gas with the specific heat ratio $\gamma=(5+d_r+d_v)/(3+d_r+d_v)$,
\begin{equation}
\begin{aligned}
    \rho_d &= \frac{(\gamma+1)Ma^2}{2+(\gamma-1)Ma^2},\quad u_d = \frac{\rho_u}{\rho_d} u_u, \quad T_d = \frac{(2+(\gamma-1)Ma^2)(2\gamma Ma^2-(\gamma-1))}{(\gamma+1)^2Ma^2}.
\end{aligned}
\end{equation}
The boundary conditions at both ends are specified by these corresponding fixed equilibrium states, as the large domain ensures the flow is near equilibrium at the boundaries. 

The physical space is discretized using a uniform mesh with spacing $\Delta x = 0.25 L_0$. The velocity space is reduced to two dimensions and truncated to $[-15, 15]\times[-12, 12]$, with $64$ and $48$ uniformly discrete points in the two directions, respectively. For radiation transport, the polar angle $\theta \in [0,\pi]$ and the azimuth angle $\phi \in [0, 2\pi]$ are discretized into $48$ and $32$ uniform cells, respectively.

\begin{figure}[t]
    \centering
    {\includegraphics[width=0.4\textwidth,trim = 10 10 10 40, clip = true]{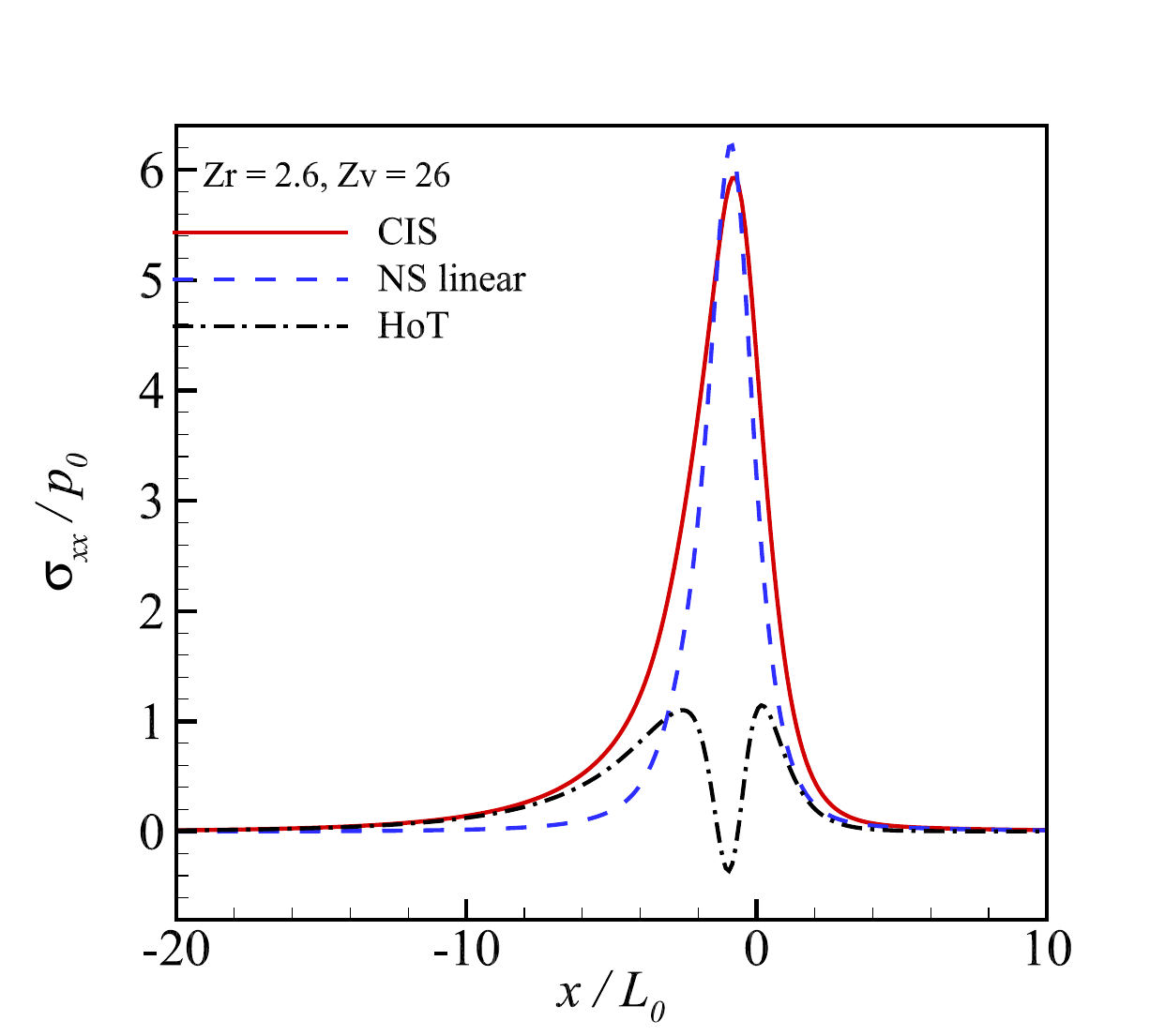}}
    {\includegraphics[width=0.4\textwidth,trim = 10 10 10 40, clip = true]{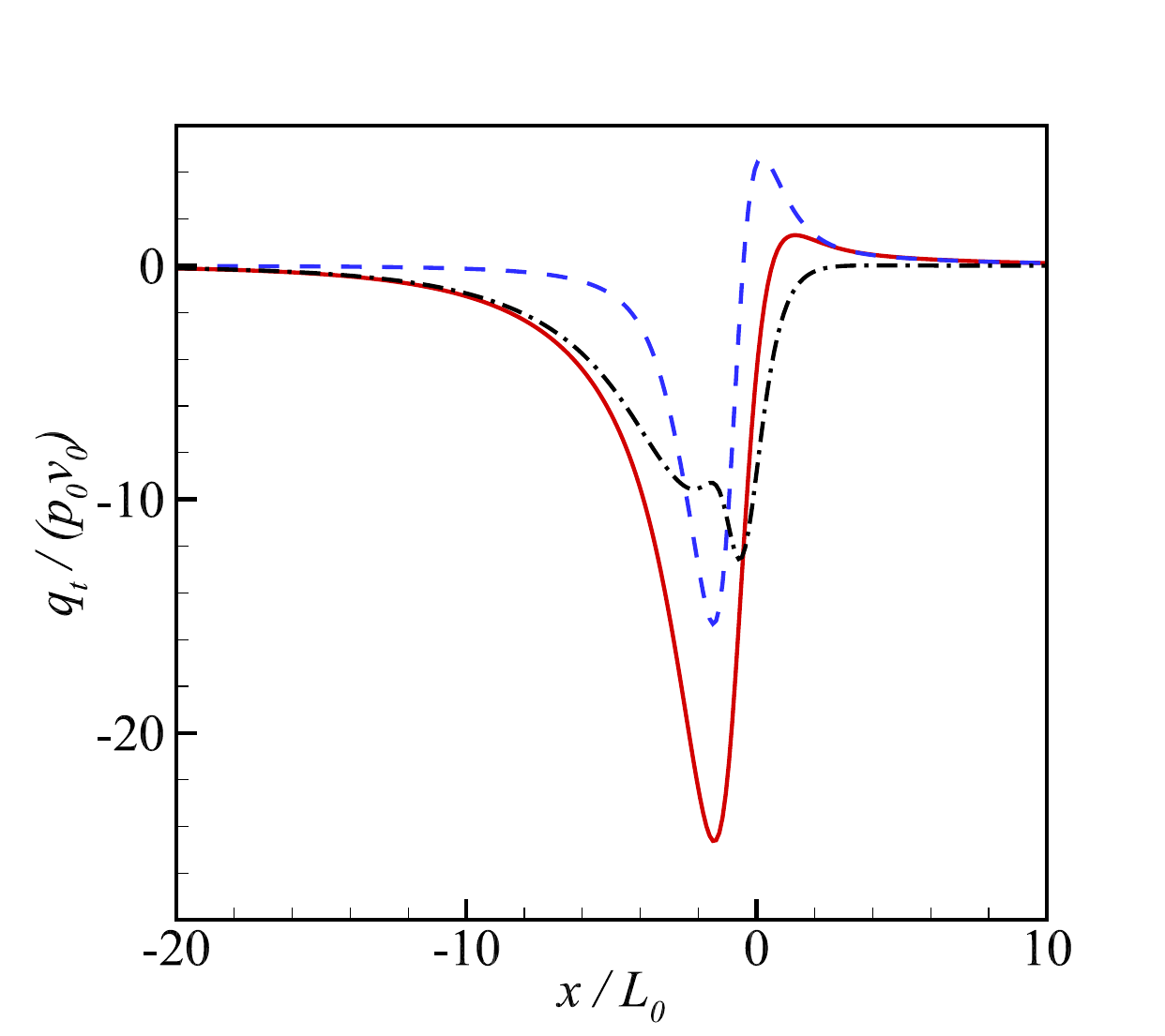}}
    {\includegraphics[width=0.4\textwidth,trim = 10 10 10 40, clip = true]{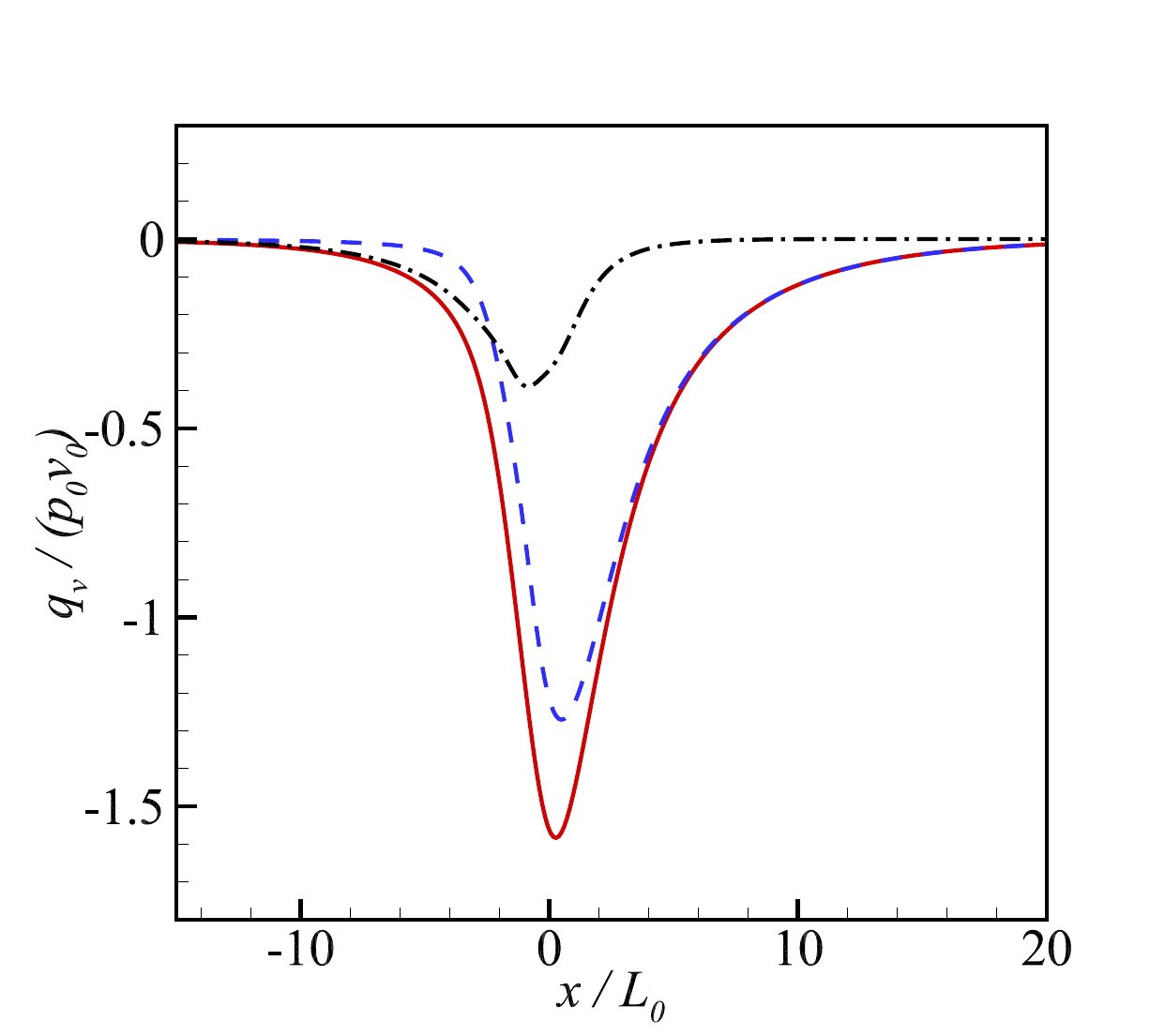}}
    {\includegraphics[width=0.4\textwidth,trim = 10 10 10 40, clip = true]{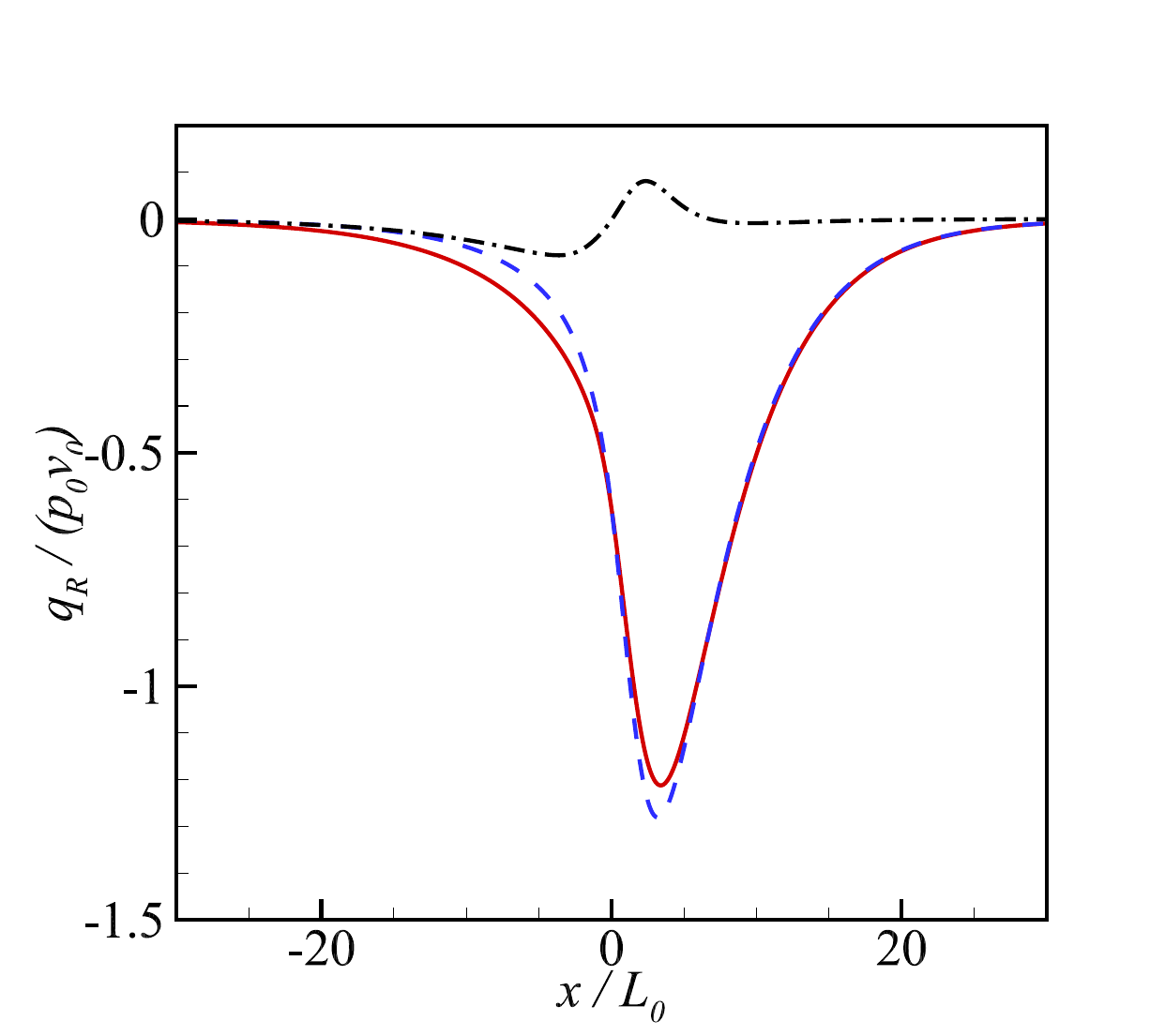}}
    \caption{The contributions of NSF linear constitute relation and higher order terms to the full shear stress and heat fluxes, for the normal shock wave at $\text{Ma} = 5$, .}
    \label{fig:nswMa5_HoT}
\end{figure}

The accuracy of the kinetic model adopted here has been previously validated by comparisons with DSMC results for normal shock wave cases \cite{li2023kinetic}. In this work, the numerical solution obtained by CIS using fine enough grids is adopted as the reference. Fig.~\ref{fig:nswMa5_macro} presents the macroscopic variable profiles obtained by both GSIS (shown as symbols) and CIS (shown as lines). It is observed that GSIS produces results in excellent agreement with the CIS reference for different collision numbers. The shock wave exhibits a distinct structure characterized by sharp increases in density and temperature over a few mean free paths, followed by a gradual approach to the downstream values due to the slow relaxation between the translational and internal modes. This type of profile corresponds to the Type B structure described in \cite{taniguchi2014thermodynamic}. Furthermore, the radiation temperature is observed to rise at a position about $40 L_0$ upstream of the shock center and becomes nearly identical to the vibrational temperature after passing the shock center. This indicates that for shock waves with stable radiation sources established upstream and downstream, the radiation temperature is higher than the gas temperature upstream. Due to the large vibrational collision number $Z_v$, the vibrational energy mode primarily interacts with the photons via the radiative transition process, absorbing energy from the radiation field, before gradually approaching the translational temperature through inelastic gas-gas collisions.

\begin{figure}[t]
    \centering
    {\includegraphics[width=0.5\textwidth,trim = 10 120 10 220, clip = true]{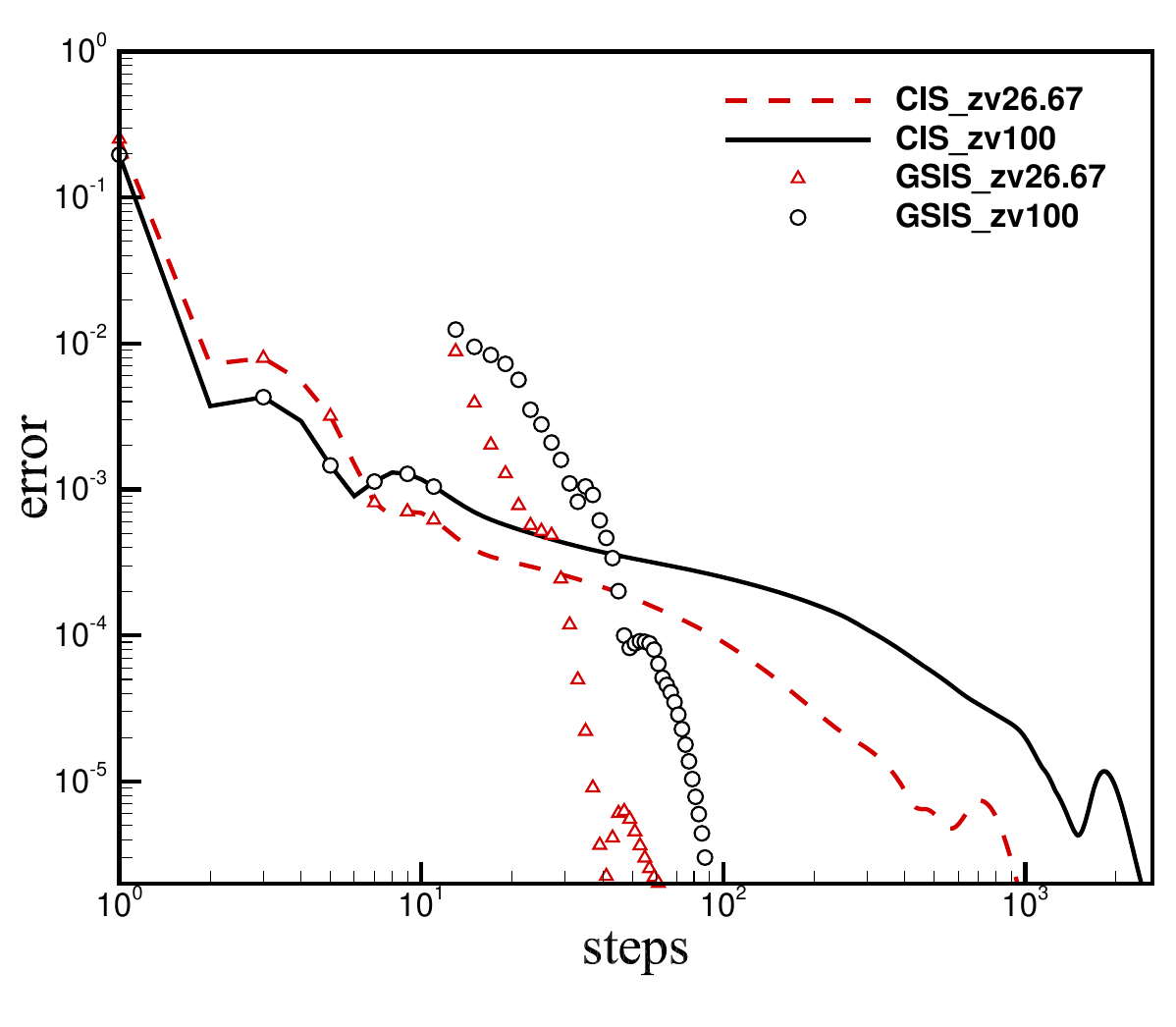}}
    \caption{Error decay history of CIS and GSIS for the simulations of the normal shock wave at three different collision numbers $Z_r = 2.6, 8$ and $Z_v = 26, 100$, when $\text{Ma}=5$. }
    \label{fig:nswMa5_error}
\end{figure}

Fig.~\ref{fig:nswMa5_HoT} illustrates the contributions of HoTs to the shear stress and heat fluxes, which essentially represent non-equilibrium effects beyond the NSF description. The figure clearly demonstrates that the numerical values of the HoTs constitute a significant portion of the total stress and heat flux within the shock structure. Particularly in the upstream pre-shock region, the HoTs are substantially larger than the NSF linear constitutive parts. This implies that the equivalent transport coefficients are severely underestimated by the NSF approximation in this region, which consequently fails to accurately capture the non-equilibrium flow structure. It should also be noted that these HoTs exhibit pronounced nonlinear characteristics. Capturing such complex non-equilibrium effects is challenging when relying on simply extending linear or nonlinear constitutive relations. In contrast, the GSIS method directly constructs these nonlinear high-order terms numerically from the mesoscopic kinetic equation, thereby avoiding the inherent robustness issues associated with solving high-order macroscopic equations. As a result, the GSIS provides a more accurate and robust description of non-equilibrium effects across the entire flow field.

Fig.~\ref{fig:nswMa5_error} presents the comparison of error decay as a function of the iteration steps between GSIS and CIS for two sets of collision numbers. To reach the convergence criterion of $2\times 10^{-6}$, GSIS requires 68 and 90 steps for $Z_v=26$ and $Z_v=100$, respectively, while CIS requires 991 and 2566 steps. The initial error decay for both methods is identical during the first ten steps, because the GSIS iteration includes an initial preconditioning phase using the CIS solver. It is evident that GSIS achieves substantial acceleration factors of ${14.5}$ (for $Z_v=26$) and ${28.5}$ (for $Z_v=100$) in these two cases, and this speedup will be significantly more pronounced as the collision number $Z_v$ further increases. In addition, the residual decay slope in GSIS remains nearly constant throughout the iterations, as the macroscopic synthetic equations enable the disturbance to be rapidly transmitted across the entire domain, especially to the slow-relaxation downstream region. Notably, since the computational cost of the macroscopic solver in one-dimensional problems is almost negligible compared to the mesoscopic step, the reduction in GSIS computational cost is essentially equivalent to the dramatic decrease in the number of iteration steps.

\subsection{Lid-driven cavity flow}

\begin{figure}[t]
    \centering   
    \subfloat[]{\includegraphics[width=0.42\textwidth,trim = 30 10 10 50, clip = true]{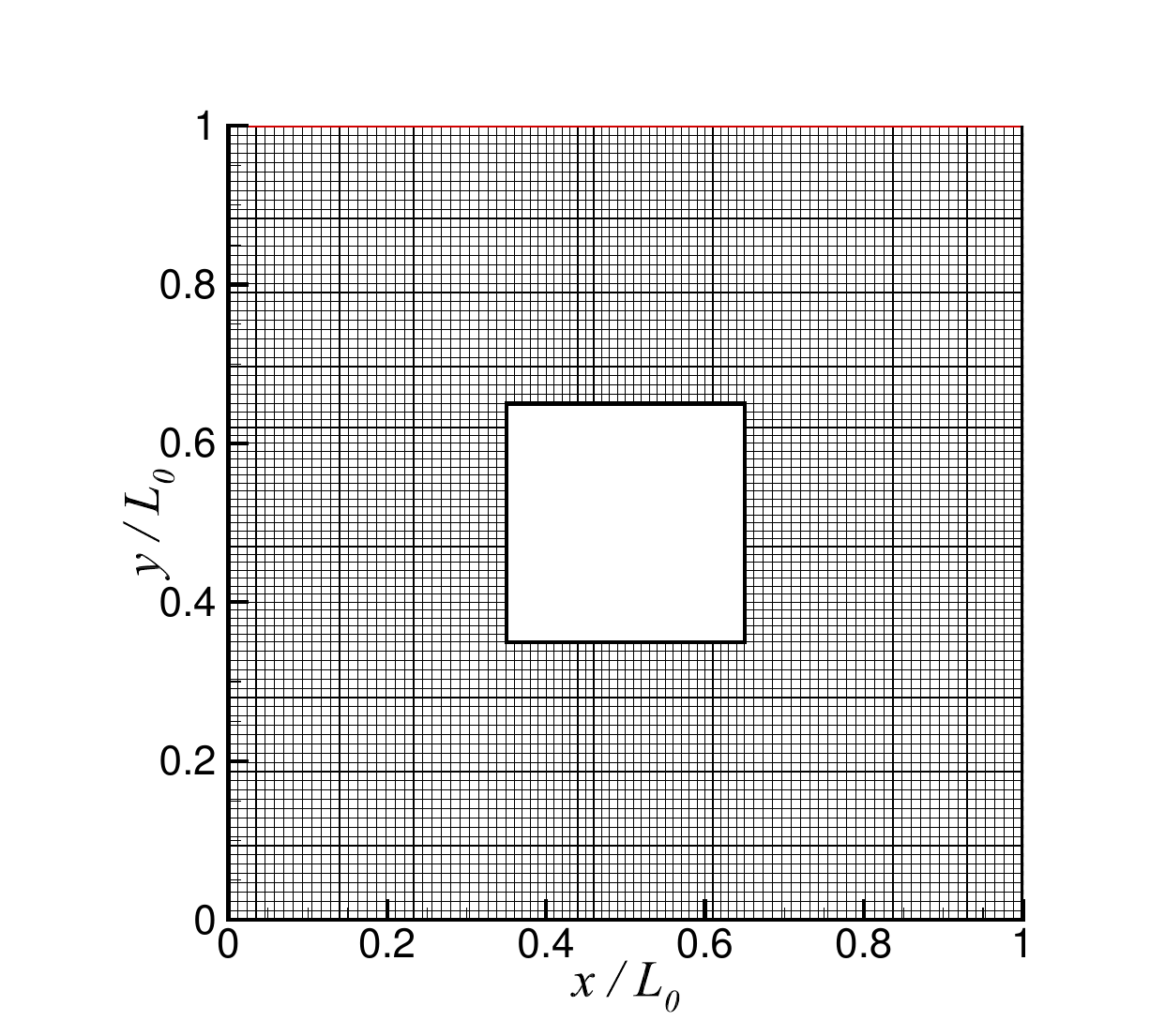}\label{fig:cavity_mesh}} 
    \subfloat[]{\includegraphics[width=0.42\textwidth,trim = 10 10 10 50, clip = true]{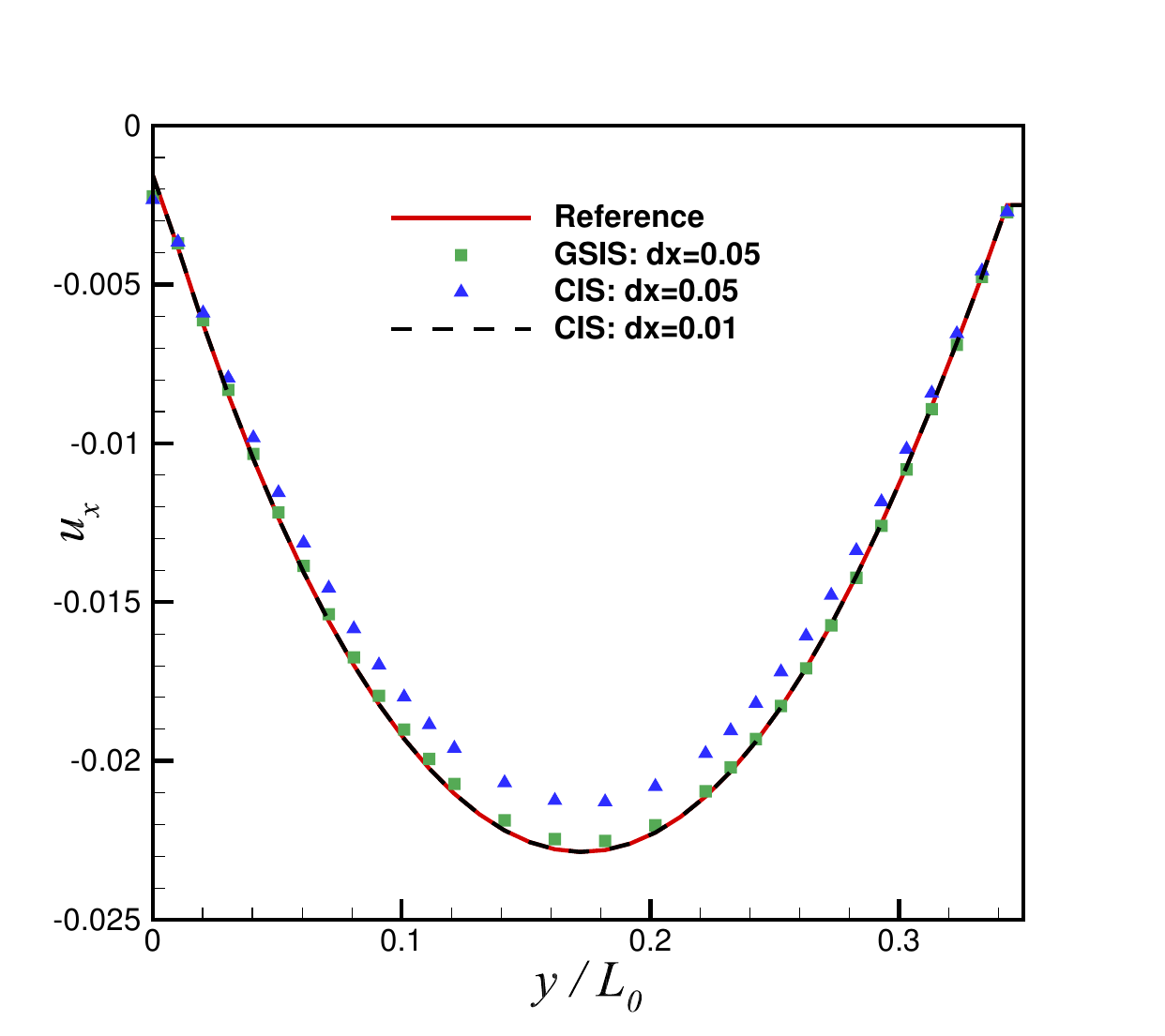}} \\
    \subfloat[]{\includegraphics[width=0.42\textwidth,trim = 10 10 10 50, clip = true]{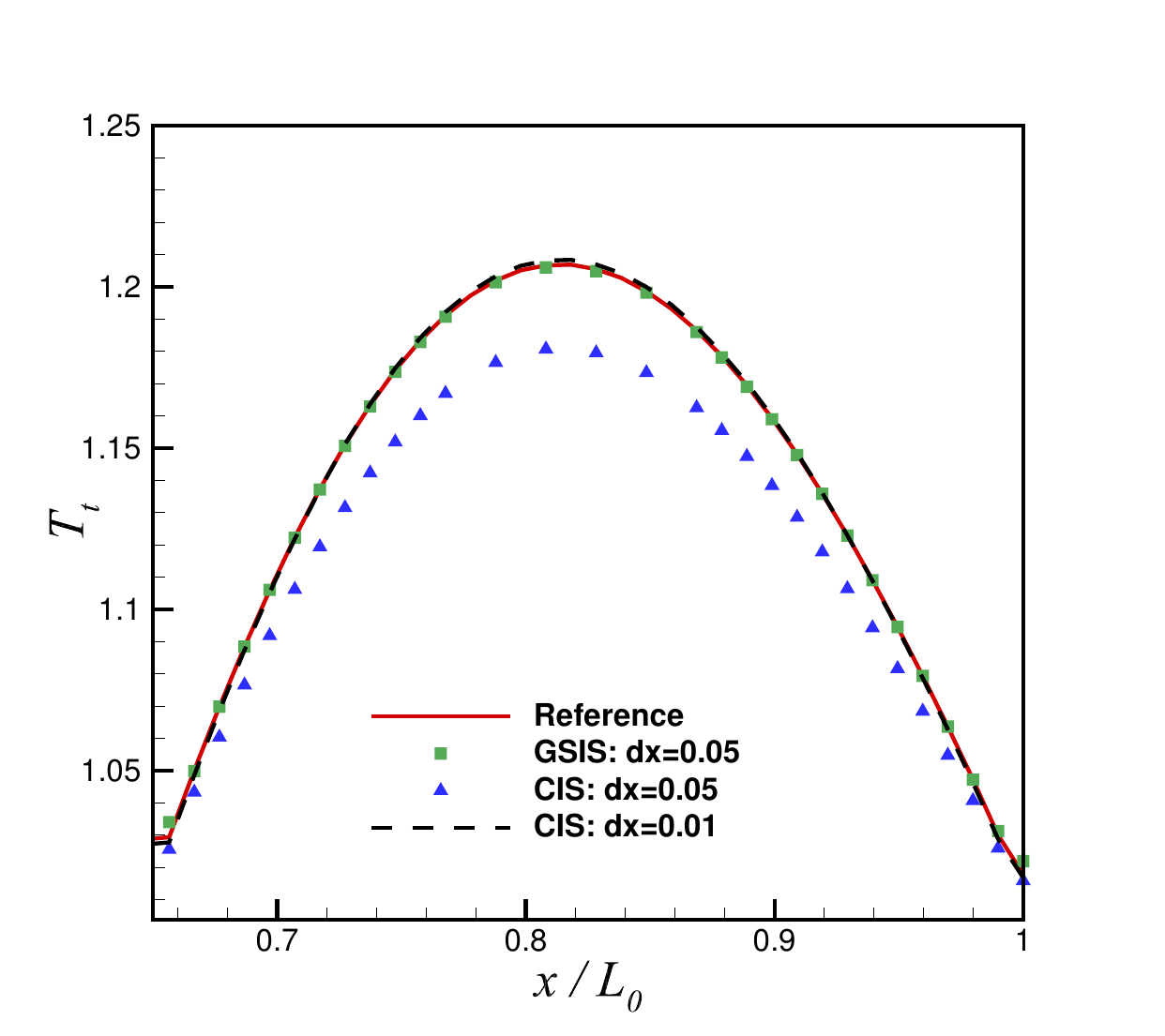}}
    \subfloat[]{\includegraphics[width=0.42\textwidth,trim = 10 10 10 50, clip = true]{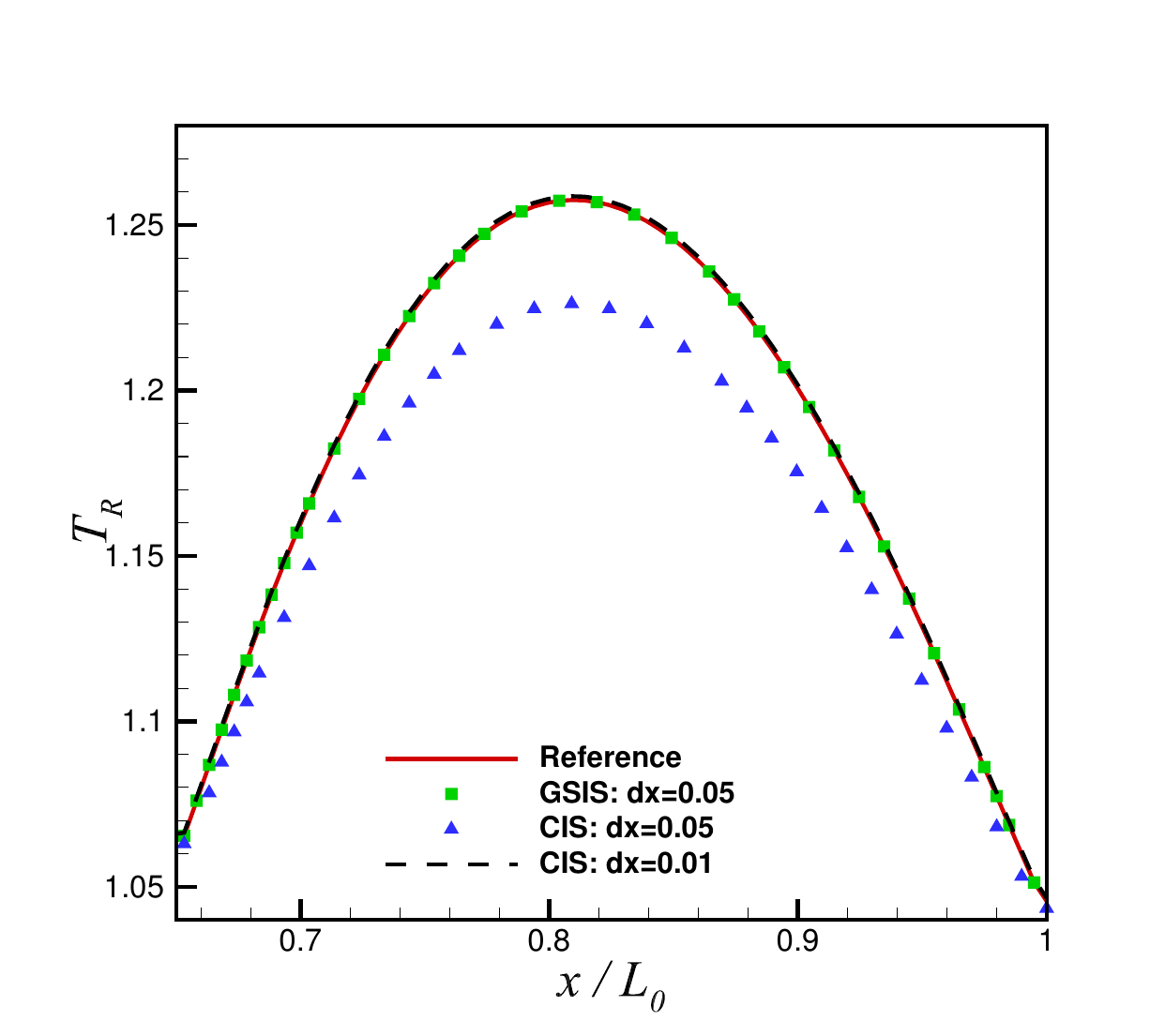}}
    \caption{(a) The computational mesh for the two-dimensional simulations of a lid-driven cavity flow. (b-d) Comparisons of the dimensionless flow velocity along $x=0.5,y\in[0,0.35]$ and translational and radiative temperatures along $y=0.5,x\in[0.65,1]$ computed by CIS and GSIS on two meshes ($\Delta x= 0.05L_0$ and $0.01L_0$). The reference solution is obtained on a finer mesh with $\Delta x=0.005L_0$.}
    \label{fig:cavity_velocity_temperature}
\end{figure}

We examine a lid-driven cavity flow, which a stringent benchmark for verifying the asymptotic-preserving property of the numerical method, since the temperature and velocity distributions away from the moving lid are highly sensitive to numerical dissipation. The configuration consists of a square domain of side length $L_0$ containing a concentric inner square of side $0.3L_0$ (Fig. \ref{fig:cavity_mesh}). Flow is induced by the top lid moving in the $x$-direction with velocity $u_w=0.18v_0$ at temperature $3T_0$, while all other walls remain stationary at temperature $T_0$. The gas Knudsen number is set to $\Kn_{\text{gas}}=0.005$, making the flow in the near-continuum regime where asymptotic-preserving capability is crucial. The photon Knudsen number is $\Kn_{\text{photon}}=1$ and relative radiation intensity is $\sigma_R=0.005$ to model moderate radiation coupling.


To systematically probe grid sensitivity, we employ two structured uniform meshes: a coarse mesh with $\Delta x= 0.05L_0$ (approximately $10 \lambda$, where $\lambda$ is the mean free path) and a fine mesh with $\Delta x= 0.01L_0$ (around $2 \lambda$). A reference solution is computed on a much finer mesh ($\Delta x= 0.005L_0$) where both GSIS and CIS converge to identical results. The velocity space is truncated to $[-15, 15]\times[-12, 12]$ and discretized uniformly with $50\times40$ grids. 

Fig. \ref{fig:cavity_velocity_temperature} compares the velocity profile along $x=0.5$ for $y\in[0,0.35]$ and the translational and radiative temperatures along $y=0.5$ for $x\in[0.65,1]$. On the coarse mesh ($\Delta x= 10 \lambda$), GSIS predictions coincide precisely with the reference solution, confirming its asymptotic-preserving property even when grid cells vastly exceed the mean free path. In contrast, CIS suffers from excessive numerical dissipation on the same mesh, and fails to produce accurate velocity and temperature fields. Only when $\Delta x$ is reduced to $0.01 L_0$, does CIS achieve satisfactory accuracy.

The efficiency gain is also dramatic: GSIS reaches convergence in 32 iterations (including 10 CIS preconditioning steps), whereas CIS requires 6700 iterations. Since the cost of 100 inner macroscopic steps in each GSIS iteration is comparable to one kinetic step, the overall speedup is two orders of magnitude at small Knudsen numbers. Therefore, this case demonstrates that GSIS retains accuracy and achieves rapid convergence on coarse meshes where conventional method fails.

\subsection{Supersonic flow over cylinder}

One of the most important situations in which the radiative energy interacts strongly with the rarefied gas flow is the propagation of a radiative shock wave passing vehicles. To evaluate the capability of GSIS in modeling this type of problem, a two-dimensional hypersonic rarefied flow past a cylinder is simulated. The freestream Mach number is fixed at $\text{Ma} = 5$ with reference pressure $p_0$. Both the freestream and the isothermal wall temperatures are set to $0.5 T_0$, where $T_0$ is equal to the reference vibrational temperature $T_v^{\text{ref}}$.

\begin{figure}[t]
    \centering
    \subfloat[Whole mesh]{\includegraphics[scale=0.38,trim={10 10 10 10},clip = true]{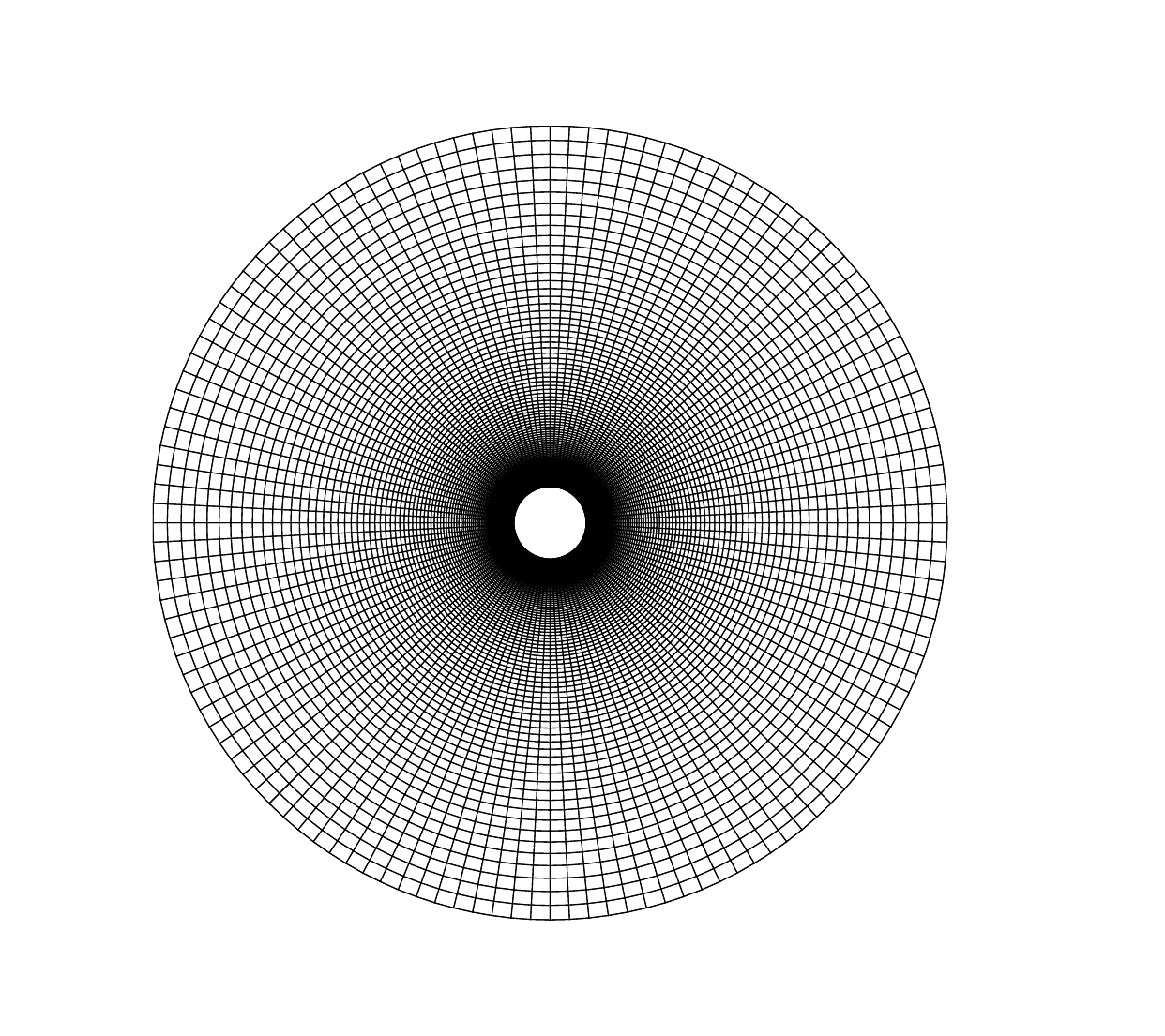}}
    \subfloat[local enlargement]{\includegraphics[scale=0.38,trim={10 10 10 10},clip = true]{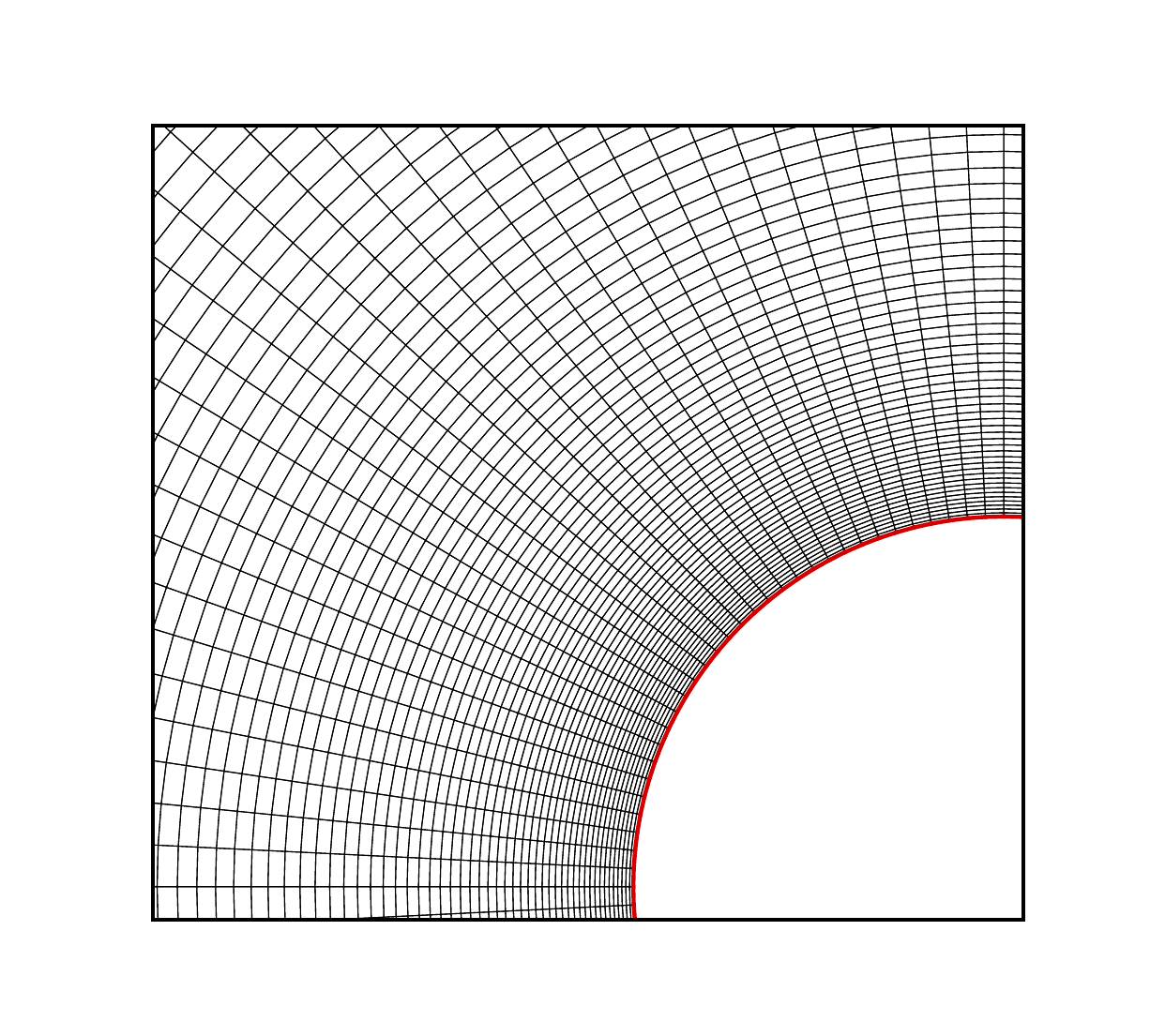}}
    \caption{The computational mesh for the two-dimensional simulations of supersonic flow over cylinder.}
    \label{fig:cylindermesh}
\end{figure}

The simulation extensively explores the multiscale and gas-radiation coupling nature of the problem by varying the key governing parameters. The gas Knudsen number $\mathrm{Kn}_{\mathrm{gas}}$ is varied widely from $1$ (rarefied) down to $0.001$ (near-continuum). Simultaneously, the photon Knudsen number $\mathrm{Kn}_{\mathrm{photon}}$ ranges from $0.1$ (optically thick) to $100$ (optically thin), and the relative radiation strength ${\sigma}_R$ is varied between $0.1$ and $10$.

The computational domain, with a diameter of $D = 6L_0$ ($L_0$ is the diameter of the cylinder), is discretized using structured quadrilateral cells of $400 \times 60$ (Fig. \ref{fig:cylindermesh}). To ensure accurate resolution, the mesh is highly refined near the cylinder surface, with a minimum spacing of $\Delta h = 0.002 L_0$. The reduced two-dimensional velocity space of the gas is truncated to $[-15, 15]\times[-7, 7]$ and uniformly discretized with $90 \times 50$ velocity points. Finally, the solid angle for photon transport is divided into $64 \times 48$ uniformly spaced angular cells.

 \begin{figure}[t]
     \centering
     {\includegraphics[scale=0.38,trim={10 10 10 30},clip = true]{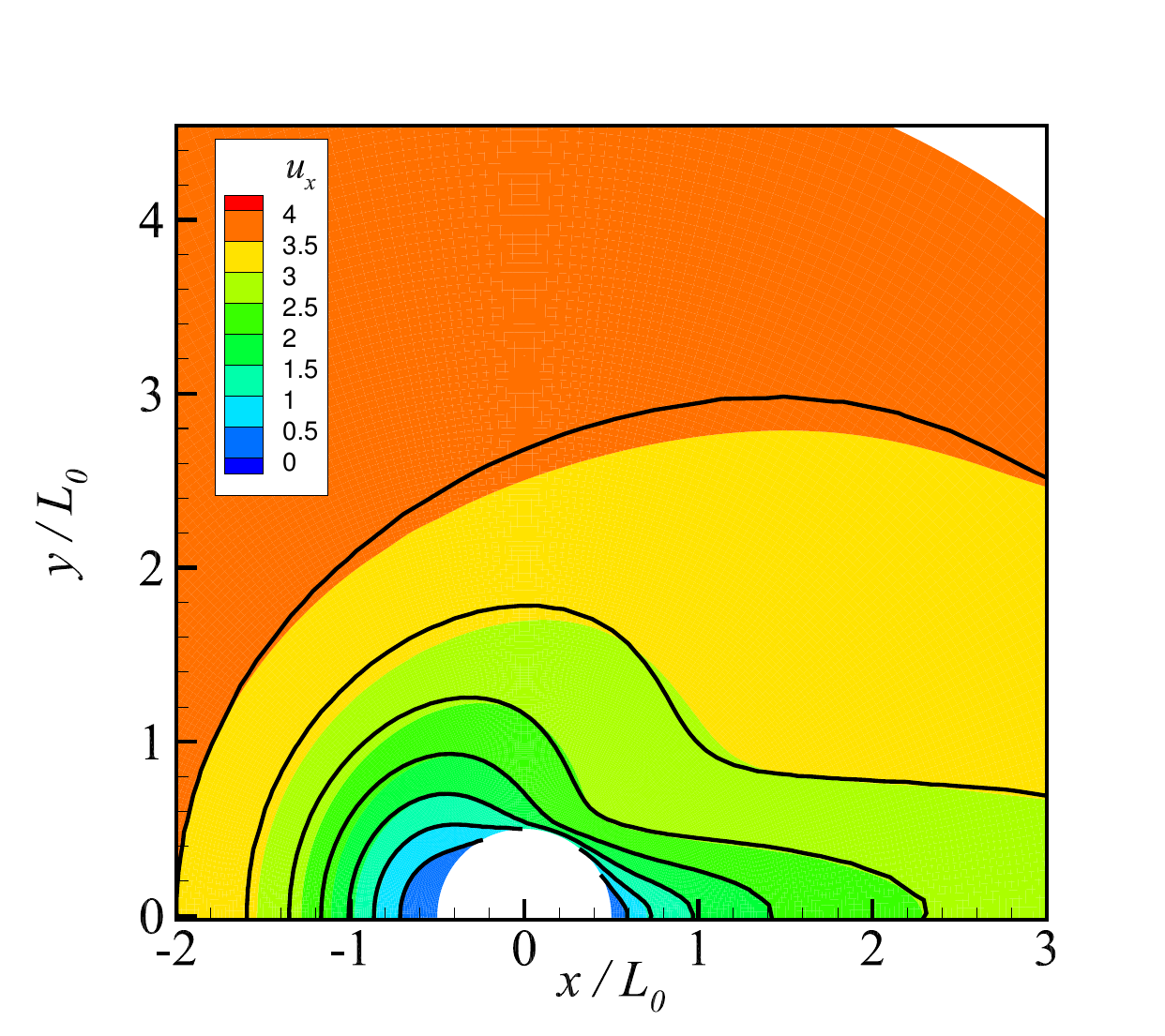}}
     {\includegraphics[scale=0.38,trim={10 10 10 30},clip = true]{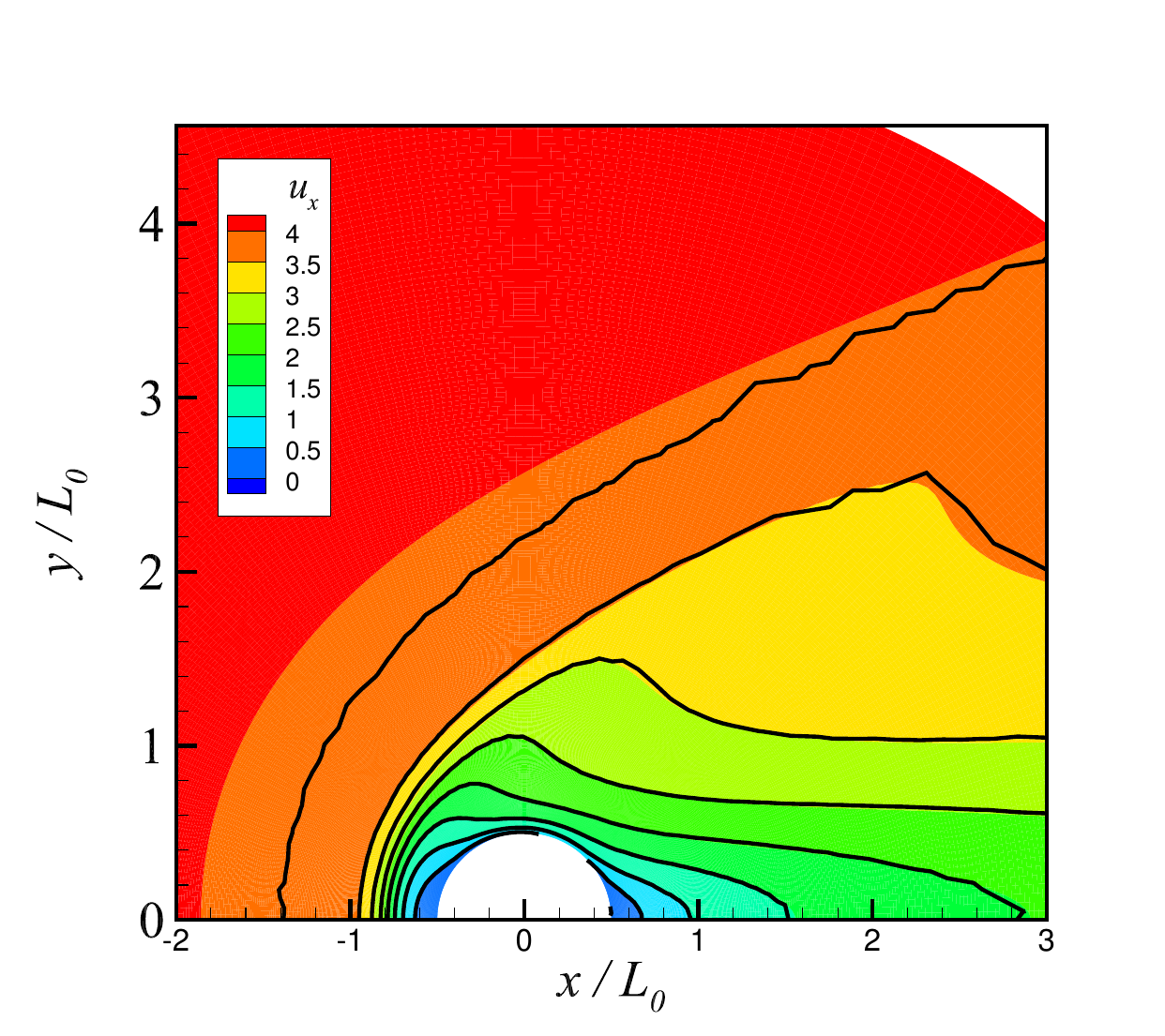}}
     \\
     {\includegraphics[scale=0.38,trim={10 10 10 30},clip = true]{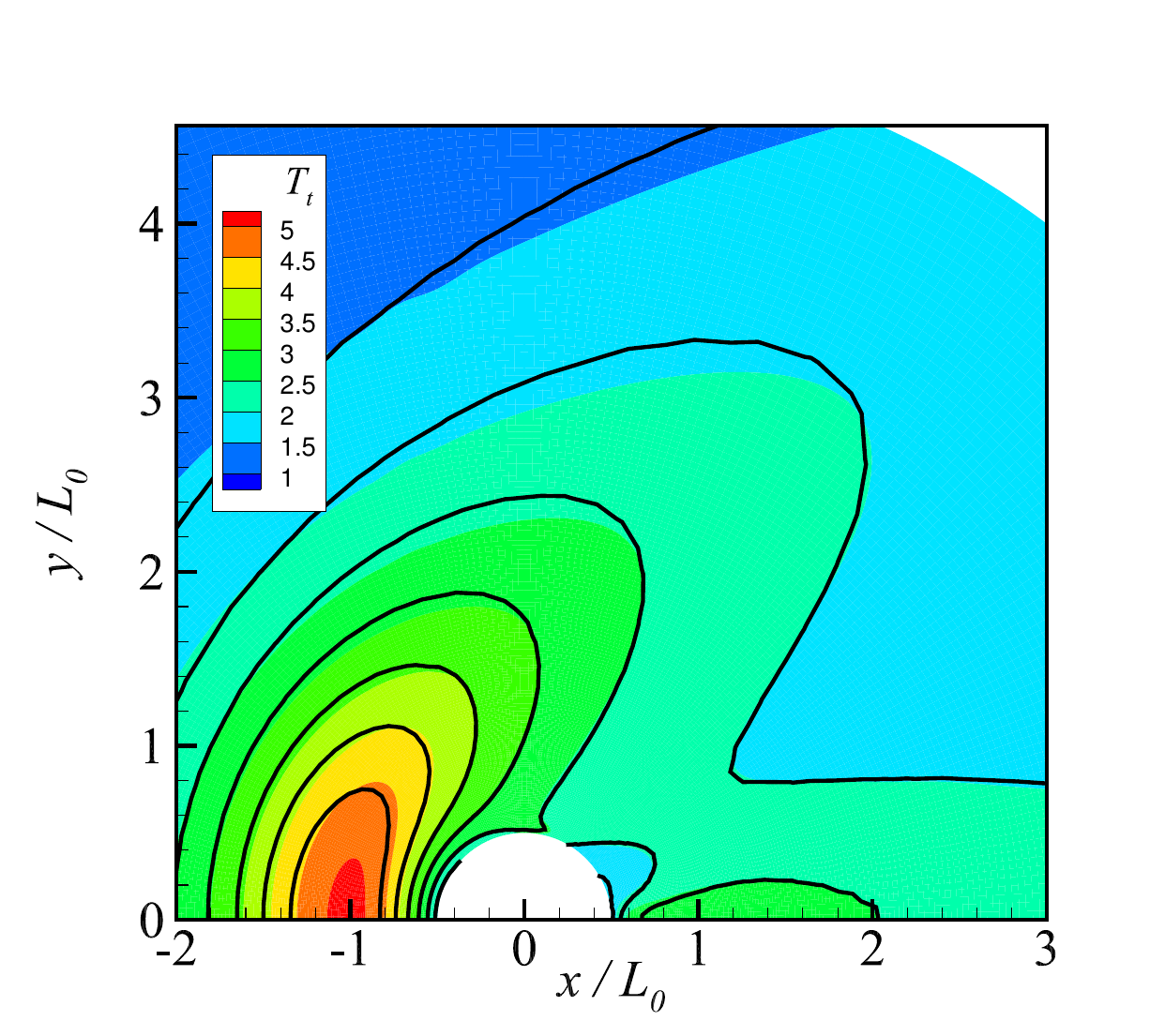}}
     {\includegraphics[scale=0.38,trim={10 10 10 30},clip = true]{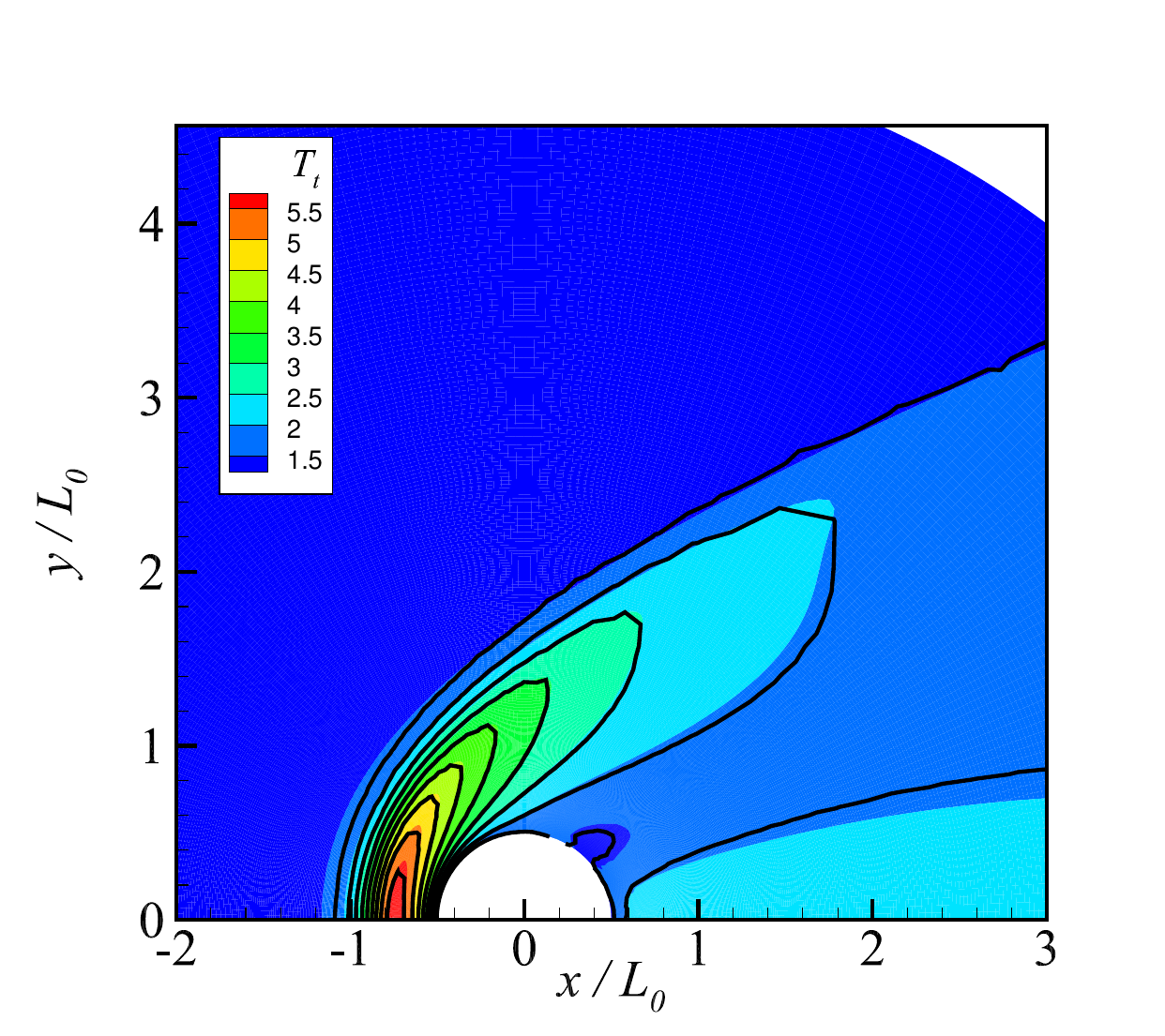}}
     
     \caption{Comparisons of the dimensionless horizontal velocity $u_x$ and translational temperature distribution between the results of DSMC (black lines) and GSIS (background contours) for a supersonic flow over cylinder at $\text{Ma}=5$. The gas Knudsen number is $\mathrm{Kn}_{\mathrm{gas}} = 1$ (left column) and $0.1$ (right column). }
     \label{fig:cylindercenterline}
 \end{figure}

 \begin{figure}
    \centering
    {\includegraphics[scale=0.39,trim={10 10 10 10},clip = true]{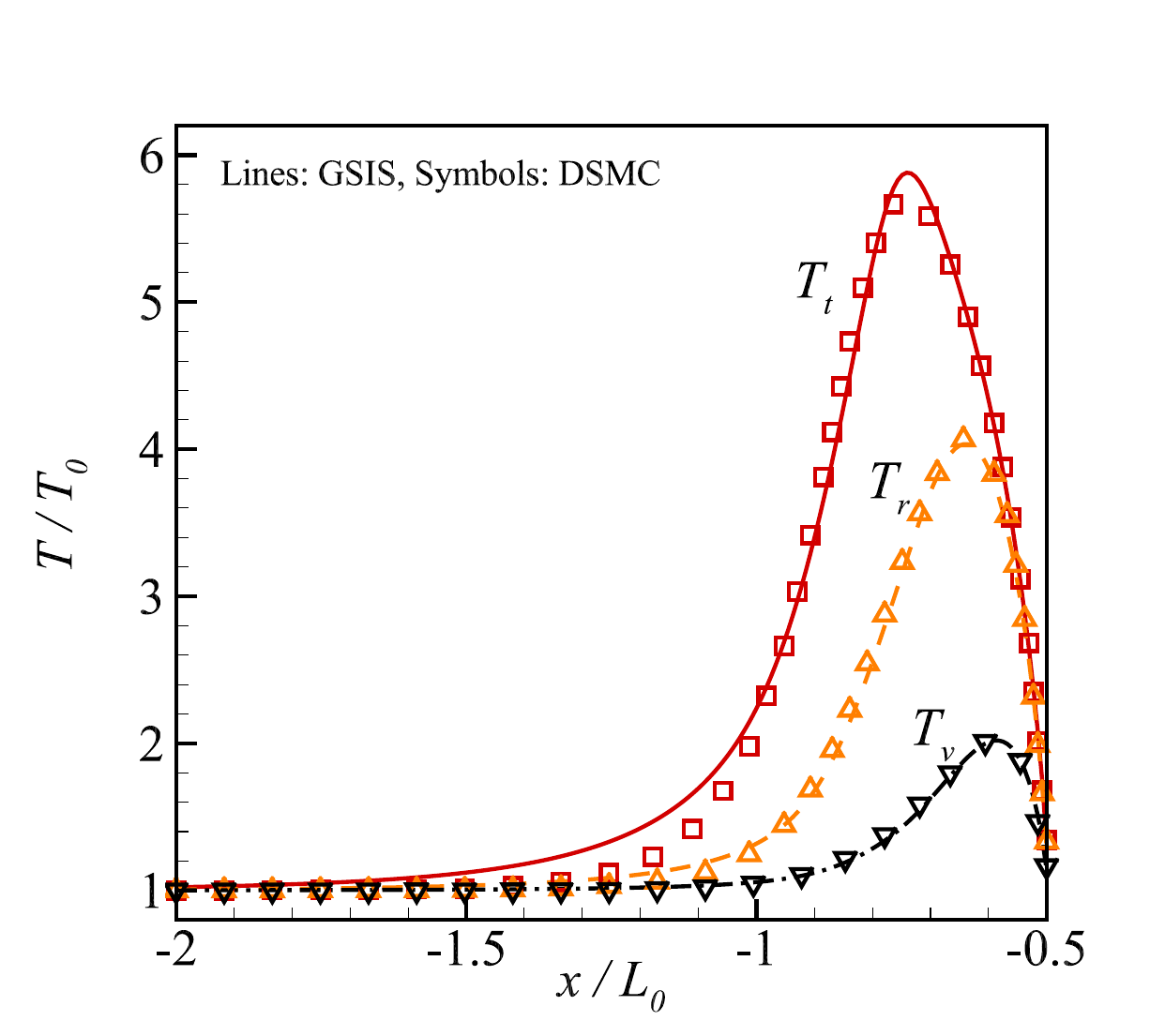}}
    {\includegraphics[scale=0.39,trim={10 10 10 10},clip = true]{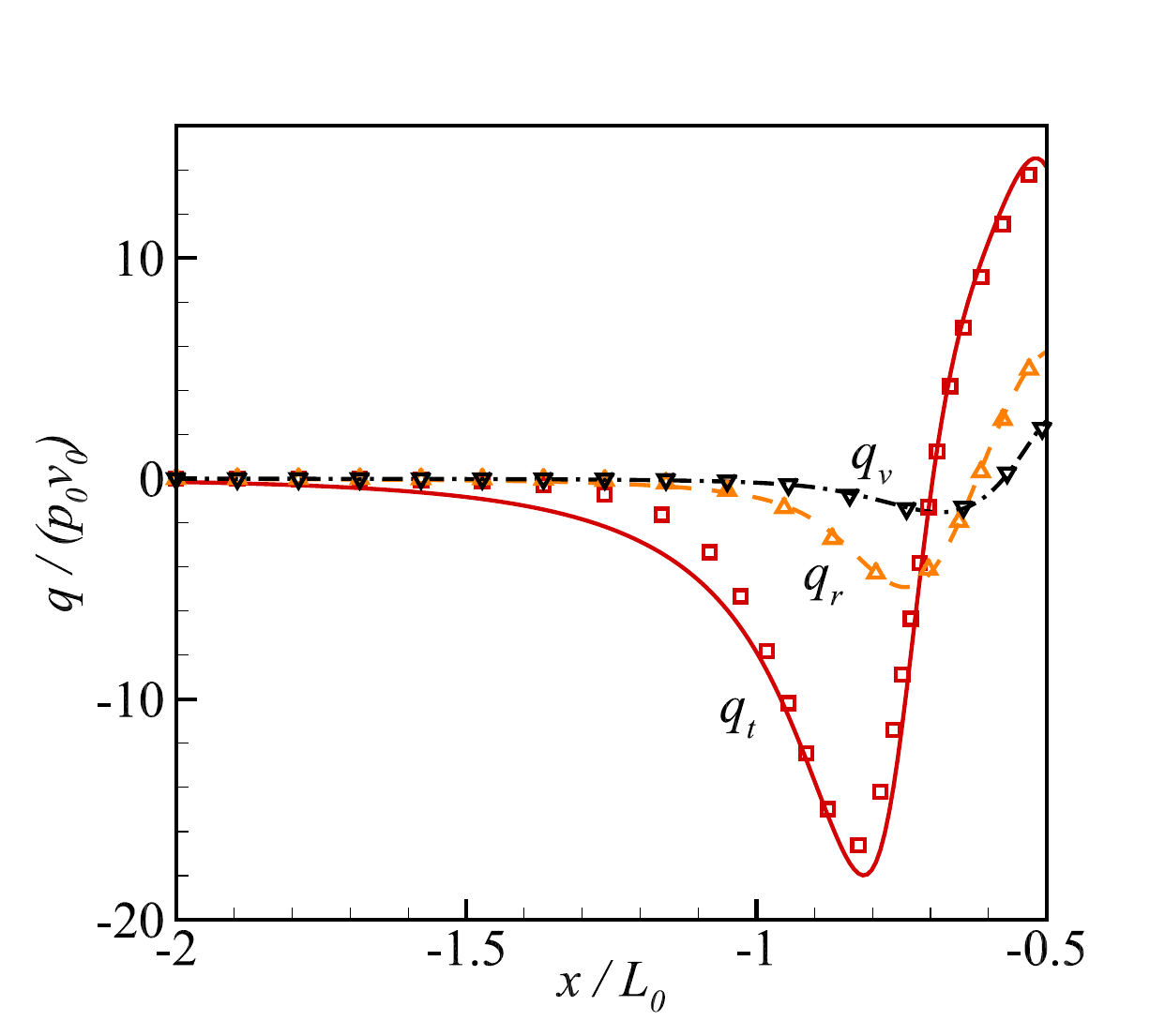}}
    \caption{Comparisons of the dimensionless temperature and heat flux of each gas internal mode between the results obtained by GSIS (lines) and DSMC (symbols) when $\text{Ma}=5$ and $\mathrm{Kn}_{\mathrm{gas}} = 0.1$.}
    \label{fig02:cyd_rykov_line}
\end{figure}

Fig.~\ref{fig:cylindercenterline} compares the GSIS results with the reference solutions obtained using the DSMC method (open-source code SPARTA) under non-radiative conditions. Dimensionless velocity and temperature fields are presented for $\text{Ma} = 5$ at $\mathrm{Kn}_{\mathrm{gas}} = 0.1$ and $1$. In general, good agreement between GSIS and DSMC is achieved, which validates the accuracy of the present method in simulating rarefied hypersonic flows. Furthermore, Fig.~\ref{fig02:cyd_rykov_line} shows the distributions of temperature and heat flux along the stagnation line when $\mathrm{Kn}_{\mathrm{gas}} = 0.1$. Overall, the GSIS predictions are in close agreement with the reference DSMC solution. A noticeable deviation occurs upstream of the shock, where the GSIS gives a higher translational temperature and heat flux. This can be attributed to the tendency of the BGK-type kinetic model to produce an early temperature rise before the shock.

Table \ref{tab:cyd_cmp_step} summarizes the iteration costs for both CIS and GSIS across a wide range of flow regimes, spanning three orders of magnitude variation in the gas Knudsen number. For the CIS method, the number of iteration steps required to reach convergence increases dramatically as $\mathrm{Kn}_{\mathrm{gas}}$ decreases towards the near-continuum regime. Furthermore, for a fixed $\mathrm{Kn}_{\mathrm{gas}}$, the CIS convergence also degrades with a higher relative radiation strength ${\sigma}_R$ and a lower photon Knudsen number $\mathrm{Kn}_{\mathrm{photon}}$. This trend is consistent with the convergence analysis presented in Section \ref{sec:convergence}. This degradation arises from the stronger gas-radiation coupling (smaller $\mathrm{Kn}_{\mathrm{photon}}$), which effectively restricts the allowable numerical time step. In contrast, GSIS demonstrates a significant computational advantage by achieving steady-state solutions within approximately 100 iterations for all tested cases, regardless of the gas rarefaction or radiation coupling strength. This remarkable efficiency highlights the effective role of the macroscopic synthetic equations in rapidly propagating information across the domain. For example, in the case of $\mathrm{Kn}_{\mathrm{gas}} = 0.01$, $\mathrm{Kn}_{\mathrm{photon}} = 1000$ and ${\sigma}_R = 0.1$ (a typical situation in a non-equilibrium hypersonic flow), CIS requires $9582$ steps compared to only $25$ for GSIS, yielding an iteration-step acceleration ratio of ${383.3}$. This acceleration becomes even more significant in the near-continuum regime: for $\mathrm{Kn}_{\mathrm{gas}} = 0.001$, where CIS has a residual decay of only $10^{-5}$ after $32499$ steps, GSIS converges in just $16$ steps, implying an iteration-step acceleration ratio exceeding ${2000}$.

\begin{table}[!h]
\centering
\caption{Convergence step and speedup ratios in CIS and GSIS for the hypersonic flow passing a cylinder at different flow regimes $\mathrm{Kn}_{\mathrm{gas}}$, optical thickness $\mathrm{Kn}_{\mathrm{photon}}$ and relative radiation strength ${\sigma}_R$. The convergence criterion is $\epsilon=10^{-6}$. All simulations are executed on a 64-core CPU.}
\label{tab:cyd_cmp_step}
\begin{threeparttable}  
\begin{tabular}{c c c c c c c c c c}
\hline
\multirow{2}{*}{~$\mathrm{Kn}_{\mathrm{gas}}$~}  &\multicolumn{3}{c}{ ~$\mathrm{Kn}_{\mathrm{photon}}=1000~, {\sigma}_R=0.1$}  &\multicolumn{3}{c}{ ~$\mathrm{Kn}_{\mathrm{photon}}=10~, {\sigma}_R=1$} &\multicolumn{3}{c}{ ~$\mathrm{Kn}_{\mathrm{photon}}=1~, {\sigma}_R=10$}\\ \cmidrule(r){2-4} \cmidrule(r){5-7} \cmidrule(r){8-10}
~  & CIS\tnote{1} & GSIS\tnote{2} & ratios\tnote{3} & CIS & GSIS & ratios &CIS &GSIS & ratios\\ \hline 
$1$& 73      &21 &3.5 & 71  &52  &1.4 & 127  &51 & 2.5\\ 
$10^{-1}$& 499 &24 &20.8 & 519    &102 & 5.1 & 555    &74 & 7.5\\ 
$10^{-2}$& 9582 &25 &383.3 & 10514    &62 & 169.6 & 12420     &45 & 276 \\ 
$10^{-3}$& $\geq$32499 &16 &$\geq$2031 & -    &29 & - & -     &18 & -\\
\hline
\end{tabular}

\begin{tablenotes}
 \footnotesize
 \item[1]  When $\mathrm{Kn}_{\mathrm{gas}} = 10^{-3}, \mathrm{Kn}_{\mathrm{photon}}=1000, {\sigma}_R=0.1$, the residual of CIS only decayed to $10^{-5}$ after $32499$ iterations. Therefore, the iteration numbers for the remaining two cases when $\mathrm{Kn}_{\mathrm{gas}} = 10^{-3}$ are not reported here.
 \item[2] The reported GSIS iteration steps include a 10-step CIS preconditioning phase. Each GSIS iteration consists of one solution step of the mesoscopic equation and 600 inner iterations of the macroscopic synthetic equations.
 \item[3] The ratios represent the acceleration factor in terms of iteration steps. Due to the inclusion of the macroscopic solver, the computational cost per GSIS iteration is approximately $2.1$ times that of a CIS iteration. The overall computational speedup is thus calculated by dividing the iteration-step ratio by this cost factor.
\end{tablenotes}
 \end{threeparttable}
\end{table}

To illustrate the effects of gas-radiation coupling, simulations are performed for conditions with high Mach number $\text{Ma} = 10$ and radiation strength ${\sigma}_R = 0.015$ when $\mathrm{Kn}_{\mathrm{gas}} = 0.1$ and $\mathrm{Kn}_{\mathrm{photon}} = 0.5,100$. It should be noted that the specific parameters adopted here are not intended to represent specific physical conditions, but rather to serve as a challenging scenario to test the numerical stability of the proposed GSIS algorithm under highly coupled, non-equilibrium conditions.

\begin{figure}[t]
    \centering
    \subfloat[]{\includegraphics[scale=0.37,trim={10 10 10 50},clip = true]{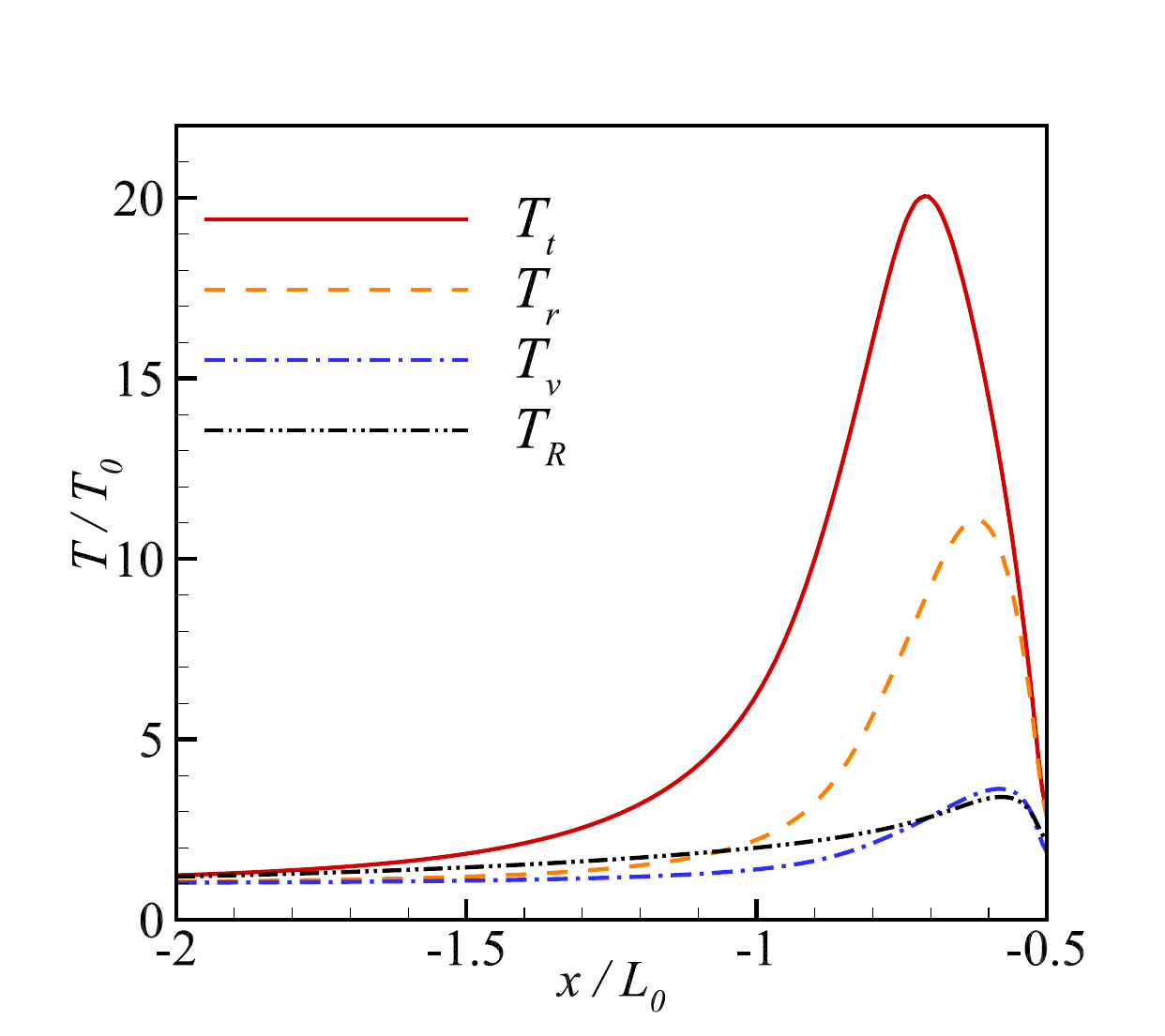}\label{fig:cyd_cmp_temp}}
    \subfloat[]{\includegraphics[scale=0.4,clip = true]{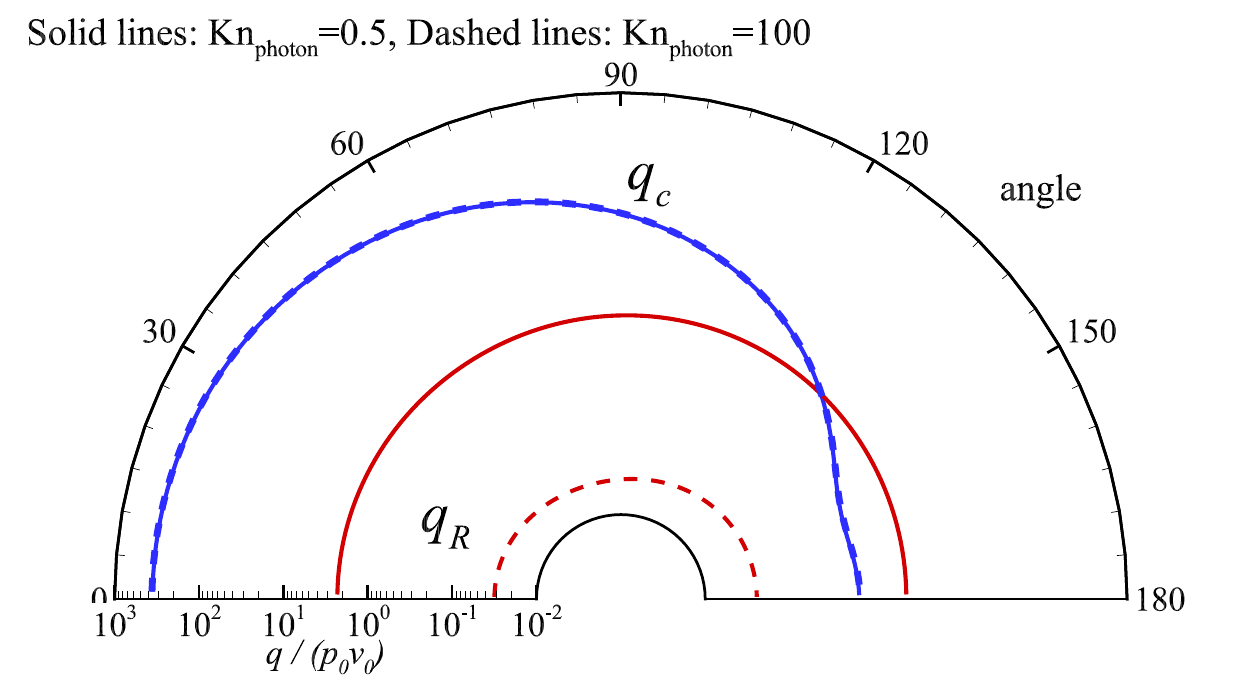}\label{fig:cyd_cmp_flux}}
    \caption{Hypersonic flow over cylinder with radiation effect when $\mathrm{Kn}_{\mathrm{gas}} = 0.1$, $\text{Ma}=10$ and ${\sigma}_R = 0.015$. (a) The temperature distributions along the stagnation line when $\mathrm{Kn}_{\mathrm{photon}} = 0.5$. (b) The heat fluxes from the convective $q_c$ and radiative $q_R$ components along the cylinder surface for two distinct photon Knudsen numbers ($\mathrm{Kn}_{\mathrm{photon}} = 0.5$ and $100$).}
    \label{fig:cyd_cmp}
\end{figure}

Fig.\ref{fig:cyd_cmp_temp} presents the temperature distributions along the stagnation line when $\mathrm{Kn}_{\mathrm{photon}} = 0.5$. Due to the strong shock compression, the internal energy relaxation, as well as the gas-photon coupling, the maximum temperatures exhibit a distinct hierarchy: the translational temperature reaches the highest value $T_t \approx 20.1T_{0}$, followed by rotational $T_r \approx 11.1T_{0}$, vibrational $T_v \approx 3.6T_{0}$, and finally the radiative temperature $T_R \approx 3.4T_{0}$. Upstream of the shock ($x \le -1$), the radiative temperature exceeds the internal gas temperatures, clearly indicating a net energy transfer from the radiation field to the gas in the pre-shock region. The fact that the radiative temperature peak is nearly identical to that of the vibrational mode highlights the direct role of the vibrational mode in the energy exchange with the photon field.

Fig.\ref{fig:cyd_cmp_flux} compares the heat fluxes from the convective and radiative components along the cylinder surface for two distinct photon Knudsen numbers ($\mathrm{Kn}_{\mathrm{photon}} = 0.5$ and $100$). The parameter $\mathrm{Kn}_{\mathrm{photon}}$ characterizes the coupling strength between gas molecules and photons under reference conditions.
\begin{enumerate}
\item Weak Coupling ($\mathrm{Kn}_{\mathrm{photon}} = 100$): When the gas-photon interaction is weak, the radiative intensity is generally weak and the corresponding heat flux remains low and nearly uniform along the entire surface. Crucially, the radiative heat flux is significantly lower than the convective heat flux across the entire wall.
\item Strong Coupling ($\mathrm{Kn}_{\mathrm{photon}} = 0.5$): As $\mathrm{Kn}_{\mathrm{photon}}$ decreases, the interaction strength increases significantly, leading to a much higher overall intensity of the radiation field. This results in a pronounced radiative heat flux, especially in the leeward region. This radiative effect is strong enough that, around $60^{\circ}$ downstream of the rear stagnation point, the radiative heat flux even exceeds the convective heat flux, clearly highlighting the enhanced radiation effects under strong coupling conditions.
\end{enumerate}

\subsection{Hypersonic flow around an Apollo reentry capsule}

To evaluate the efficiency and robustness of GSIS in complex three-dimensional simulations of multiscale non-equilibrium flows, we consider the hypersonic flow over an Apollo reentry capsule. The geometry of the capsule has a length of $3.431$ m and a maximum cross-sectional diameter of $3.912$ m. The simulation adopts a free-stream Mach number of $\text{Ma}=5,~15$, with a freestream temperature of $T_{\infty}=142.2$ K and an isothermal wall temperature of $T_{w}=500$ K.

The physical domain is discretized using $162,000$ hexahedral cells, while the mesh is highly refined near the wall, with the height of the first layer grid set at $10^{-4}$ m. The discretization of the velocity space utilizes an unstructured mesh tailored to the spherical velocity domain, truncated at a radial range of $20 v_0$, involving a total of ${27,704}$ discrete velocity points. For the photon distribution function, spherical coordinates are adopted, with ${72}$ and ${48}$ uniformly distributed grid points in the polar and azimuthal directions, respectively.

\begin{figure}[h]
    \centering
    \subfloat[Mesh overview]{\includegraphics[scale=0.35,clip = true]{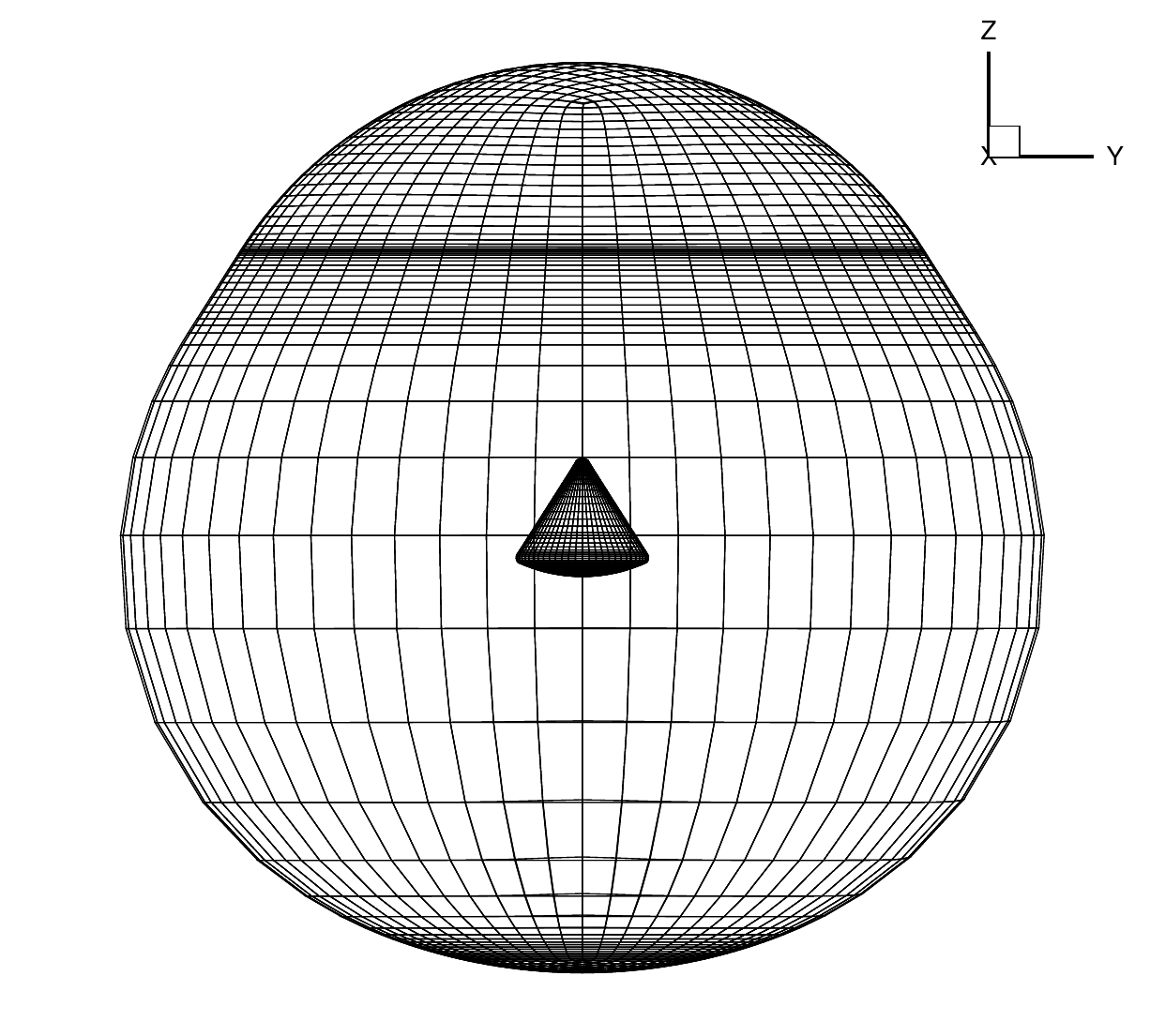}}
    \subfloat[Local enlargement]{\includegraphics[scale=0.35,clip = true]{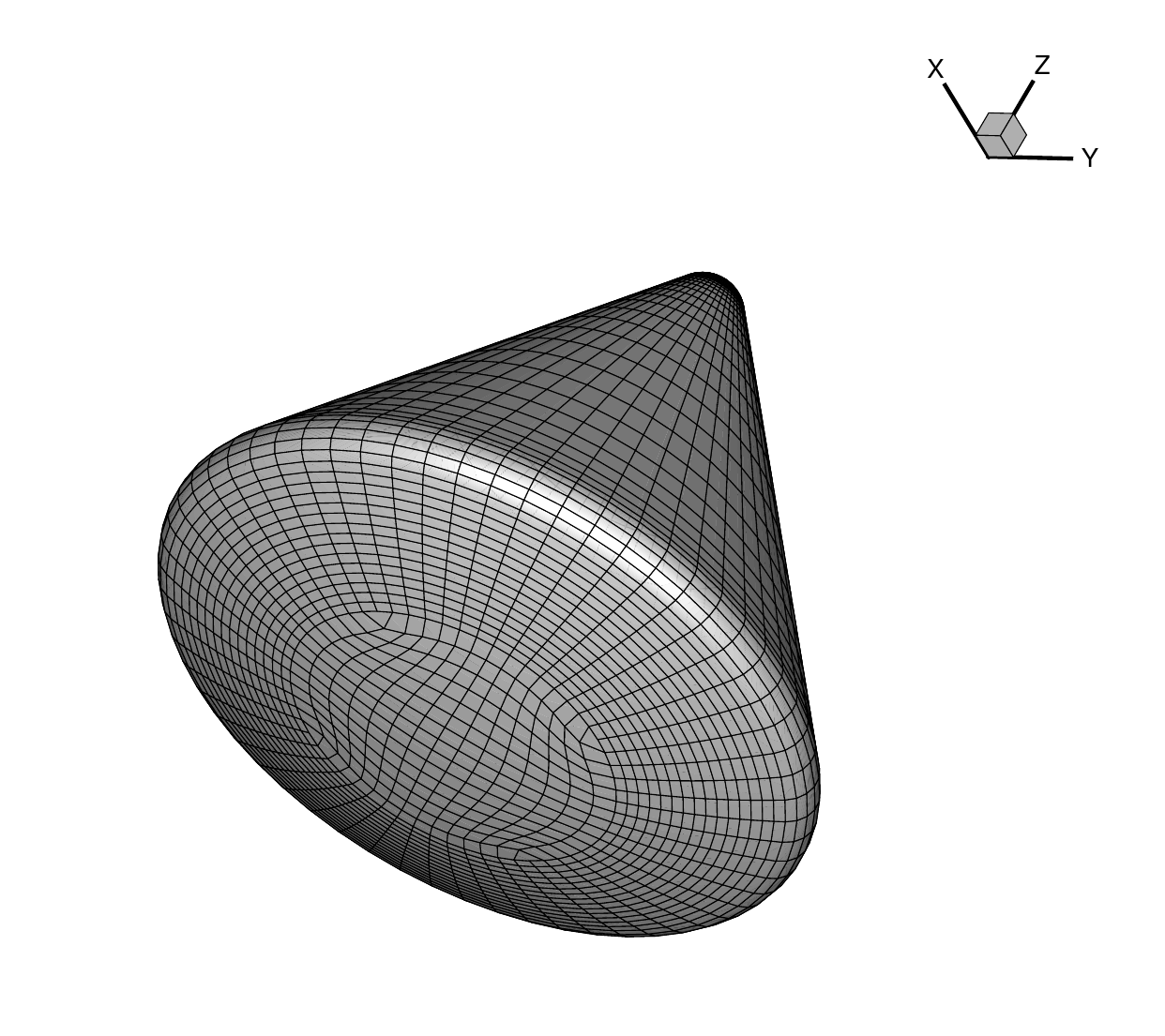}}
    \caption{Computational domain and mesh (hexahedral cells) of the flow over an Apollo capsule.}
    \label{fig03:apollomesh}
\end{figure}

\begin{figure}[t]
    \centering
    \subfloat[Local Mach number]{\includegraphics[scale=0.8,clip = true]{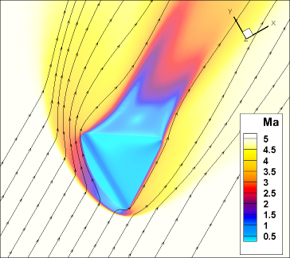}}\;
    \subfloat[Translational temperature]{\includegraphics[scale=0.8,clip = true]{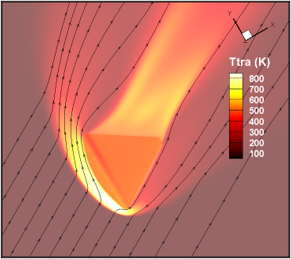}}\\
    \subfloat[Velocity]{\includegraphics[scale=0.48,trim={0 0 20 0},clip = true]{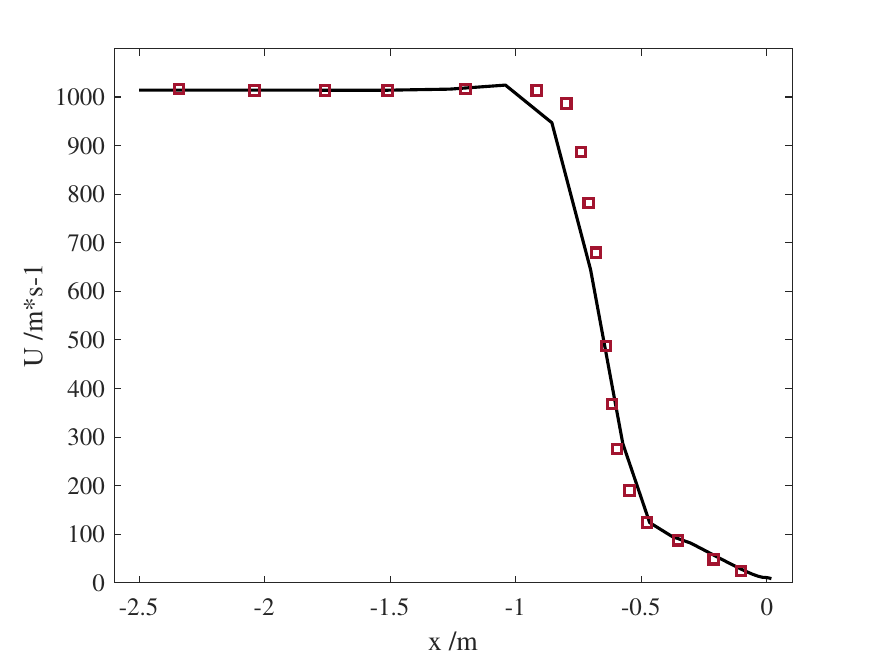}\label{fig:apollo_gsis_ugkwp_u}}
    \subfloat[Temperature]{\includegraphics[scale=0.48,clip = true]{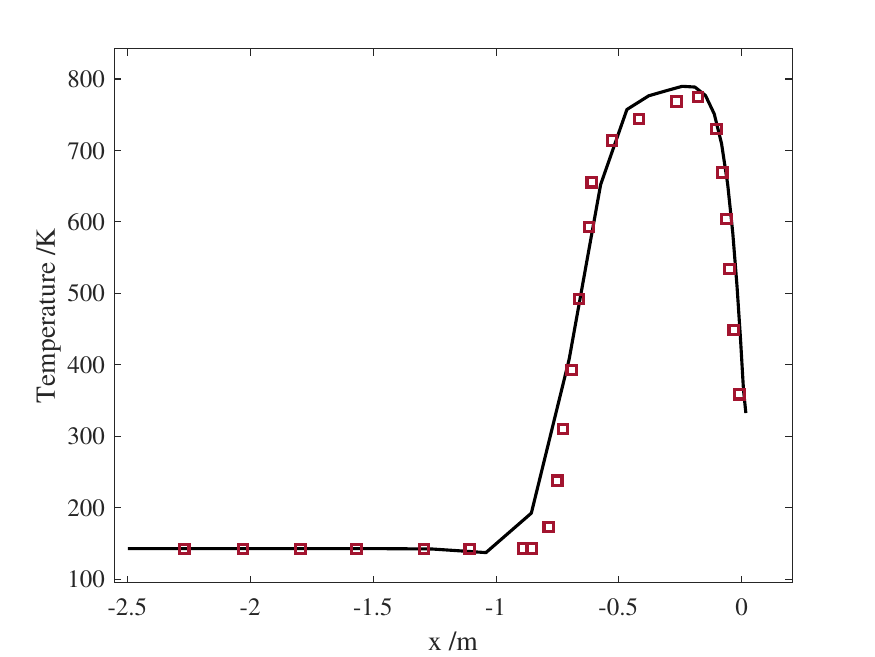}\label{fig:apollo_gsis_ugkwp_T}}
    \caption{Hypersonic flow with a freestream temperature of $T_{\infty}=142.2$ K, $\text{Ma}=5$ and $\mathrm{Kn}_{\mathrm{gas}} = 0.001$ over the Apollo capsule at an angle of attack of ${30^{\circ}}$ with an isothermal wall temperature of $T_{w}=500$ K . The flow field of (a) local Mach number and (b) translational temperature; Comparison of the (c) velocity and (d) temperature between the GSIS (lines) and AUGKWP (symbols) \cite{wei2023adaptive}. The origin $x=0$ is located at the center point of the windward surface.
    }
    \label{fig03:apoll_rykov_contours}
\end{figure}

The accuracy of the GSIS method is first evaluated by considering a non-radiative hypersonic flow at an angle of attack of ${30^{\circ}}$ with a near-continuum gas Knudsen number of $\mathrm{Kn}_{\mathrm{gas}}=0.001$. The numerical solution is compared against results obtained by the adaptive unified gas-kinetic wave-particle (AUGKWP) method \cite{wei2023adaptive}. The flow field is initially generated using the Euler solver with $10,000$ iterations, and GSIS required only ${29}$ iterations from this initial state to reach the convergence criterion. Fig.~\ref{fig03:apoll_rykov_contours} shows the velocity and translational temperature distributions in the flow field (the symmetry plane) and over the capsule surface. A strong bow shock is observed, where the maximum translational temperature reaches almost $800$ K. Fig.~\ref{fig:apollo_gsis_ugkwp_u} and \ref{fig:apollo_gsis_ugkwp_T} further compare the distributions of macroscopic quantities along the windward centerline. The GSIS results are in good agreement with the AUGKWP data \cite{wei2023adaptive}. Slight deviations are observed in the region of $x\in[-1,-0.5]$, which is typically attributed to the BGK-type kinetic model adopted in GSIS.

\begin{figure}[t]
    \centering
    \includegraphics[scale=0.4,clip = true]{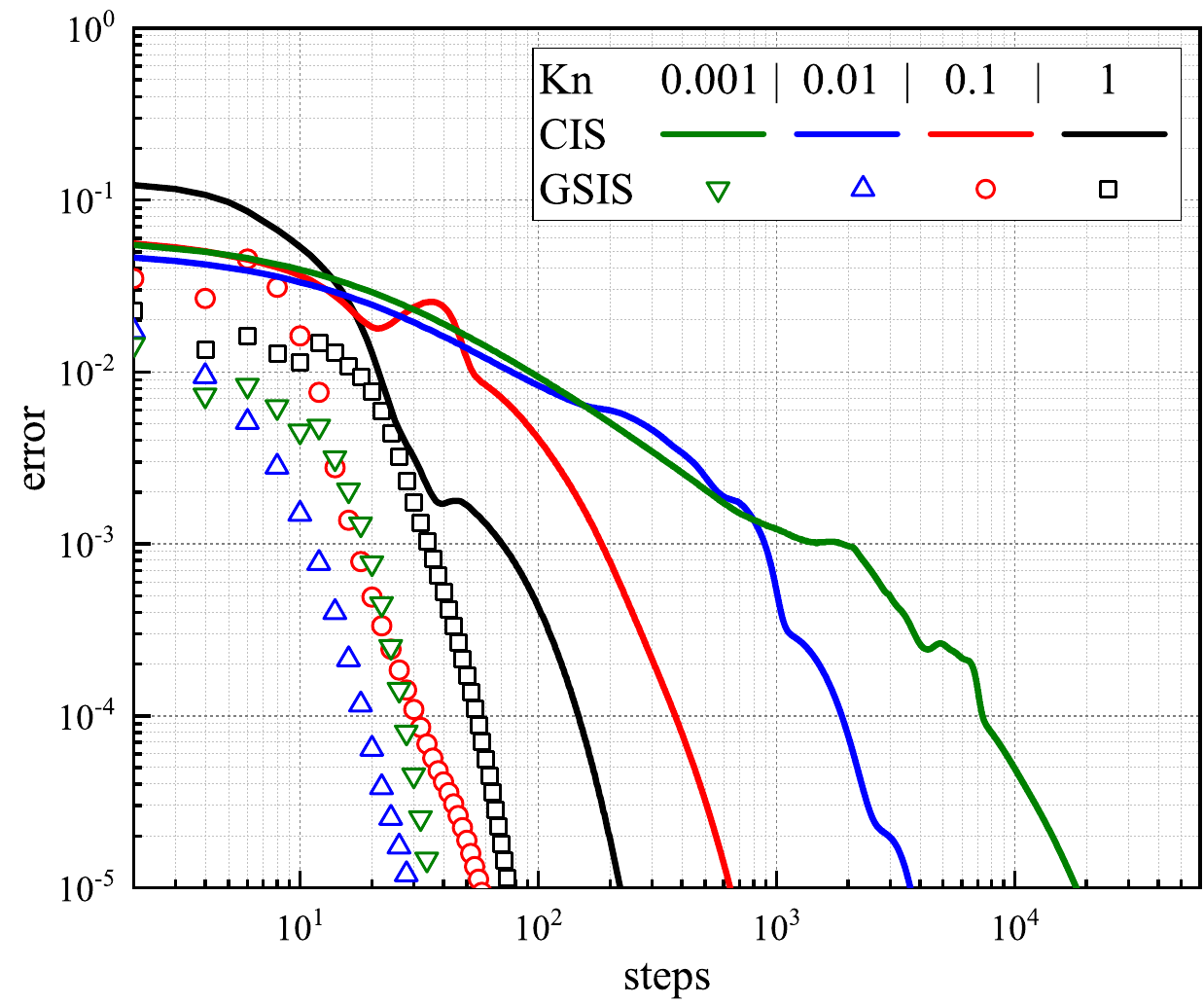}
 \caption{Comparison of residual convergence history between GSIS and CIS for the Apollo capsule flow at various gas Knudsen numbers when $\mathrm{Kn}_{\mathrm{photon}} = 10$, ${\sigma}_R = 0.045$ and $\text{Ma}=15$. The iteration counts exclude the initial $20$ CIS preconditioning steps. }
    \label{fig03:apollo_step_cmp}
\end{figure}

\begin{table}[!h]
    \centering
    \caption{Convergence steps, wall-clock CPU time and overall speedup ratios for GSIS and CIS in the hypersonic flow over the Apollo capsule across varying gas Knudsen numbers ($\mathrm{Kn}_{\mathrm{photon}} = 10$, ${\sigma}_R = 0.045$ and $\text{Ma}=15$). All simulations were performed using 160 CPU cores.}
\begin{threeparttable}

\begin{tabular}{c p{3em} p{6em} p{3em} p{6em} p{8em} }\toprule
\multirow{2}{*}{$\mathrm{Kn}_{\mathrm{gas}}$}   & \multicolumn{2}{c}{CIS} & \multicolumn{2}{c}{GSIS} \ & ~ \\ \cmidrule(r){2-3} \cmidrule(r){4-5}
~ & steps & CPU time (h) & steps\tnote{1} & CPU time (h) & {speedup ratios\tnote{2}}\\ \hline 
1 & 119 & 0.39 & 75 & 0.42 & 1.6 (0.92) \\ 
0.1 & 657 & 1.72 & 58 & 0.33 & 11.3 (5.2)  \\ 
0.01 & 3657 & 9.54 & 29 & 0.18 & 126.1 (53.0) \\ 
0.001 & 18131 & 47.34 & 36 & 0.22 & 503.6 (215.2)  \\ 
\bottomrule
\end{tabular}
\begin{tablenotes}
 \footnotesize
 \item[1] The reported steps include the initial 20 CIS steps for preconditioning. Each GSIS iteration involves one mesoscopic equation solution and $400$ inner iterations of the macroscopic synthetic equations.
 \item[2]  The ratios give the acceleration factors based on the iteration step count and the total computational wall-clock time (shown in parentheses). For example, 503.6 (215.2) means that GSIS required $503.6$ times fewer iteration steps, leading to a $215.2$ times reduction in the total time. The computational time does not include overhead such as memory allocation and initial velocity distribution function setup. 
\end{tablenotes}
\end{threeparttable}
    \label{tab03:apollo_hour_cmp}
\end{table}

\begin{figure}
    \centering
    \subfloat[Translational temperature]{\includegraphics[scale=0.3,clip = true]{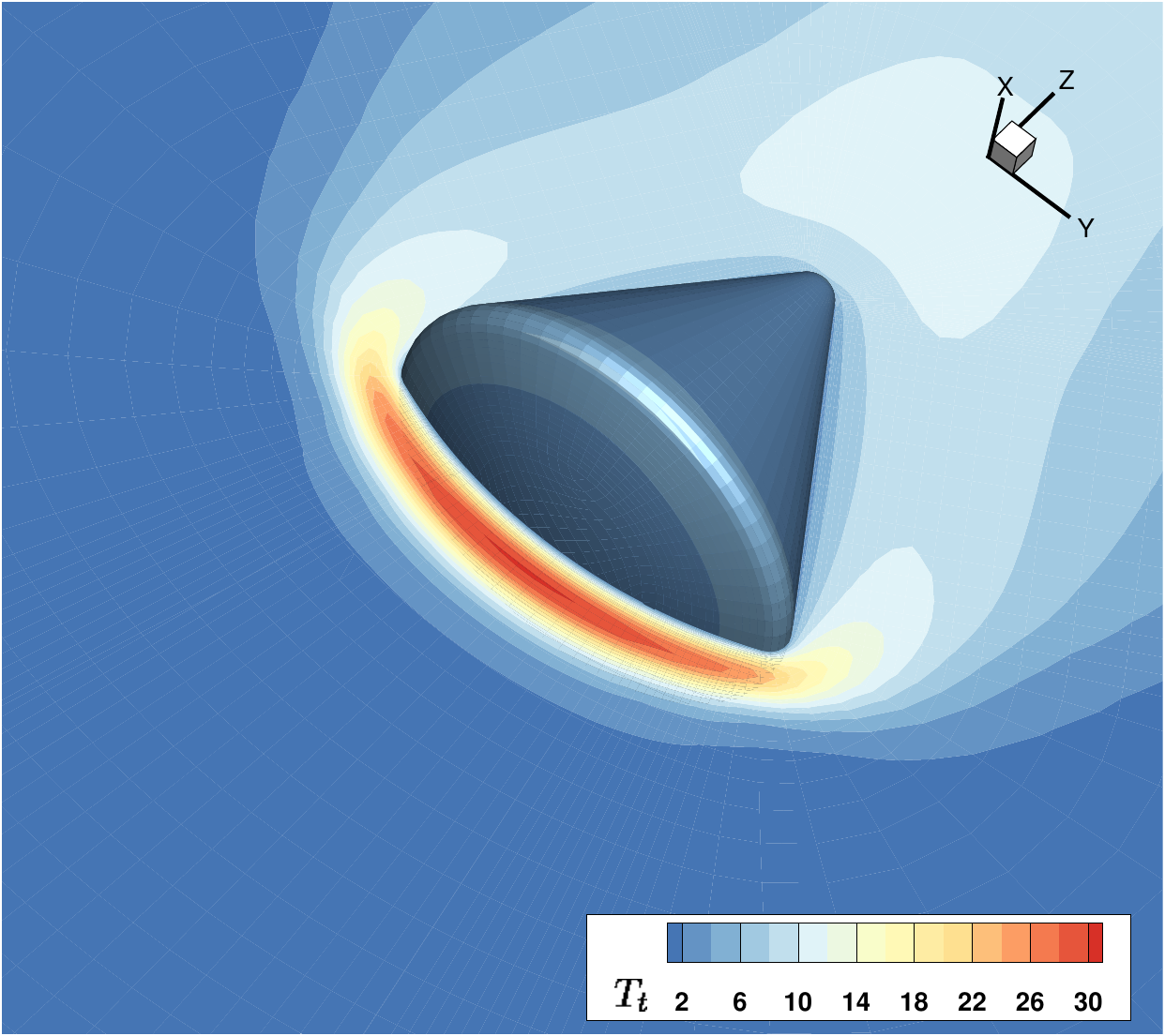}}\;
    \subfloat[Rotational temperature]{\includegraphics[scale=0.3,clip = true]{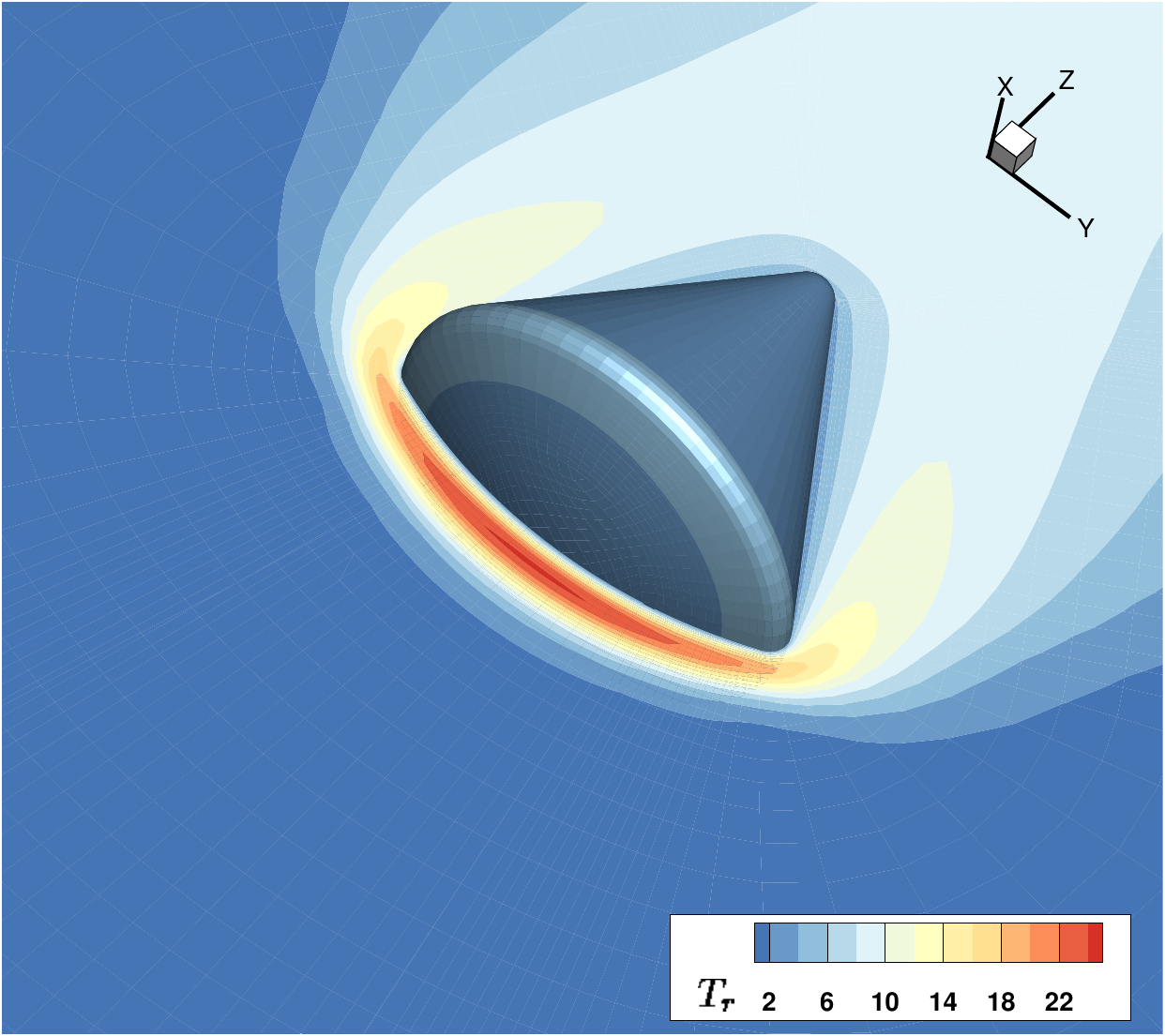}}\\
    \subfloat[Vibrational temperature]{\includegraphics[scale=0.3,clip = true]{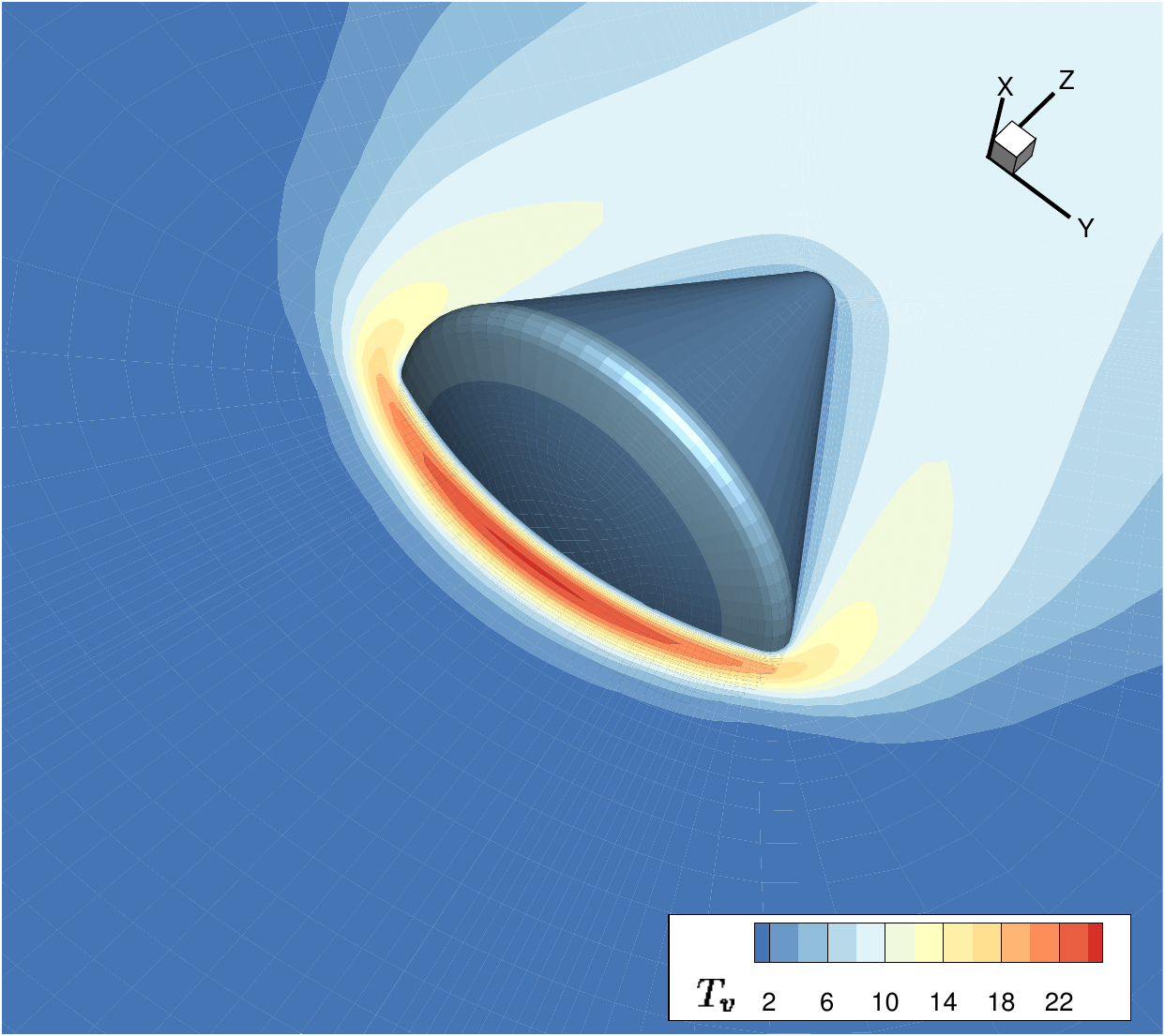}}\;
    \subfloat[Radiative temperature]{\includegraphics[scale=0.3,clip = true]{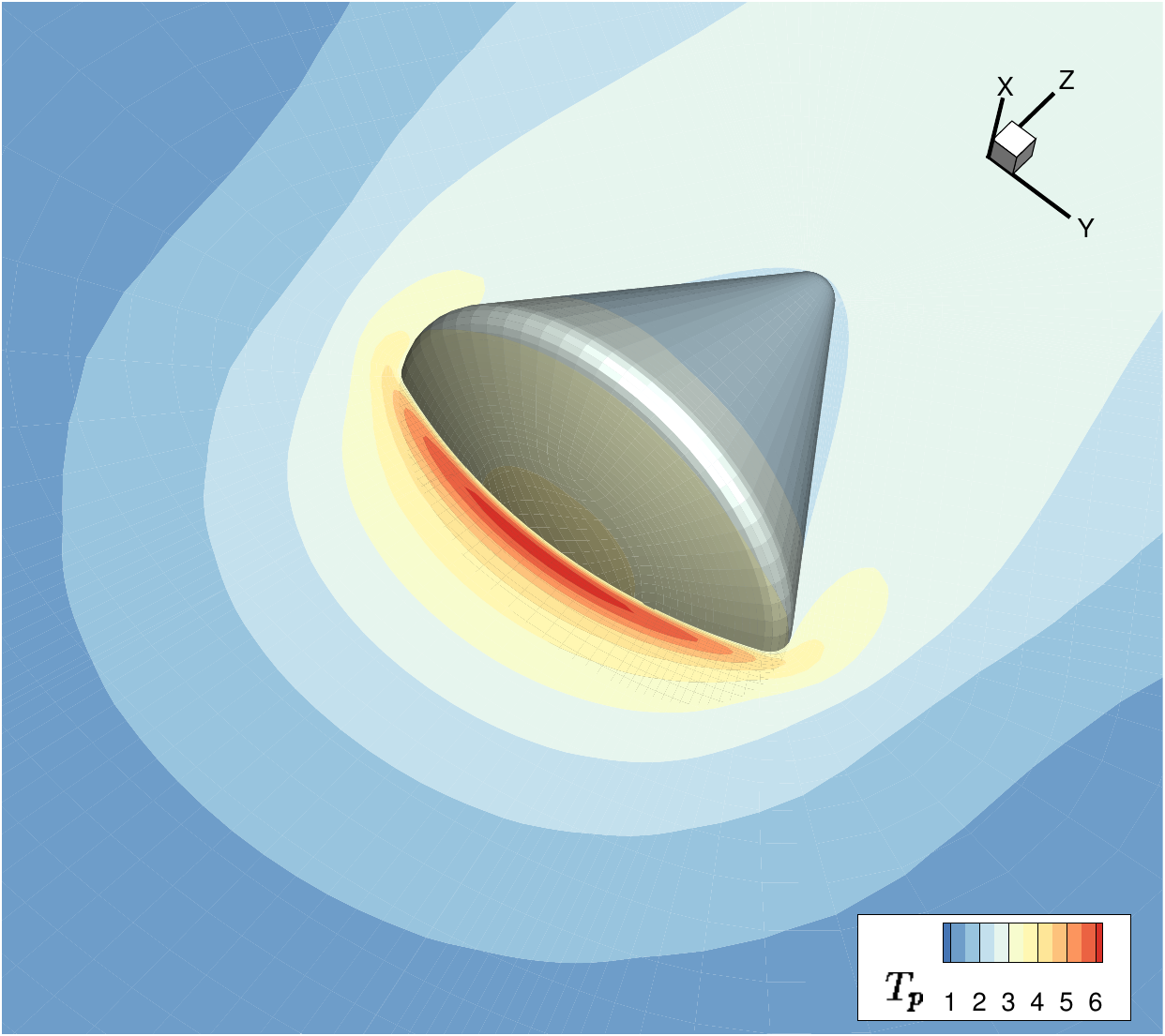}}\\
    \subfloat[Convective heat flux \\ (component in flow direction)]{\includegraphics[scale=0.3,clip = true]{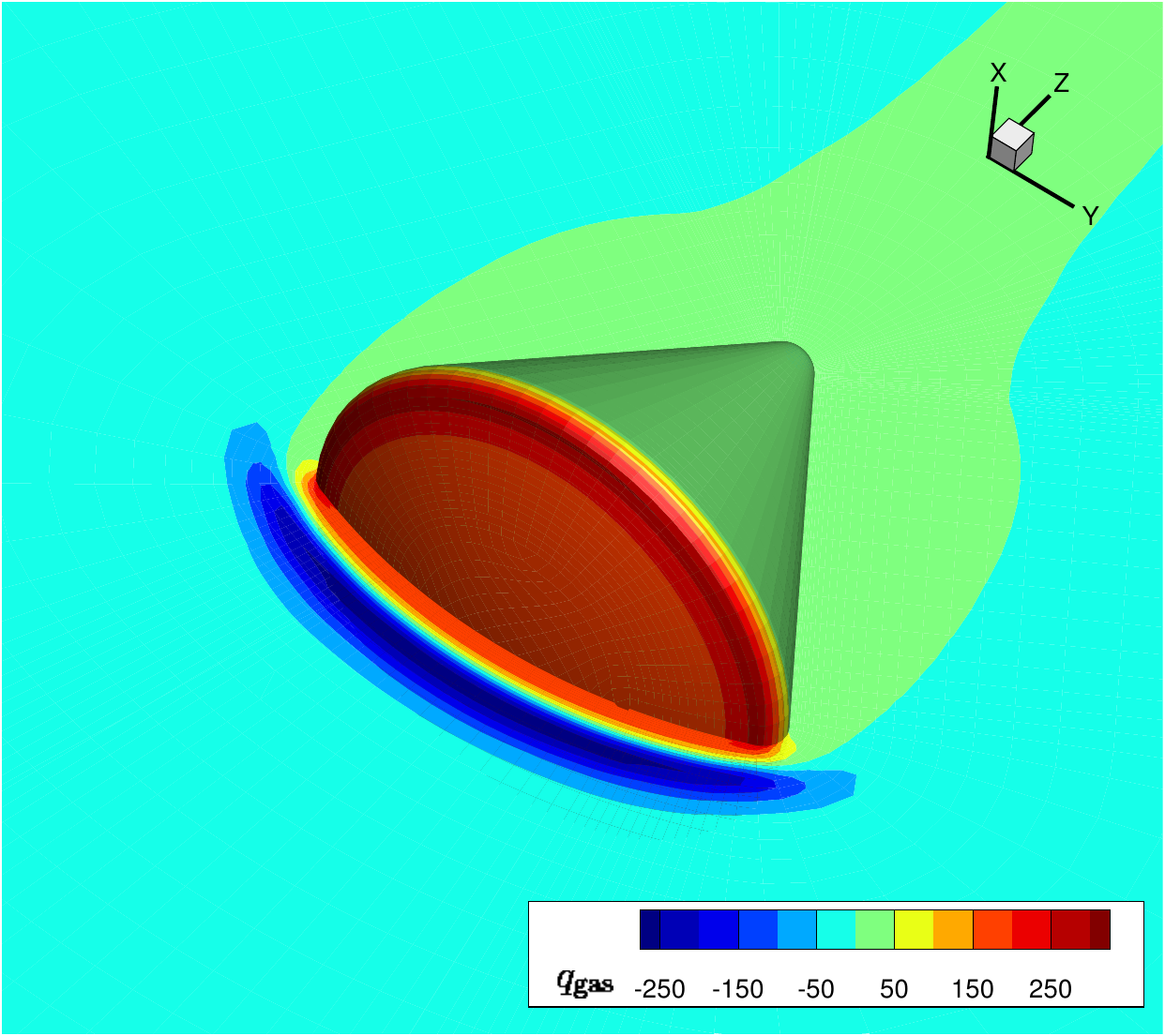}}\;
    \subfloat[Radiative heat flux \\ (component in flow direction)]{\includegraphics[scale=0.3,clip = true]{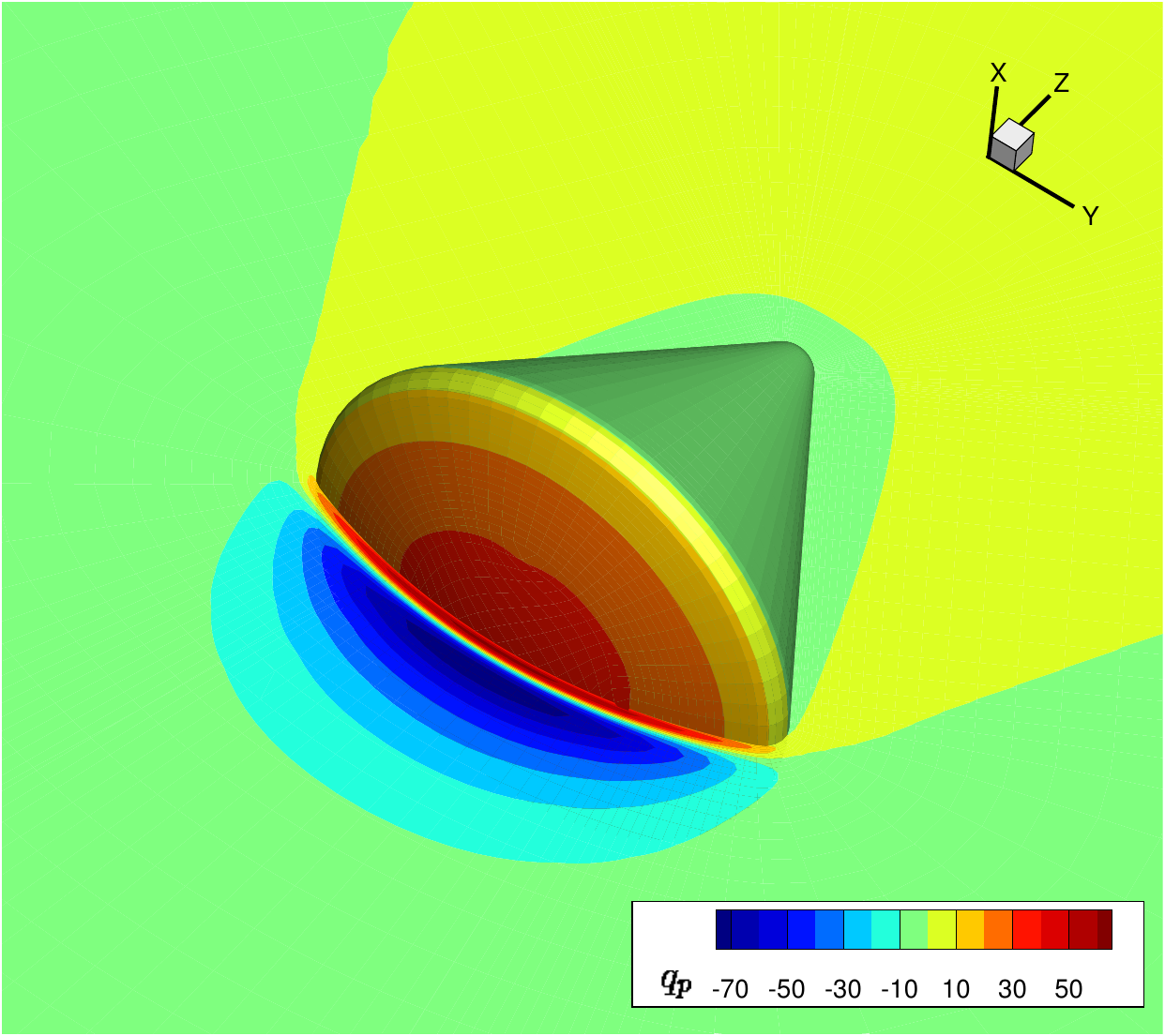}}
    \caption{The distribution of temperatures and heat fluxes  (component in flow direction) under the flow conditions $\mathrm{Kn}_{\mathrm{gas}}=0.5,~ \text{Ma}=15, \mathrm{Kn}_{\mathrm{photon}}=100, \sigma_R = 0.015$.}
    \label{fig03:apoll_rad_t}
\end{figure}

The performance of GSIS for modeling complex non-equilibrium radiative gas flows is further examined under the conditions of $\mathrm{Kn}_{\mathrm{photon}} = 10$, ${\sigma}_R = 0.045$ and $\text{Ma}=15$. Fig.~\ref{fig03:apollo_step_cmp} presents the residual convergence histories for both CIS and GSIS across a wide range of gas Knudsen numbers, where the iteration steps exclude the $20$ initial CIS steps used for preconditioning. Consistent with the previous two-dimensional case, the convergence speed for CIS drops significantly as the flow approaches the near-continuum regime, whereas the GSIS method remains almost unaffected and maintains its rapid convergence throughout. Table~\ref{tab03:apollo_hour_cmp} summarizes the iteration steps and wall-clock CPU time required for convergence on $160$ cores. The data confirm that the CIS iteration steps increase by over two orders of magnitude (from $119$ to $18,131$) when $\mathrm{Kn}_{\mathrm{gas}}$ decreases from $1$ to $0.001$. Conversely, GSIS maintains its efficiency, consistently obtaining converged solutions within ${100}$ iteration steps over this entire range of Knudsen numbers. In terms of total computational cost, the wall-clock time per GSIS iteration is approximately twice that of the conventional CIS scheme, due to the macroscopic solver, and thus the total CPU time cost is slightly higher when $\mathrm{Kn}_{\mathrm{gas}}=1$. However, the significant reduction in the total number of iterations allows the GSIS to achieve an overall wall-clock time speedup of up to approximately ${215}$ times in the near-continuum regime.

Fig.~\ref{fig03:apoll_rad_t} illustrates the influence of radiation effects on the flow field as well as the heat load on the surface of the capsule. The gas translational temperature rises dramatically across the strong shock wave, which is subsequently transferred to the rotational and vibrational internal modes through internal energy relaxation processes. The resultant high vibrational temperature then causes significant radiation effects through radiation transitions. Consequently, within the high-temperature shock layer, the presence of radiative emission acts as a cooling mechanism for the gas, reducing the overall gas enthalpy and limiting the peak temperatures. In the region from the high-temperature shock layer to the windward surface of the capsule, the heating of the capsule is predominantly governed by the convective heat flux. Although the radiative heat flux is substantial in magnitude, its contribution remains a smaller percentage of the total heat load compared to the convective component near the stagnation point.

A particularly interesting observation is made upstream of the shock wave. Due to radiative transport, the energy of the radiation field is slightly higher than the bulk gas temperature in a region relatively far ahead of the shock. In this pre-shock zone, where the gas temperature has not yet begun to rise significantly, the radiation field acts as a heat source, preheating the gas. This phenomenon occurs because the high-energy photons emitted from the shock layer are highly penetrating. The mean free path of these photons is significantly larger compared to the thickness of the shock layer. These photons propagate upstream, effectively transporting energy far into the cold, undisturbed freestream gas before the gas is compressed and heated by the shock wave itself. This radiative energy transfer causes the radiative preheating.

In the expansion region around the capsule's shoulder, where the flow accelerates and pressure drops, the gas translational energy reduces rapidly. However, the vibrational temperature remains at a relatively higher level due to the slow internal energy relaxation rate. This sustained high $T_v$ maintains a relatively high radiative emission. Consequently, the relative importance of the radiative heating effect becomes more pronounced on the capsule's shoulder and afterbody, similar to observations made in previous cases of supersonic flow over two-dimensional cylinders.

%% file: Conclusions.tex
\section{Conclusion}\label{sec:conclusion}

The accurate and efficient simulation of high-temperature gas flows with self-consistent radiation remains a persistent challenge, constrained by the simultaneous manifestation of flow multiscale phenomena (ranging from continuum to rarefied regimes) and radiative transport multiscale phenomena (spanning optically thick to optically thin regimes). We developed and implemented a highly efficient general synthetic iterative scheme tailored for simulating multiscale rarefied gas dynamics coupled with high-temperature radiation. The core ingredient lies in the two-way coupling between the mesoscopic kinetic and macroscopic synthetic descriptions. The solutions from the mesoscopic kinetic equations provide exact moment-based  closure terms for the macroscopic equations. The solutions from the macroscopic synthetic equations (density, velocity, temperatures of various internal energy modes, and radiative energy density) directly guide the rapid evolution of the distribution functions towards the steady state. 

A rigorous Fourier stability analysis is conducted to compare the performance of the conventional mesoscopic algorithm with the proposed GSIS in the implicit solution of steady-state gas-radiation coupled problems. It is found that conventional methods, which rely on the explicit coupling of macroscopic quantities between the two physical fields, suffer from extremely slow convergence in the near-continuum regime and may diverge when strong radiation effects are coupled. In contrast, the GSIS method, by incorporating the solution of the synthetic equations within the iterative loop, demonstrates a significant enhancement in computational efficiency under near-continuum and optically thick conditions. The spectral radius of the iteration matrix consistently remained less than 0.5, indicating highly accelerated convergence. On the other hand, we also demonstrate that this framework exhibits desirable asymptotic-preserving property, thus circumventing the prohibitive mesh-resolution constraints in conventional kinetic methods. GSIS recovers the macroscopic limit of radiative NSF equations on coarse grids as Knudsen numbers of both gas and photon approach zero, enabling practical simulations with spatial cells orders of magnitude larger than the mean free path.

We validated the proposed GSIS algorithm by simulating challenging problems incorporating high-temperature gas radiation effects, including the normal shock wave, lid-driven cavity flow, flow past a cylinder, and three-dimensional flow around an Apollo reentry capsule. The proposed scheme consistently obtained steady-state solutions within 100 iterations for all the above cases, where the gas Knudsen number varied from $10^{-3}$ to $1$, the photon Knudsen number varied from $0.1$ to $1000$, and the relative radiation strength ranged from $0.01$ to $10$. In the near-continuum regime, the GSIS achieved an efficiency advantage of two orders of magnitude over the conventional approach. Furthermore, studies on the radiative heating of hypersonic flow past bodies provided useful physical insights, revealing preheating of the upstream flow due to radiative energy transport, and highlighting the important contribution of radiative heating to the total heat flux in the expansion and leeward region.

In conclusion, this synthetic-accelerated method provides a unified and highly effective framework for modeling non-equilibrium radiative transport phenomena in high-temperature rarefied gas flows. It provides an advancement in computational tools necessary for designing and analyzing of vehicles undergoing planetary atmospheric reentry, and exploring extreme-environment applications where non-equilibrium radiation governs the physics.

\section*{Declaration of competing interest}
The authors declare that they have no known competing financial interests or personal relationships that could have appeared to influence the work reported in this paper.

\section*{Acknowledgments}
This work is supported by the National Natural Science Foundation of China (12572381). Special thanks are given to the Center for Computational Science and Engineering at the Southern University of Science and Technology.


%% file: Appendix_sourceterm_treatment.tex
\section{Implicit treatment of source terms in macroscopic equations}\label{appendix:source_term}

Taking the 2D two-temperature macroscopic equation \eqref{eq:macroscopic_equation} as an example, the conservative variables can be obtained as:
\begin{equation}
\begin{aligned}
\bm{W}&=
\begin{bmatrix}
\rho \\
\rho u_x \\
\rho u_y \\
e\\
e_{r}\\
e_{v}\\
e_{R}
\end{bmatrix},\quad 
\bm{F}_c=
\begin{bmatrix}
\rho u_n\\
\rho u_xu_n+n_xp_{t}\\
\rho u_yu_n+n_yp_{t}\\
u_n (e_{gas}+p_{t})\\
u_n e_{r}\\
u_n e_{v}\\
0
\end{bmatrix}, \quad 
\bm{F}_v=
\begin{bmatrix}
0\\
n_x\sigma_{xx}+n_y\sigma_{xy}\\
n_x\sigma_{yx}+n_y\sigma_{yy}\\
n_x\Theta_{x}+n_y\Theta_{y}\\
n_xq_{x,r}+n_yq_{y,r}\\
n_xq_{x,v}+n_yq_{y,v}\\
n_xq_{x,R}+n_yq_{y,R}
\end{bmatrix},\\
\medskip
\bm{Q}&=
\left[
0,0,0,0,{\frac{e_{tr}-e_{r}}{Z_r\tau}},
{\frac{e_{tv}-e_{v}}{Z_v\tau}}-{\frac{e_{vR}-e_{R}}{\Kn_{\mathrm{photon}}}},
{\frac{e_{vR}-e_{R}}{\Kn_{\mathrm{photon}}}}
\right]^{\mathsf{T}},
\end{aligned}
\end{equation}
with
$\Theta_{x} =u_x\sigma_{xx}+u_y\sigma_{xy}+q_{x}$ and
$\Theta_{y} =u_x\sigma_{yx}+u_y\sigma_{yy}+q_{y}$. 
Here, $u_n = u_x n_x + u_y n_y$ is defined as the scalar product of the macro-velocity vector and the unit normal vector of the face. The total energy in the macroscopic system is obtained by adding its internal energy to its kinetic energy. 

We provide an approximation procedure for the source term Jacobi matrix in the Cartesian coordinate system. It should be noted that the rotational collision numbers $Z_r$ and degree of freedom $d_v$ used here are all constants. If those parameters are temperature dependent, a new derivation would be required. The Jacobi matrix $T$ represents the derivative of the source term $Q$. The matrix can be further approximated by only keeping its main diagonal elements, which leads to diagonalization of the matrix as follows:
\begin{equation}
\begin{aligned}
\mathbf{T} = \text{diag} \left( \frac{\partial \mathbf{Q}}{\partial \mathbf{W}} \right) &= \text{diag} \left(0,0,0,0, \frac{\partial Q(5)}{\partial W(5)}, \frac{\partial Q(6)}{\partial W(6)}, \frac{\partial Q(7)}{\partial W(7)}\right),
\end{aligned}
\end{equation}
where
\begin{equation}
\begin{aligned}
\frac{\partial Q(5)}{\partial e_R}&=\frac{\partial \left(\frac{d_r}{2}\frac{\rho(T_{tr}-T_{r})}{Z_r\tau}\right)}{\partial \left(\frac{d_r}{2}\rho T_{r}\right)}=\frac{1}{Z_r}\frac{\partial \left(\frac{\rho(T_{tr}-T_{r})}{\tau}\right)}{\partial \left(\rho T_{r}\right)}\\
&=\frac{1}{Z_r}\left[\frac{\partial \left(\rho T_{tr}-\rho T_{r}\right)}{\partial (\rho T_{r})}\cdot\frac{1}{\tau}-\frac{\rho(T_{tr}-T_{r})}{\tau^2}\cdot\frac{\partial \tau}{\partial T_{t}}\cdot\frac{\partial T_{t}}{\partial (\rho T_{r})}\right]\\
&=-\frac{3}{(3+dr)Z_r\tau}+\frac{2(\omega-1)(T_{tr}-T_{r})}{3Z_r\tau T_{t}},
\end{aligned}
\end{equation}
\begin{equation}
\begin{aligned}
\frac{\partial Q(6)}{\partial e_v}&=\frac{\partial \left(\frac{d_v}{2}\frac{\rho(T_{tv}-T_{v})}{Z_v\tau}\right)}{\partial \left(\frac{d_v}{2}\rho T_{v}\right)}=\frac{1}{Z_v}\frac{\partial \left(\frac{\rho(T_{tv}-T_{v})}{\tau}\right)}{\partial (\rho T_{v})}\\
&=\frac{1}{Z_v}\left[\frac{\partial \left(\rho T_{tv}-\rho T_{v}\right)}{\partial (\rho T_{v})}\cdot\frac{1}{\tau}-\frac{\rho(T_{tv}-T_{v})}{\tau^2}\cdot\frac{\partial \tau}{\partial T_{t}}\cdot\frac{\partial T_{t}}{\partial (\rho T_{v})}\right]\\
&=-\frac{3}{(3+dv)Z_v\tau}+\frac{2(\omega-1)(T_{tv}-T_{v})}{3Z_v\tau T_{t}},
\end{aligned}
\end{equation}
\begin{equation}
\begin{aligned}
\frac{\partial Q(7)}{\partial e_R} &= \frac{\partial }{\partial e_R}\left(\frac{e_{vR}-e_R}{\Kn_{\mathrm{photon}}}\right)\\
&=\frac{1}{\Kn_{\mathrm{photon}}}\left(\frac{\partial e_{vR}}{\partial e_R}-1\right)=\frac{1}{\Kn_{\mathrm{photon}}}
\left(\frac{\partial }{\partial e_R}e_R\left(\frac{T_R}{T_v}\right)^4-1\right)\\
&=\frac{({T_R}/{T_v})^4-1}{\Kn_{\mathrm{photon}}}.
\end{aligned}
\end{equation}

These derivative terms can be expressed as:
\begin{equation}
\frac{\partial T_t}{\partial e_R}=-\frac{2}{3\rho},\quad \frac{\partial (\rho T_{tr})}{\partial ( \rho T_r)} = \frac{d_r}{3+d_r}, \quad \frac{\partial (\rho T_{tv})}{\partial ( \rho T_v)} = \frac{d_v}{3+d_v}, \quad \frac{\partial \tau}{\partial T_t}=(\omega-1)\frac{T_t^{\omega-2}}{\rho}.
\end{equation}

The approximate Jacobi matrix indicates that the non-principal diagonal elements are all zero, and the solution of the rotational energy relaxation equation is approximated by decoupling.